\documentclass[pra,onecolumn,superscriptaddress,nofootinbib]{revtex4}
\usepackage[a4paper, left=2cm, right=2cm, top=2cm, bottom=2cm]{geometry}

\usepackage[english]{babel}
\usepackage[utf8]{inputenc}
\usepackage[T1]{fontenc}
\usepackage{amsthm}
\newtheorem{thm}{Theorem}
\newtheorem{cor}{Corollary}[thm]
\newtheorem{lem}{Lemma}[thm]

\newtheorem{prop}{Proposition}[thm]
\newtheorem{eg}{Example}[thm]

\newtheorem{fact}{Fact}

\usepackage{comment}

\usepackage{amsmath}
\usepackage{graphicx}
\usepackage[colorinlistoftodos]{todonotes}
\usepackage[colorlinks=true, allcolors=blue]{hyperref}
\usepackage{bbm,microtype,mathrsfs,amsmath,amssymb,color,amsthm,graphicx,cleveref,bm}
\usepackage{tikz-cd}
\usepackage{xcolor}
\usepackage{braket}
\usepackage{natbib}

\newcommand{\BE}{\mathbb{E}}
\newcommand{\CF}{\mathcal{F}}

\newcommand{\CL}{\mathcal{L}}

\newcommand{\CO}{\mathcal{O}}

\newcommand{\BP}{\mathbb{P}}

\newcommand{\lV}{\lVert}
\newcommand{\rV}{\rVert}

\newcommand*{\tr}{\mathrm{Tr}}

\newcommand{\vertiii}[1]{{\left\vert\kern-0.25ex\left\vert\kern-0.25ex\left\vert #1 
    \right\vert\kern-0.25ex\right\vert\kern-0.25ex\right\vert}}

\begin{document}
\title{Concentration of OTOC and Lieb-Robinson velocity in random Hamiltonians}
\author{Chi-Fang Chen
\\\small Institute for Quantum Information and Matter 
\\\small  California Institute of Technology, Pasadena, CA, USA
}

\begin{abstract}
    The commutator between operators at different space and time has been a diagnostic for locality of unitary evolution. Most existing results are either for specific tractable (random) Hamiltonians (Out-of-Time-Order-Correlators calculations), or for worse case Hamiltonians (Lieb-Robinson-like bounds or OTOC bounds).
    
    In this work, we study commutators in \textit{typical} Hamiltonians. Draw a sample from any zero-mean bounded independent random Hamiltonian ensemble, time-independent or Brownian, we formulate concentration bounds in the spectral norm and for the OTOC with arbitrary non-random state.
    Our bounds hold \textit{with high probability} and scale with the sum of interactions squared. Our Brownian bounds are compatible with the Brownian limit while deterministic operator growth bounds must diverge. We evaluate this general framework on short-ranged, 1d power-law interacting, and SYK-like k-local systems and the results match existing lower bounds and conjectures. Our main probabilistic argument employs a robust  matrix martingale technique called uniform smoothness and may be applicable in other settings.

\end{abstract}

\maketitle
\date{}							


\section{Introduction}
The notion of dynamic locality in Hamiltonian often boils down to the commutator between operators at different space and time. For a 1-D nearest-neighbour Hamiltonian, a bound was first formulated by Lieb and Robinson and later sharpened to the form ~\cite{Lieb1972,Hastings_koma,Bravyi2006,chen2019operator, D_A_graph_GU_LUcas,PRXQuantum.1.010303} \vspace{-0.2cm}
\begin{align}
    \lV [O_0(t),A_r]\rV \le \frac{(2h t)^r}{r!}\le \CO(e^{vt-r}),\label{eq:LRbound}\vspace{-0.2cm}
\end{align}
where $O_X(t): = e^{iHt}O_0e^{-iHt}$ is the time-evolved operator, $h=\lV H_{i,i+1}\rV$ the 2-body interaction strength, $r$ the lattice distance between local operators $O_0,A_r$, and $\lV\cdot \rV := \lV\cdot \rV_\infty$ the spectral norm or the operator norm. Reminiscent of causality between events in relativistic systems, the commutator decreases (super-)exponentially outside the "light cone" $r-vt>0, v = 2eh$, which operationally constrains the speed of information propagation between distant local sites. Many other notions of locality, dynamic or static, can be derived from the Lieb-Robinson bound such as the decay of correlation and tensor network description of ground state~\cite{Bravyi2006,PhysRevB.73.085115}, efficient classical or quantum simulation of Hamiltonian evolution~\cite{haah2020quantum}, the stability of topological order~\cite{Bravyi_2010}, etc. The Lieb-Robinson bound is also appreciated for its generality: arbitrary nearest neighbor (including time-dependent) Hamiltonian must be constrained by Eq.~\eqref{eq:LRbound} and thus its consequences. Though, the celebrated generality is at the same time a \textit{worst case} analysis via recursive applications of the triangle inequality of the spectral norm\cite{Lieb1972,Bravyi2006,chen2019operator,PRXQuantum.1.010303}. Certain time-independent Hamiltonian\footnote{Consider XY model, it essentially becomes a 1-D quantum walk~\cite{Konno_2005}, where explicit late time calculation is possible. } can be fine-tuned to match this worst case velocity in late times up to small prefactors $(v=4h < 2eh )$. The above line of thought carries over to other systems whenever a Lieb-Robinson-like bound holds, such as power-law interacting systems~\cite{strictlylinear_KS,alpha_3_chenlucas,Tran_2019_polyLC}.

Another intimately related quantity is the \textit{Out-of-time-Order-Correlator}(OTOC)~\cite{stephen14}
\begin{align}
   \tr\bigg( \rho [O(t), A ]^\dagger [O(t), A] \bigg),
\end{align}
where $O, A$ are initially 'simple' operator and $\rho$ the Gibbs state at some temperature (or the maximal mixed state for convenience). This form of commutator is central to quantum many-body chaos~\cite{Maldacena_2016} and recently attracts attentions from high energy physicists. There, a correlator of a thermal state is arguably more physical and even measurable in lab~\cite{powerlaw_experiment}. Though, most models are nothing like local spin chains and hardly solvable, such as the renowned Sachdev-Ye-Kiteav~\cite{Sachdev_1993,SYK} model for holography or field theoretical models that requires perturbative/diagramatic/large-$N$/asymptotic treatments~\cite{Maldacena_2016,Gu_2017}. The dependence on background state further makes the calculation specific to the model. In comparison, the Lieb-Robinson bounds aim for an \textit{inequality} that concerns the fine-tuned Hamiltonians and the worst (presumably unphysical) input state. This general (worst) case analysis makes it model unspecific and a generally applicable subroutine, at least for spin lattices. Recently, with Lieb-Robinson-like mindset and techniques some bounds was derived for OTOC with the maximally mixed state~\cite{Kuwahara_OTOC,hierarachy}.


The central theme of this work concerns general bounds for \textit{typical} Hamiltonians, which is not addressed by the above worst case bounds or model-specific calculation. Though, quantum system in the wild are typically non-integrable or without a transparent effective theory, so to prove anything we may have to interpret typicality differently. A common strategy is through statistics: study (time-independent) Hamiltonians with parameters drawn at \textit{random} under the constraint of some universality class such as nearest-neighbor, long-ranged, or with certain conserved quantities and symmetries. Now, typicality becomes well-defined if the desired quantity \textit{concentrates}, i.e. the vast majority of random samples enjoy a similar characterization. The idea of 'randomizing' Hamiltonian dated back to random matrix models for nuclear Hamiltonians (See, e.g.~\cite{nuclear_random}) and recently resurfaced in holography and many-body chaos (See, e.g.~\cite{saad2019jt,stanford2020jt}).
In general though, analytic control of time-independent evolution of random Hamiltonians have been out of reach.  

One common and fruitful alternative is to study random \textit{time-dependent} Hamiltonian (discrete random circuits or continuous Brownian circuits). Time-dependent circuits need not capture the static locality such as ground state correlation, but still it often serves as a proxy for dynamic locality for generic (chaotic) \textit{time-independent} Hamiltonian. 
Some insightful toy models include the discrete 1D-local Haar random unitary (See, e.g.~\cite{Brand_o_2016,hunterjones2019unitary,dalzell2020random}) and continuous all-to-all interacting systems~\cite{brownian_SYK,Lashkari_2013,S_nderhauf_2019}. Time-dependent random circuits in itself is a natural model for stochastic noises in quantum systems and an interesting mathematical subject (See e.g.,~\cite{Onorati_2017} and the reference therein for the continuum regime).

Now that we have a random unitary ensemble in mind (time-independent or not), we can sharpen our quests in statistical terms as follows. Throughout this work we will keep the distinctions to emphasize they in fact deserve different bounds.

\begin{figure}[t]
    \centering
    \includegraphics[width=0.9\textwidth]{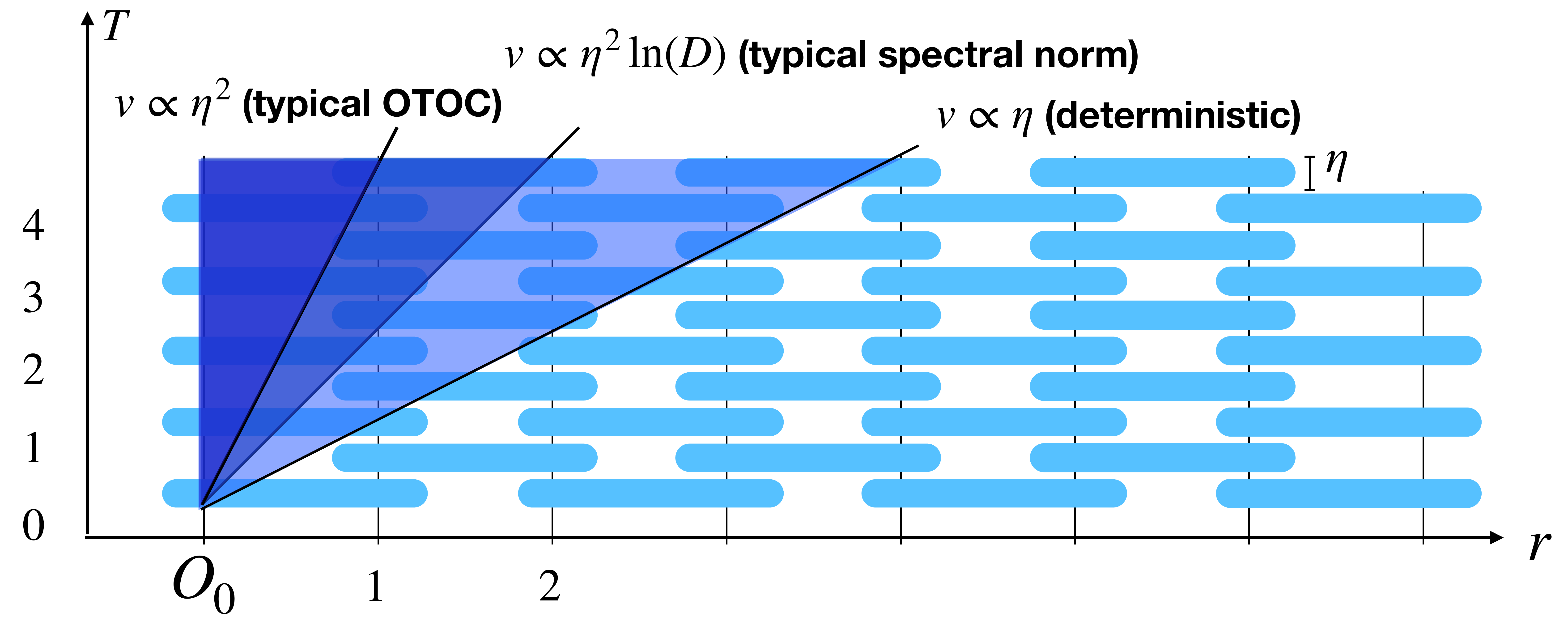}
    \caption{
    (a): The discrete Brownian circuit. Each block is an independent random unitary of strength $\le\eta$. This submission derived a signaling velocity for a typical sample, much slower than the worst case bound.
    }
    \label{fig:notions_of_light_cones}
\end{figure}

\begin{itemize}
    \item \textit{The expectation}. Taking an average may not yield the precise quantity of interest, but direct calculation is more accessible and offers valuable intuition. For example some analytic leverage in holography was gained by taking an ensemble average over random models~\cite{saad2019jt,stanford2020jt,xu2020sparse}. In the context of operator growth one may take average of unitary evolutions and arrive at a mixed channel, for which Lieb-Robinson bounds are also available~\cite{Poulin_2010}. Though, formally a channel is not equivalent to the behavior of the typical cases. This would mean we re-sampling a fresh Hamiltonian each time the circuit is applied and remove the zero mean fluctuations. 
    \item \textit{The typical value}. This is the goal of this work, which may or may not \textit{concentrate} near the average. Unfortunately, analytic results often rely on the fine-tuned structure of the model and very rarely is there a general bound. On the discrete extreme, properties of Haar random circuits can be calculated for the particular tasks such as scrambling~\cite{brown2013scrambling} and a lower bound on the design growth\footnote{As it considers arbitrary polynomial moments and not only the first few, we put it in this category. } in brick-wall circuits~\cite{Brand_o_2016,hunterjones2019unitary}; in the Brownian continuum limit, the analogous design growth lower bounds was also proven~\cite{Onorati_2017}. 
    For our purposes consider the spectral norm $\lV [O(t),X]\rV$, and there has not been a well-defined general bound for Brownian systems. Deterministic Lieb-Robinson bounds are not compatible with the Brownian limit and diverge.
    \item \textit{The typical value, over a background state}. In physical settings there often is an underlying background state or initial state to evolve from, such as a Gibbs state or a pure state of low complexity or entanglement. To include state dependence with randomization though, it often requires model specific calculations~\cite{S_nderhauf_2019}, without a general bound. \footnote{In a precursor of this work~\cite{chen2020quantum}, it was pointed out unitary evolving from an arbitrary non-random state already deserves better bounds.} 
    For our purpose, consider the OTOC. The (essentially only) available general tool in the Brownian limit would be stochastic differential equations. Often only in fine-tuned models can it be solved for the OTOC at infinite temperature~\cite{brownian_SYK} or specialized quantities~\cite{Lashkari_2013}. In a case study of time-independent random SYK model, bounds were obtained by specialized combinatorics~\cite{chen2019operator,Lucas_2020}. To our knowledge there is no general tool--other than the deterministic bounds-- for OTOC in random time-independent Hamiltonians.
\end{itemize}
\textbf{Sketch of main results.}\\
We formulate concentration bounds on the commutator for zero mean random Hamiltonians, time-independent or time-dependent, for spectral norm or OTOC. In each case, we give explicit plug-and-play formula for arbitrary systems as a sum over interaction paths $\Gamma$ between the operators $O_0, A_r$, akin to nested commutator expansions. For instance, for infinite temperature OTOC in time-independent random Hamiltonians we obtain the following bounds.
\begin{thm}[Sketch]\label{thm:informal}
Consider a random time-independent Hamiltonian $H = \sum H_X$, where terms $H_X$ are independent, zero mean $\BE [H_X] =0$, and bounded $\lV H_X\rV\le b_X$.
Then the OTOC \footnote{The norm we are using here is the Forbenius norm, i.e. the OTOC over the maximally mixed state.} can be bounded by a weighted sum over self-avoiding paths of interactions $\Gamma=\{X_\ell,\ldots,X_1\}$

\begin{align}
 \BE\left[ \tr\bigg( \rho [O_0(t), A_r ]^\dagger [O_0(t), A_r] \bigg) \right]
&\lesssim \sum_{\Gamma}  \int\limits_{t>t_\ell>\cdots >t_1>0}  \mathrm{d}t_\ell \cdots \mathrm{d}t_1\  \frac{e^{\beta_\ell(t-t_\ell)}}{\beta_\ell} b_{X_\ell}^2 \cdots \frac{e^{\beta_1(t_2-t_1)}}{\beta_1}b_{X_1}^2,
\end{align}
where $\beta_k$ are tunable parameters and $\lesssim$ supresses absolute constants.
\end{thm}
The novelty of this bound is that interaction strengths appear in squares $b^2_X$, entailing the incoherence across different terms $H_X$. Analogously, we obtain a formula for Brownian (random time-dependent) systems that in addition incorporates temporal fluctuation (i.e. doing Ito's calculus). See Sec.~\ref{sec:prelim} for backgrounds on self-avoiding paths, and see Sec.~\ref{sec:mainresults} for concentration statements in an appropriate statistical language. Our general bounds are evaluated on several systems (Sec.~\ref{sec:results_systems}), and they either match existing lower bounds or conjectures, or give predictions, highlighted as follows.  \\ 
\textbf{Example: Brownian $1d$ spin chain(Sec.~\ref{sec:shortranged})}. \\
Consider a 1d nearest-neighbor spin chain in brick wall of depth $2T$, with single site dimension $D$, and each unitary is generated by independent, bounded and zero mean Hamiltonian $U^{\tau}_{n,n+1} = e^{-iH^{\tau}_{n,n+1}\xi}$, $\lV H^{(T)}_{n, n+1} \rV\xi \le \eta, \BE[H^{(T)}_{n,n+1}] = 0$ (Fig.~\ref{fig:notions_of_light_cones}). We proved that with failure probability $\delta$:
\begin{align}
    \left\lV[O_0(t), A_r ]\right\rV  &\lesssim  \left( [\ln(1/\delta)+r \ln(D)] \frac{\eta^2T}{r^2} \right)^{r/2} \stackrel{r\rightarrow \infty}\sim \left( \ln(D)\frac{\eta^2T}{r} \right)^{r/2},\label{eq:quick_LR_random}
\end{align}
where constants are suppressed by $\lesssim$. For OTOC on arbitrary non-random state $\rho$,\footnote{the first inequality is a Cauchy Schwartz.} 
\begin{align}
    \tr\bigg( \sqrt{\rho} [O_0(T), A_r ]^\dagger \sqrt{\rho}[O_0(T), A_r] \bigg) \le \tr\bigg( \rho [O_0(T), A_r ]^\dagger [O_0(T), A_r] \bigg) \lesssim  \frac{1}{\delta}\left(\frac{\eta^2T}{r}\right)^{2r} \stackrel{r\rightarrow \infty}\sim \left(\frac{\eta^2T}{r}\right)^{2r}.
\end{align}
Some highlights:
\begin{itemize}
    \item It depends \textit{quadratically}  $\eta^2$ on the Hamiltonian strength, more stringent than the linear deterministic Lieb-Robinson bounds (for discrete circuits) when gates are weak $\eta\ll 1$
    \begin{equation}
       \lV[O_0(T), A_r ]\rV, \frac{\lV[O_0(T), A_r ]\rV_2}{\lV I\rV_2} \stackrel{det}{\lesssim} \left(\frac{\eta T}{r}\right)^{r}, v_{det} \propto \eta.
    \end{equation}
     In the continuum limit $\eta=a\xi \rightarrow 0$, the natural notion of Brownian 'time' is the variance $\xi^2T:=\tau$. The deterministic Lieb-Robinson velocity diverges, while our bound remains well-defined. The improvement comes from only demanding a bound that holds \textit{with high probability}. For finite distances the worst cases can saturate the traditional Lieb-Robinson bounds, but asymptotically we have $v_{typ} = \CO(\eta^2\ln(D))$ and $v_{typ} = \CO(\eta^2)$ \textit{almost surely}.
     \item The OTOC bound works for any non-random state, where as the spectral norm (worst state) bound has a extra factor of logarithm of local dimension $\ln(D)$. This distinction is a feature of high dimension probability: it matters whether we optimize the state \textit{after} revealing the random sample of unitary. The spectral norm bounds are for absolute safety, but the OTOC bounds may be more realistic. Note that we did not prove for finite temperature Gibbs states, which correlate with the random unitary.
    \item Both bounds are ballistic. We only attempt to bound typical wave front but not the diffusive broadening near the wave front due to backward movers~\cite{hydro_OTOC,PhysRevX_xu_swingle,operator_spread_RUC}. Note we are not targeting many-body-localized or disordered systems (e.g.~\cite{Burrell_2007,Burrell_2009}), where the light cone is logarithmic in $t$ in certain models.
\item This result is robust that we only need the Brownian Hamiltonian to be zero mean \textit{conditioned on the past}. This can be more carefully noted by the conditional expectation $\BE_{T-1}[H^{(T)}_{n,n+1}] = 0$. See Sec.\ref{sec:proofs} for introduction to martingales.
\end{itemize}
Altogether, we formulate the first appropriate Brownian generalization of the Lieb-Robinson bounds incorporating the stochastic nature. This immediately applies to tasks where the Lieb-Robinson bounds are a subroutine such as digital quantum simulation~\cite{haah2020quantum}. Heuristically, for nearest neighbor (time-independent) chaotic Hamiltonians, we may hope the Brownian Lieb-Robinson velocity to be more physical (with some estimate of the coherence time $\xi$) than the deterministic bounds. 

\begin{figure}[t]
    \centering
    \includegraphics[width=0.9\textwidth]{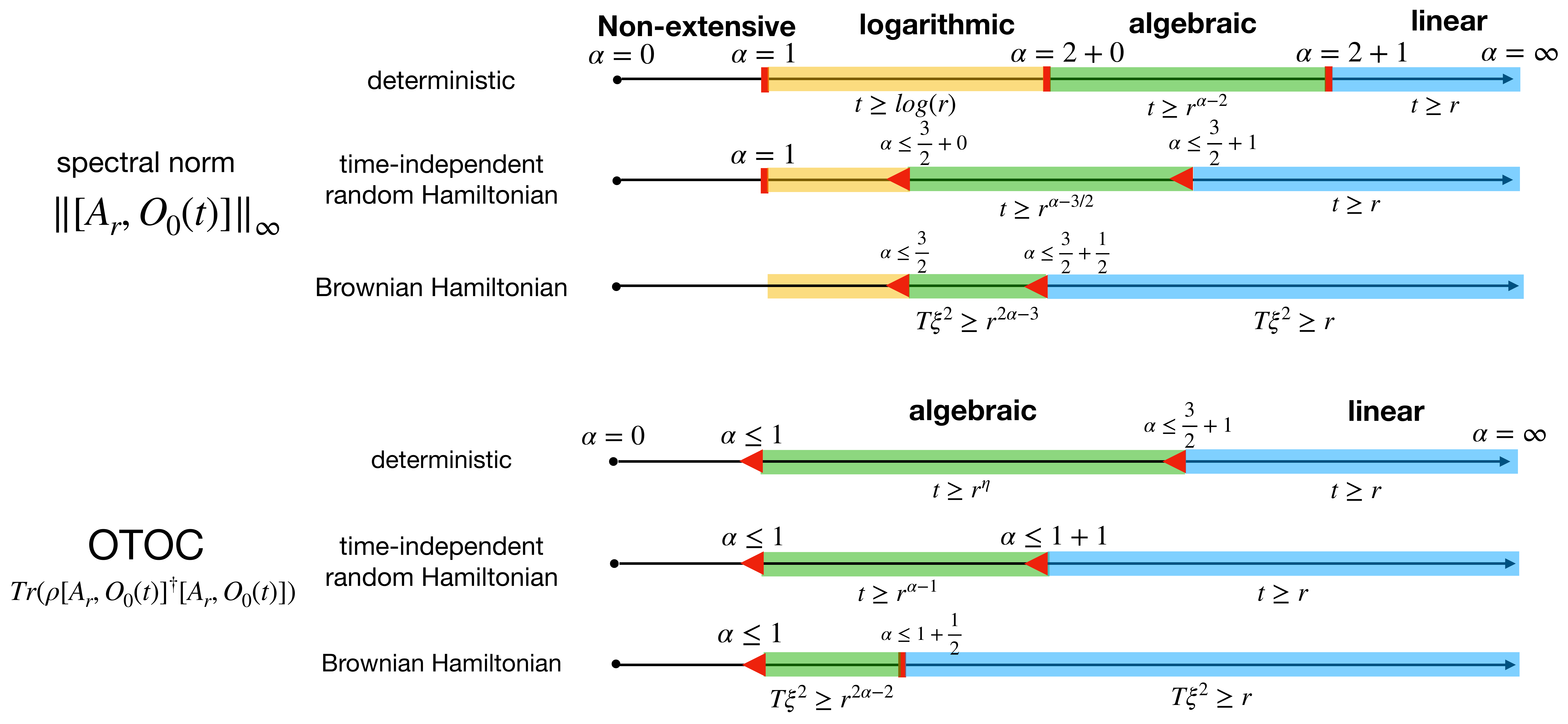}
    \caption{Phases of operator growth in $d=1$ power-law interacting systems and the (bounds on) $\alpha$ at the transitions. The vertical bar are the optimal values of $\alpha$ with saturating operator spreading protocols or experimental/numerical observation; the left pointing triangles denotes a bound. As of today, the deterministic Lieb-Robinson bounds is nearly completed with upper and lower bounds~\cite{alpha_3_chenlucas,strictlylinear_KS,tran2021optimal,hierarachy}, except for the (unphyisical) $\alpha <1$ non-extensive regime, which is also not the focus of this paper. The deterministic OTOC bounds are still in development~\cite{hierarachy,Kuwahara_OTOC}. In this work, we fill in the Brownian phases, one of which is consistent with recent simulation~\cite{long_range_brownian,Levy} $\alpha =1.5$. Indeed, the stochastic effects in space and in time make operator spreading more constrained. The higher dimension $d>1$ cases has more missing pieces but expected to extrapolate from $a+b$ to $ad+b$, for example $1+1/2\rightarrow 1\cdot d+1/2$. 
    }
    \label{fig:power law phase}
\end{figure}

\textbf{Example: SYK-like random k-local systems(Sec.~\ref{sec:k-local}).}\\
The renowned (random) SYK model and also spin liquids reside in the class of random complete $k$-local system
\begin{equation}
    H^{(T)
    }_{k,N} :=\sum_{i_1< \ldots <i_k \le N}  \sqrt{\frac{J^2 (k-1)!}{kN^{k-1}}} H^{(T)}_{i_1\cdots i_k}, 
\end{equation}
 where each term is normalized\footnote{In some versions it is a Gaussian. That is actually an implication of our result on the bounded ones: duplicate the same interaction $H^{(T)}_{i_1\cdots i_k}$ with Rademacher coefficients and call the central limit theorem.} $\lV H^{(T)}_{i_1\cdots i_k} \rV\le 1$, independent, and zero mean $\BE[H^{(T)}_{i_1\cdots i_k}]=0$. The ideal models are time-independent yet we allocate the superscript for the Brownian cases, which are also extensively studied. These systems are not geometrically local like a spin chain but only that each interacting terms acts on a finite number ($k$) of sites. Even with all-to-all interaction, they are not expected to lose information arbitrarily fast. It was proposed by Susskind and Sekino~\cite{Sekino_2008} that
\begin{align}
    t_{scr} = \Omega(\log(N)) \ \ \ (\text{The fast scrambling conjecture}),
\end{align}
where the scrambling time was not mathematically defined and $N$ is the number of sites or degrees of freedom. The original argument came from thought experiments for black holes and time-discrete k-local models~\cite{Sekino_2008}, and was later discussed in~\cite{stephen14,Maldacena_2016, Lashkari_2013,chen2019operator,Lucas_2020}.
For scrambling taken as the two halves becoming entangled, a matching lower bound was found in a particular Brownian 2-local model~\cite{Lashkari_2013}.
In time-independent SYK-model, matching upper bounds on growth of infinite temperature OTOC were obtained with specialized, intensive combinatorics~\cite{Lucas_2020}. Using Theorem~\ref{thm:sum_over_paths}, we obtain one page proofs of fast scrambling for both time-independent and Brownian $k$-local systems\footnote{We should not compare the constant too harshly as the Brownian time is not really comparable to the ordinary time.},
\begin{align}
     \tr\bigg( \rho [O_0(t), A_r ]^\dagger [O_0(t), A_r] \bigg) &\le   \frac{k-1}{N\delta} e^{4Jt}  &\text{(time-indep.)}\\ 
     \tr\bigg( \rho [O_0(\tau), A_r ]^\dagger [O_0(\tau), A_r] \bigg) &\le   \frac{k-1}{N\delta}e^{\frac{17}{2}J^2\tau}, &\text{(Brownian.)}
\end{align}
for arbitrary non-random state\footnote{Once again, the finite temperature Gibbs states correlates with the random unitary and does not apply here!} $\rho$ and failure probability $\delta$. Our Brownian upper bound is new. We emphasize our contribution being the generality and simplicity.\\
\textbf{Example: $1d$ power-law interacting spin chain (Sec.~\ref{sec:longrange}).}\\
We obtain bounds for both time-independent and time-dependent random Hamiltonians. Their distinction can be illustrated in the $1d$ power-law interacting spin chain, whose Hamiltonian is \begin{equation}
     H^{(T)} :=\sum_{i} \frac{1}{|i-j|^\alpha} H^{(T)}_{ij},
\end{equation}
where the random terms are bounded $\lV H^{(T)}_{ij}\rV\le 1$, independent, and zero mean $\BE[H^{(T)}_{ij}]=0$, and the superscript is allocated for time-dependence if needed. With presence of long-ranged-but-weak and short-ranged-but-strong interaction, these models interpolate between local($\alpha=\infty$) and all-to-all interacting($\alpha=0$). In the deterministic case in $1d$, the intermediate 'phases' (and the critical values of $\alpha$ where the transition occurs) was proven to be a linear light cone regime ($\alpha >3$)\cite{alpha_3_chenlucas,strictlylinear_KS,tran2021optimal}, an algebraic/sub-linear light cone ($2<\alpha <3$), and then a logarithmic light cone ($\alpha< 2$)~\cite{Bravyi2006}. In this work we study the transition between these dynamical phases for typical zero mean random Hamiltonian, time-indep or not, spectral norm or OTOC. See Fig.~\ref{fig:power law phase} for a summary of bounds we obtained. We also derive exponential tail bounds in each setting. For example, in time-independent random Hamiltonian, the OTOC concentrates
\begin{align}
     {\tr\bigg( \rho [O_0(t), A_r ]^\dagger [O_0(t), A_r] \bigg)} &\lesssim {\ln(1/\delta)}(\frac{t}{r^{\min {(1,\alpha-1)}}} )^2 &\text{(time-indep.)} \\
     {\tr\bigg( \rho [O_0(\tau), A_r ]^\dagger [O_0(\tau), A_r] \bigg)} &\lesssim {\ln(1/\delta)}{\frac{\tau}{r^{\min{(1,2\alpha-2)}}}}, &\text{(Brownian.)}\label{eq:brownlongOTOC}
\end{align}
 for any non-random background state and for failure probability $0<\delta <\frac{1}{e}$, where the ${\ln(1/\delta)}$ dependence reflects an exponential tail. $\lesssim$ suppresses absolute constants. Remarkably, our Brownian prediction~\eqref{eq:brownlongOTOC} of light cone transition from linear, algebraic, to subalgebraic ($\alpha =1; 3/2$) matches the simulation results for chaotic (time-independent non-random)~\cite{Levy} and Brownian (time-dependent random)~\cite{long_range_brownian} systems.


\textbf{Sketch of proof}.\\
The idea behind the proof boils down to the sum of random matrices. Our technical contribution is to neatly account for a nonlinear function, the time-evloved commutator, of many random matrices using a robust analytic framework from \textit{matrix martingales}. It allows us to obtain bounds on the expected moments that imply strong concentration by a Markov's inequality. Let us quickly estimate the sum of random matrices and leave the matrix martingale methods, self-contained in Section~\ref{sec:proofs}, for readers seeking for other applications. 


The essential mechanism can captured in the operator growth from a region to another $[A,e^{iHt}Oe^{-iHt}]$ due to evolution of random Hamiltonian $\sum H^{(\tau)}_k$(time-dependent) and $\sum H_k$(time-independent), and suppose they are supported on a $D$-dimension Hilbert space $dim(O\otimes A)=D$. 
The first order contribution is a sum
\begin{align}\label{eq:linearized of error}
    t \sum_k [A,[iH_k,O]]&:= \sum_ktX_{k} &(\text{time-indep.})\\
    \sum_\tau \sum_k [A,[i\xi H_k,O]]&:=\sum_\tau \sum_k \xi X^{\tau}_{k},  &(\text{Brownian.})
\end{align}  
where $\lV H^{(\tau)}_k\rV_\infty \le b_k, \BE[H_k] = 0$.  Then, the sum behaves differently in the typical and worst case settings.\\
The \textbf{worst case bound} uses the triangle inequality to get a linear sum
\begin{align}\label{eq:worst growth}
\lV \sum_ktX_{k} \rV_\infty &\le t\sum_k b_k &(\text{time-indep.})\\
\lV \sum_{T'} \sum_k \xi X_k\rV_\infty &\le \sum_k b_k\xi T. &(\text{Brownian.})
\end{align}
\begin{figure}[t]
    \centering
    \includegraphics[width=0.9\textwidth]{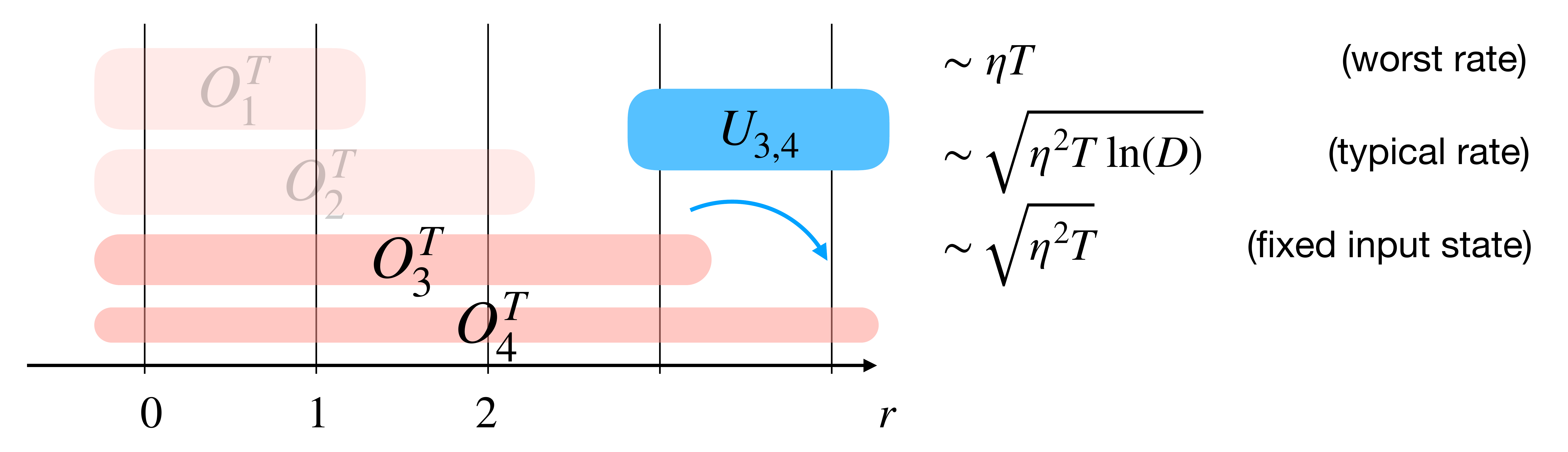}
    \caption{
    Proof idea for nearest neighbor Brownian spin chain. At each time step, decompose the operator according to how far away it has reached. After a time step, operator can grow towards adjacent sites due to local unitary evolution. For example, $O_3$ is rotated and adds to $O_4$ due to $U_{3,4}$. The noises across different times do not add up coherently and require ~ $(\eta^2)^{-1}$ steps to fully advance one site further, contrasting with the linear triangle inequality bounds. The different statistical settings (spectral norm,  fixed-input) further adorn the rate.
    }
    \label{fig:singlehop}
\end{figure}
The \textbf{typical case bound in spectral norm} can be obtained by a matrix concentration result (See Example~\ref{ex:sum_matrices}, e.g.~\cite{tropp2015introduction})
\begin{align}
 \label{eq:typical growth op}
\BE \lV t\sum_k X^{}_{k} \rV_\infty &{\lesssim} 
\ t\sqrt{\ln(D)\sum_kb_k^2} & (\text{time-indep.})\\
 \BE \lV \sum_{T'} \sum_k \xi X^{T'}_{k} \rV_\infty &{\lesssim} 
 \sqrt{\xi^2 T\ln(D)\sum_k b_k^2}, &(\text{Brownian})
\end{align}
where $\lesssim$ suppresses absolute constants and gaussian tail bounds are also available.\\
The \textbf{typical case bound for a non-random state }can be obtained by direct variance calculation
\begin{align}
    \BE {\tr\left(\rho(t \sum_k X^{}_{k})^\dagger(t \sum_k X^{}_{k})\right)} & \le t^2{\sum_k b_k^2 },&(\text{time-indep.})\\
    \BE {\tr\left(\rho(\sum_{T'} \sum_k \xi X^{{T'}}_{k} )^\dagger(\sum_{T'} \sum_k X^{{T'}}_{k})\right)} & \le  {\xi^2T\sum_k b_k^2} &(\text{Brownian})\label{eq:typical growth otoc},
\end{align}
where Gaussian concentration are also available (See Example~\ref{ex:sum_matrices}). 
The two typical case bounds capture the sum-of-squares behavior for sum of uncorrelated noises, including spatial($\sum_k$) and temporal($\sum_{T'}$) fluctuations. 
The OTOC concentration behaves analogously to the scalar case, while spectral norm concentration has an extra logarithm of dimension. This is a feature of matrix concentration inequalities and it gives rise to the $\ln(D)$ factor in Lieb-Robinson velocity in Eq.\eqref{eq:quick_LR_random}. 

\begin{figure}[t]
    \centering
    \includegraphics[width = 0.9\textwidth]{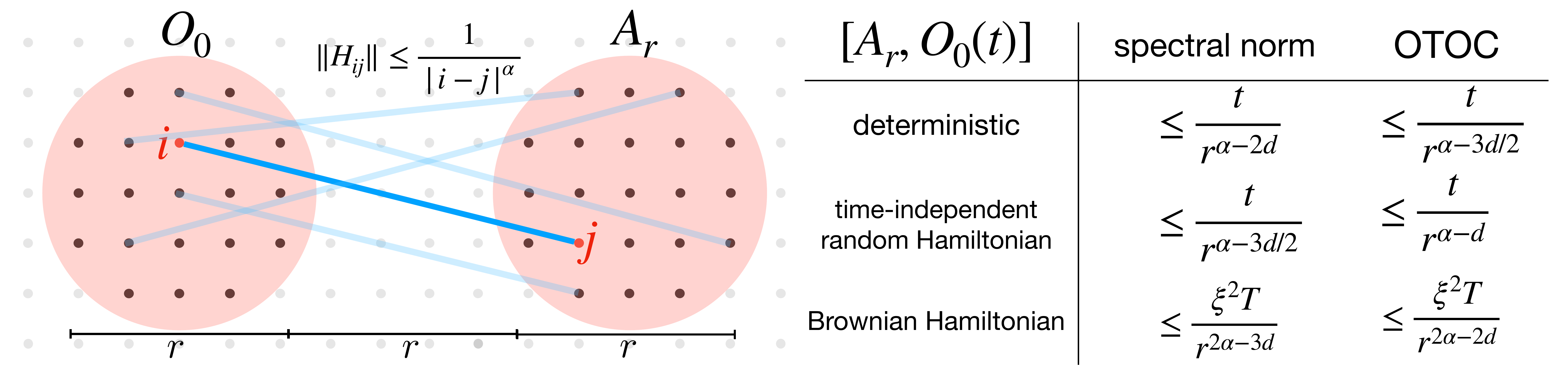}
    \caption{The contribution from long-ranged terms can be succintly characterized by the operator growth from a supersite of radius $r$ to another sphere distant $\CO(r)$ away. Intuitively this is because the short-ranged interactions smear each long ranged terms out, and see~\cite{hierarachy,strictlylinear_KS} for further justifications. This allows us to quickly calculate (the anticipated) shape of light cones and critical values of $\alpha$ according to how these random terms adds up in different settings. The prediction from this table matches the rigorous bounds we obtained in $d=1$(Fig~\ref{fig:power law phase}).}
    \label{fig:interacting sph}
\end{figure}

For our short-ranged example, only the temporal noise matters (Fig.~\ref{fig:singlehop}) because the interactions are geometrically local.\footnote{ Here, time-independent bounds is the same as deterministic bounds.} The Brownian Hamiltonian grows the operator one site further with the typical rate $\sim\eta^2 \ln(D)$ and $\sim\eta^2$. This contrasts with the linear behavior $\propto\eta$ in the triangle inequality used in Lieb-Robinson's derivation. For the 1d power-law example, we have both the short-ranged interaction and the long-ranged, all-to-all interaction. It turns out the effects of long-ranged interaction can be intuitively captured by the interaction between a two sphere of radius $r$ separated $r$ apart (Fig.~\ref{fig:interacting sph})~\cite{hierarachy,strictlylinear_KS}. The sum has spatial (and temporal if Brownian) noises and how they scale depends on whether the noise is time-dependent and whether in the spectral norm or OTOC. 

Technically, commutators are not at all an i.i.d sum. To actually derive concentration, our main analysis tool recursively uses a matrix martingale technique called \textit{uniform smoothness} (Proposition \ref{prop:sub_average_pq}), 
    \begin{prop}[{Uniform smoothness of Schatten p-norms~\cite{ricardXu16},\cite[Proposition~4.3]{HNTR20:Matrix-Product}}]
Consider random matrices $X, Y$ of the same size that satisfy
$\BE[Y|X] = 0$. When $2 \le p$,
\begin{equation}
\vertiii{X+Y}_{p}^2 \le \vertiii{X}_{p}^2  + (p-1)\vertiii{Y}_{p}^2. 
\end{equation}
\end{prop}
The expected Schatten p-norm will turn into concentration bounds by Markov's inequality \begin{equation}
    \vertiii{X}_{p} := \BE[\tr ({\sqrt{X^\dagger X}}^{p})]^{1/p} = \BE[(\sum \nu^p_i)^{1/p}].
\end{equation}

See Sec.~\ref{sec:proofs} for a minimal introduction. To complete the proof there are few technical remarks:
\begin{itemize}
    
    \item To turn the time-indepedent operator growth into a martingale, we need to carefully reorganize operator growth into a sum of self-avoiding paths~\cite{chen2019operator}. 
    In the Brownian operator growth, we also have to keep track of both the leading order zero-mean term (Eq.~\ref{eq:typical growth op}) and the second order deterministic term, as a manifestation of Ito's calculus.
    \item The OTOC calculation turns out much simpler than the spectral norm as the latter is sensitive to support dimension. The factor $\log(dim)$ grows with the support of the operator, and hence in addition to our general Theorem~\ref{thm:sum_over_paths}, it requires delicate support decomposition of the operator and then a messy union bound. This is also what made $2d$-nearest neighbor operator norm bounds not a direct consequence of our general formula. 
    \item Beyond OTOC and spectral norm, our matrix martingale tool kits are so general that \textit{any Schatten p-norm}, for $2< p <\infty$, can be bounded using identical arguments, except for a few change of constants (Proposition~\ref{prop:general_pq}). 
\end{itemize}

\textbf{Organization of the paper.}\\
We first briefly review deterministic results on general Lieb-Robinson bounds in Sec.~\ref{sec:prelim}. In Sec.~\ref{sec:mainresults}, we present the general formula for time-independent and Brownian systems. We then evaluate our bounds on several classes of systems in Sec.~\ref{sec:results_systems}. See Sec.~\ref{sec:proofs} for the detail proofs of our main theorems and Sec.~\ref{sec:detail_cal_OTOC}, Sec.~\ref{sec:detail_cal_Op} for the detailed calculation for different systems. 

\section{Preliminary: deterministic bounds}\label{sec:prelim}
\textbf{Setup and notions of operator growth.}
Consider a deterministic (time-dependent) Hamiltonian evolution
\begin{equation}
    U(t) = \exp\left(-i\sum H_X(t) dt\right)\cdots \exp\left(-i\sum H_X(0) dt \right)
\end{equation}
where the locality (set of permissible interactions $H_X$) and interaction strengths ($\lV H_X\rV\le b_X$) are the parameters of the model. Then, for physical purposes there are few common notions for the growth of local operators:
\begin{align}
    \textrm{Spectral norm, Lieb-Robinson-like bounds}: &\ \rV[O_0(t),A_r]\lV \\
    \textrm{Forbenius norm, infinite temperature OTOC}: &\ \frac{\tr( [O_0(t),A_r]^\dagger[O_0(t),A_r] )}{\tr(I)}\\
     \textrm{OTOC for an particular state}:&\ \tr(\rho  [O_0(t),A_r]^\dagger[O_0(t),A_r] ).
\end{align}
where the spectral norm studied by Lieb and Robinson is the most general bound for any input state and any Hamiltonian. The OTOC for certain state (and sometimes for certain Hamiltonian) is more realistic yet may not match worst case (Lieb-Robinson-like) bound.\\

The standard strategy to get a Lieb-Robinson-like bound comprises of a local analysis for induction and a global bound from integrating the local analysis~\cite{Lieb1972,Hastings_koma, Bravyi2006,chen2019operator,D_A_graph_GU_LUcas,PRXQuantum.1.010303}.

\begin{figure}[t]
    \centering
    \includegraphics[width=0.9\textwidth]{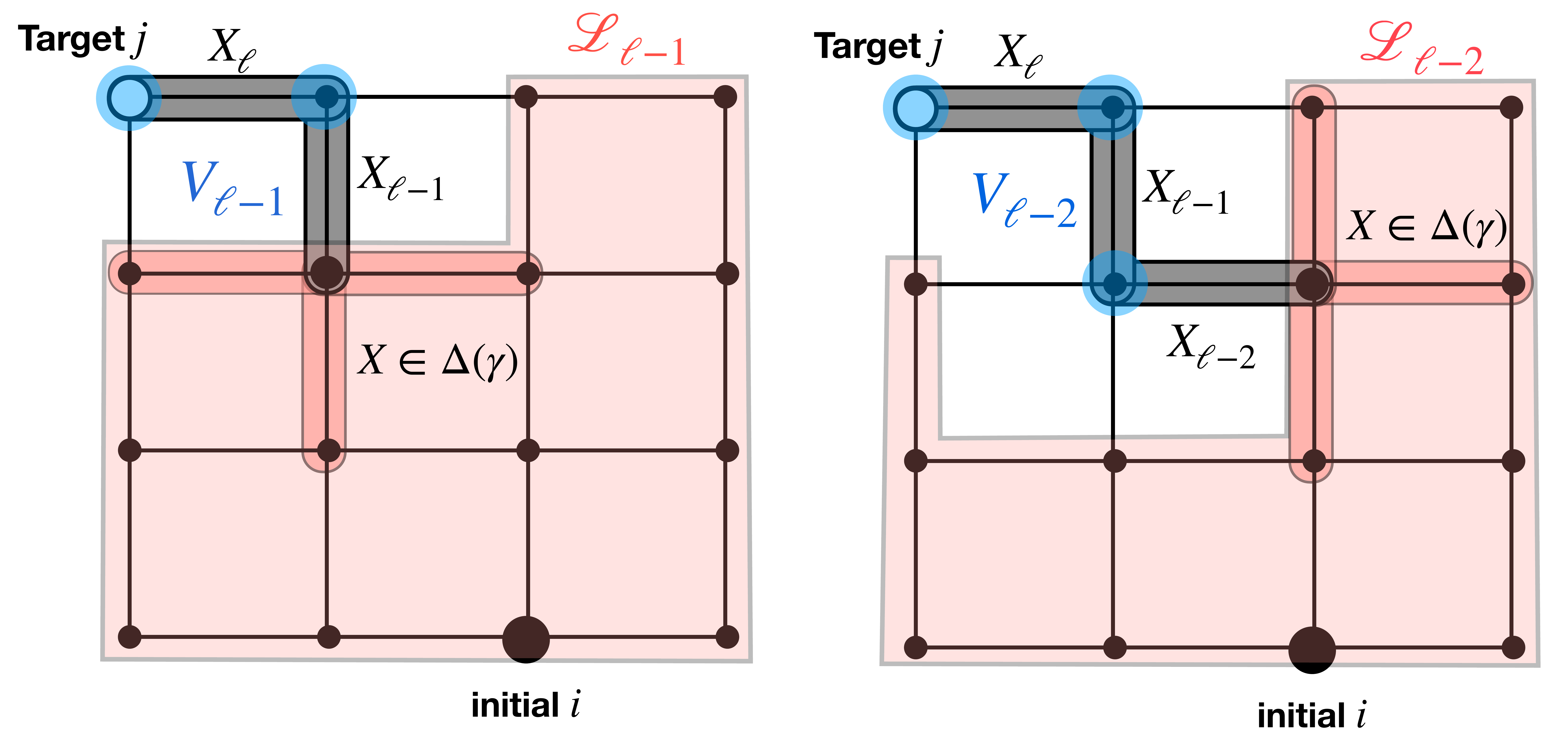}
    \caption{ The intermediate unitary evolutions $\CL_k$s and the complementing set of excluded vertices $V_k$ between hops $\cdots \CL_{X_{k+1}}e^{\CL_k}\CL_{X_{k}}\cdots$. The excluded set of vertices expands as the path $\gamma$ backtracks.
    }
    \label{fig:forbidden}
\end{figure}

\textbf{Local growth.}
In the deterministic setting, we often use a triangle inequality to bound the infinitesimal growth of an operator $A(t+dt)$ from its adjacent site $B(t)$ at a previous time step
\begin{align}\label{eq:triangle_ineq}
    \lV A(t+dt) \rV &= \lV U\left[ VA(t)V^\dagger+VB(t)V^\dagger -B(t)\right] U^\dagger\rV \\
    &= \lV \hat{U}\left[ \hat{V}[A(t)]+(\hat{V}-I)[B(t)]\right] \rV \\
    &\le \lV A(t)\rV + 2\lV V-I\rV \lV B(t)\rV,
\end{align}
where the hatted unitary $\hat{V}$ denotes a conjugation superoperator $\hat{V}[O]:=VOV^\dagger$, which gives rise to the factor of 2. In the infinitesimal limit, we can replace $\hat{V}-I$ with Linbaldian $ dt\CL[\cdot]:=[iHdt,\cdot\ ]$.\\
\textbf{The sum over paths.}
\begin{figure}[t]
    \centering
    \includegraphics[width=0.9\textwidth]{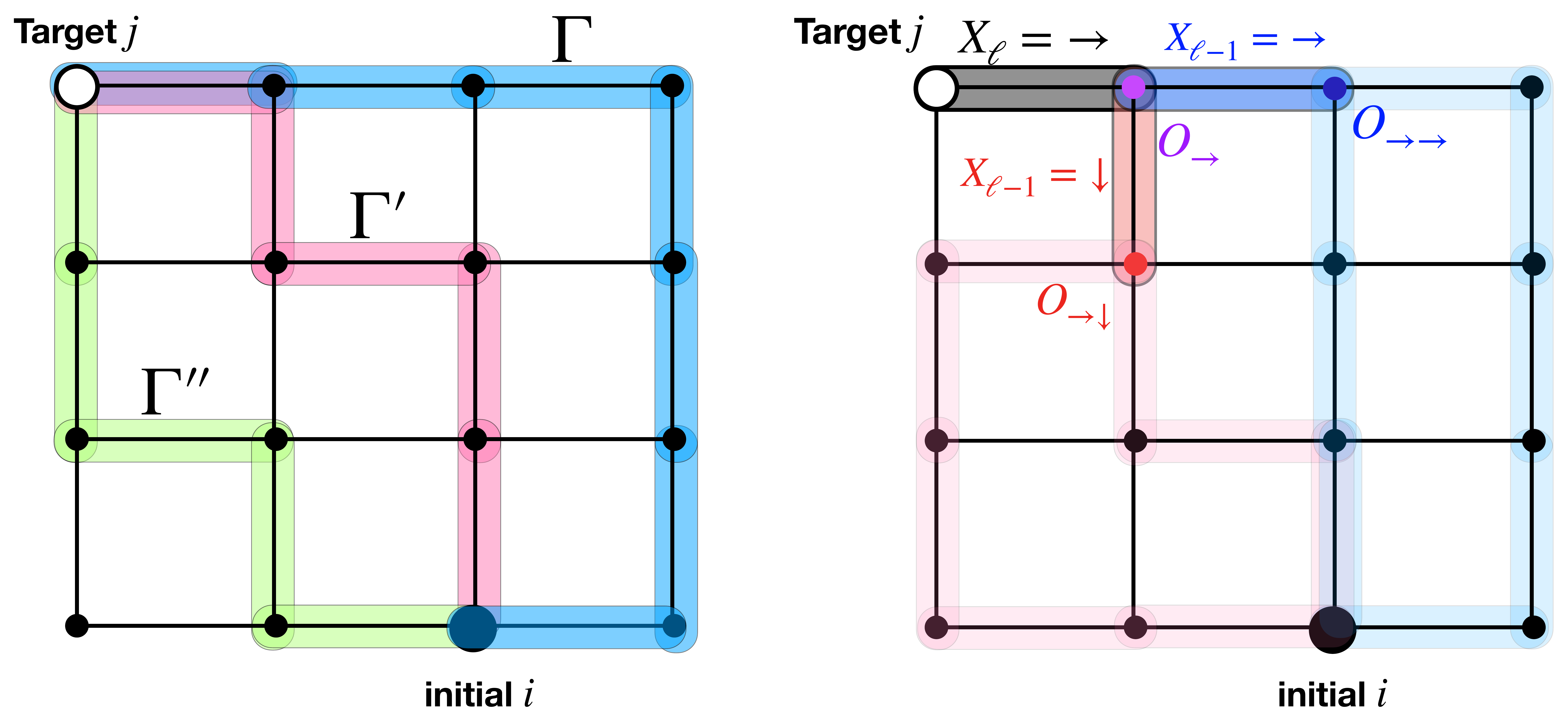}
    \caption{ (a)Decomposing operator growth as a sum of over self-avoiding paths between the initial site and the target site. Only consecutive hops are allowed to overlap. Naive expansion of the unitary will produce self-intersecting paths, which does not contribute to operator growth and will always be re-grouped to one of the self-avoiding paths. (b) The sum over paths can be enumerated in a recursive form, which is crucial for us to recursively apply martingale inequalities. Backtracking $\gamma$ from the target site, the next steps of $\gamma$ are $X\in \Delta(\gamma)$, the terms interacting $\gamma$ at only the last term $X$, which are $\rightarrow$ and $\downarrow$ in the figure. The remaining possible next-next steps are collected into operator $O_{\gamma+X}$ (or $O_{\rightarrow\rightarrow}$ in the figure). 
    }
    \label{fig:sum_over_paths}
\end{figure}
Next, we often decompose the operator growth into a recursive expansion, backtracking the paths of interactions $\Gamma = {X_\ell,\ldots,X_1} \in \mathcal{S}_{r0}$ from the target site $A_r$ to the initial site $O_0$. In particular, with careful accounting we can take these paths to be \textit{self-avoiding} (or irreducible)~\cite{chen2019operator} (Fig.~\ref{fig:sum_over_paths}.), i.e. only consecutive terms on the path $\Gamma$ can overlap, or more precisely
\begin{align}
[H_{X_i},H_{X_j}] =0\ \text{if}\ |i-j| > 1.
\end{align}
See~\cite{chen2019operator} for the details of this construction. The final general expression would be an  integral formula.
\begin{thm}[Summing over self-avoiding paths~\cite{chen2019operator}]\label{thm:det_sum_over_path}
For any time-dependent Hamiltonian $H = \sum H_X(t)$, the spectral norm of commutator between operators $O_0(t)$ and $A_r$ can be bounded by a weighted sum over self-avoiding paths of interactions $\Gamma=\{X_\ell,\ldots,X_1\}\in \mathcal{S}_{r0}$
\begin{align}
\lV [O_r,O_0(t)]\rV &= \lV \CL_re^{\CL t}[O_0]\rV = \notag \\
&= \left\lV \CL_r\sum_{\Gamma}  \int\limits_{t<t_\ell<\cdots <t_1} \mathrm{d}t_1\cdots \mathrm{d}t_\ell\  \mathrm{e}^{\mathcal{L}(t-t_\ell)} \mathcal{L}_{X_\ell}  \mathrm{e}^{\mathcal{L}_{\ell-1} (t_\ell-t_{\ell-1})} \mathcal{L}_{X_{\ell-1}}\cdots \mathrm{e}^{\mathcal{L}_1(t_2-t_1)}\mathcal{L}_{X_1}\mathrm{e}^{\mathcal{L}_0t_1} [O_0] \right\rV \\
&\le  2\lV O_r\rV \lV O_0\rV \sum_{\Gamma}  
\frac{2^\ell t^\ell}{\ell!} \prod_{k=1}^\ell \lV H_{X_{k}}\rV, 
\label{eq:interactionsum}
\end{align}
where $\ell = \ell(\Gamma)$ and the intermediate evolutions $\CL_k$ along the path are carefully chosen depending on the self-avoiding path $\Gamma$ \begin{equation}\label{neighbourhood}
\CL_k=\mathcal{L}^{(\Gamma)}_k := \mathcal{L} - \sum_{Y\in \partial V^\Gamma_k} \mathcal{L}_Y.
\end{equation}
The set vertices excluded between steps $k$ and $k+1$ on the path $\Gamma$ is a subset of the existing path 
\begin{equation}
V^\Gamma_k := \left\lbrace \begin{array}{ll} \lbrace j\rbrace &\ k=\ell(\Gamma)-1 \\ \displaystyle  \bigcup_{m=2+k}^{\ell(\Gamma)} X^\Gamma_m &\ 0 \le k<\ell(\Gamma)-1 \end{array}\right . \label{eq:VGammak}
\end{equation}
\label{self-avoiding}
\end{thm}
Roughly, we can interpret Theorem~\ref{thm:det_sum_over_path} as the integral form of the one step growth~\eqref{eq:triangle_ineq}. More carefully, the intermediate evolutions $\CL_k$ are mindfully chosen to ensure an \textit{exact} expansion of the unitary conjugation $e^{\CL t}$. The self-avoiding property might seem a minor improvement on the bound, yet this careful organization of terms turns out playing a central role in the random time-independent analysis. There, the exact expressions for each $\CL^\Gamma_k$ (i.e. the excluded vertices $V_k^\Gamma$) are absolutely crucial (Fig.~\ref{fig:forbidden}). Whereas in the last inequality~\eqref{eq:interactionsum} above, they are treated just as any unitary.
\section{Main results}\label{sec:mainresults}
Our main results are in strong parallel with the deterministic case (Sec.~\ref{sec:prelim}). We derive local growth lemmas and integral global formulas for time-independent random Hamiltonians (Sec.\ref{sec:time_indep_mainresults}) and time-dependent(i.e. Brownian) Hamiltonians (Sec.\ref{sec:brown_mainresults}). For each quantity of interest, the results give bounds on the expected p-th moments, which imply a tail bound by Markov's inequality. For the infinite temperature OTOC, we take the norm to be the p-th moments of the Hilbert-Schmidt norm
\begin{equation}
    \left(\BE[\tr(O^\dagger O)^{\frac{p}{2}}]\right)^{\frac{1}{p}}=:\vertiii{ O }_{2,p};
\end{equation}
For spectral norm, we take the p-th moment of Schatten-p norm as an upper bound
\begin{equation}
    \left(\BE[\lV O\rV^p]\right)^{1/p} \le \left(\BE[\lV O\rV^p_p]\right)^{\frac{1}{p}} =:\vertiii{ O}_{p};
\end{equation}
with the Schatten p-norm $\lV O\rV^p:=( \tr[(O^\dagger O)^{p/2}])^{1/p} = (\sum_i \nu^p_i)^{1/p}$ associated with the singular values $\nu_i$.
When there is an underlying background state, we generalize to take the p-th moment of schatten-p norm with $D_P$-dimension input subspace. 
\begin{align}
\sup_{rank(P)=D_P}\left(\BE[\lV OP\rV_p^p]\right)^{1/p} & =: \vertiii{O}_{{D_P},p}.
\end{align}
In particular, we can set the input subspace to be 1-dimensional to get moment bounds on the OTOC for arbitrary non-random input state $\rho$ (Proposition~\ref{prop:lowrank_to_mixed}).
\begin{align}
\left(\BE[\tr(\rho O^\dagger O)^{\frac{p}{2}}]\right)^{1/p}  \le \sup_{\ket{\psi}} \ \left(\BE[\bra{\psi}O^\dagger O\ket{\psi}^{\frac{p}{2}}]\right)^{1/p} =  \vertiii{O}_{{D_P=1},p}.
\end{align}
All the above upper bound the commutator $[O,A]$ by a factor of $2\lV A\rV$ as a matter of accounting\footnote{These norms are all operator ideal, see Fact.~\ref{fact:operator ideal}}.

\subsection{A general formula for time-independent random Hamiltonians}
\label{sec:time_indep_mainresults}
Consider the evolution of a time-independent Hamiltonian
\begin{equation}
    U(t) = \exp\left(-i\sum H_X t\right),
\end{equation}
where the interactions $H_X$ are independent, zero mean $\BE[ H_X] =0$, and bounded $\lV H_X\rV\le b_X$.
The analog of triangle inequality in the probabilistic setting is the following local lemma for the p-th moments. We highlight it for its potential usage in other settings. \begin{lem}[Moment bounds for random differential operator equation]\label{lem:time_indep+induction_step}
Consider a differential equation for operator-valued random variables $A(t), B_j(t)$ 
\begin{align}
    A(t+dt) & = U(dt)\left[ A(t) + \sum_j \CL_j[B_j(t)]dt \right]U^\dagger(dt),
\end{align}
generated by Linbaldians $\CL_j=[iH_j,\cdot]$ of bounded random Hermitian operators $\lV H_j\rV\le b_j$, which are zero mean and independent \textit{conditioned on} $B_j(t)$
\begin{align}
     \BE[H_j|B_1,\cdots, B_j]=0,\ \  \BP(H_{j_1},\cdots, H_{j_k}|B_1,\cdots, B_j)=\prod_j \BP(H_{j}|B_1,\cdots, B_j).
\end{align} Then the increments are added in squares
\begin{align}
    \vertiii{ A(t+dt)}_* &\le \vertiii{A(t)}_{*}+ dt\sqrt{4p \sum_j b_j^2\vertiii{B(t)_j}_{*}^2}.
\end{align}
Or it converts to a linear recursion with a tuneable parameter $\beta$
\begin{align}
    \frac{d}{dt}\left(\vertiii{ A(t+dt)}^2_*\right) & \le \beta \vertiii{A(t)}^2_{*}+ \frac{4p}{\beta}\sum_j b_j^2 \vertiii{B(t)_j}_{*}^2,
\end{align}
for expected p-th moments $\vertiii{ \cdot}_*= \vertiii{ \cdot}_{p}, \vertiii{ \cdot}_{D_P,p}$.
\end{lem}
The sum of squares between different $B_j$ is due to independence conditioned on the past, while along the time evolution, which is non-random, we only bound with a triangle inequality. The extra parameter $\beta$ to be tuned is an artifact for gaining a linear recursion via the arithmetic-geometric inequality. See Sec.~\ref{sec:proof_time_indep_induction} for the proof.

The following theorem gives a general formula for the moment bounds from recursively applying Lemma~\ref{lem:time_indep+induction_step}. We first need to rewrite the sum over paths (Theorem~\ref{thm:det_sum_over_path}) into a recursive form (Fig.~\ref{fig:sum_over_paths}).

\begin{thm}[Operator growth bounds for time-independent random Hamiltonians]\label{thm:time_indep_sum_over_path}
$\ $ \\
(A) Theorem~\ref{thm:det_sum_over_path} can be written as a system of recursive differential equations. For any path $\gamma$ backtracking some self-avoiding path $\Gamma$ from target site $r$,
\begin{align}
    \displaystyle O^{t+dt}_\gamma = e^{\CL_{\gamma}dt } \left[O^t_\gamma+ \sum_{X\in \Delta(\gamma)} \CL_{X}[O^t_{\gamma+X}]dt \right] +\CO(dt^2),
\end{align}
where $\CL_\gamma=\CL^{(\Gamma)}_k$ is the intermediate unitary associated with $\gamma$, defined in Theorem~\ref{thm:det_sum_over_path}; $\Delta(\gamma)$ is the set of possible next steps of $\gamma$. \\
(B)The commutator between local operators $O_0,A_r$ can be bounded by a weighted sum over self-avoiding paths of interactions $\Gamma=\{X_\ell,\ldots,X_1\}\in \mathcal{S}_{r0}$.
\begin{align}
\vertiii{\frac{1}{2}[A_r,O_0(t)]}_{*}^2
&\le \sum_{\Gamma} \int\limits_{t<t_\ell<\cdots <t_1} \mathrm{d}t_\ell\cdots \mathrm{d}t_1 \prod_{1\le k\le \ell} \left(e^{\beta_k(t_{k+1}-t_k)} \frac{4p b_{X_k}^2}{\beta_k} \right) \vertiii{O_0}_{*}^2,
\end{align}
for expected p-th moments $\vertiii{ \cdot}_*= \vertiii{ \cdot}_{p}, \vertiii{ \cdot}_{D_P,p}$. Also, $\beta_k$ are tunable parameters and $t_{k+1}=t$. 
\end{thm}
The upshot is that the possible next-steps $X\in \Delta(\gamma)$ are weighted in squares $b^2_X$, as hinted by Lemma~\ref{lem:time_indep+induction_step}. The proof crucially relies on the next steps $\CL_X$ being conditionally independent on all of $O^t_{\gamma+X}$, i.e. none of $O^t_{\gamma+X}$ depends on any of $X\in \Delta(\gamma)$. This is why we need the delicate self-avoiding decomposition (Theorem~\ref{thm:det_sum_over_path}). See Sec.~\ref{proof:time_indep} for the proof. Towards a final concentration, use Markov's inequality on the p-th moment bounds. The tail behavior depends on the class of Hamiltonian so we do not include here.

\subsection{A general formula for Brownian Hamiltonians}
\label{sec:brown_mainresults}
Consider a time-dependent random circuit as product of $T$ independent unitary evolutions 
\begin{equation}
    U(T) = \exp\left({\sum -iH^{T}_X \xi }\right) \cdots \exp\left({\sum -iH^1_X \xi }\right),
\end{equation}
where the interactions $H^{T'}_X$ are independent across different times ${T'}$ or different terms $X$ and bounded $\lV H^{T'}_X\rV\le b_X$. The Brownian bounds will be in strong parallel with the time-independent case except for complications due to Brownian treatment on time steps. Our bounds work for discrete time steps, but we can take the Brownian rescaling limit to eliminate finite time step effect by $\xi^2 \rightarrow 0$ holding $\tau:=T\xi^2$ fixed, which is the Brownian 'time'. Note for technical reasons\footnote{This will make uniform smoothness useless and we have to upgrade the martingale inequality to more complicated ones and make the combinatorial bounds nasty.} we do not start with a Gaussian tail as in \cite{Lashkari_2013,brownian_SYK}. We will take limit from bounded steps, which also converges in probability to Brownian motion (Donsker's Theorem, see, e.g.~\cite{revuz_yor_2005}). This saves us some functional analysis and makes our derivations more transparent. 


We once again start by replacing the triangle inequality. Note we take discrete time steps to be cautious with the Brownian rescaling limit.
\begin{lem}[Moment bounds for Brownian operator finite difference equation]\label{lem:induction_step}
Consider a finite difference equation for operator-valued random variables $A^T, B^T_j$
\begin{align}
    A^{T+1} & = \hat{U}\left[ A^{T}+(\hat{V}^{T+1}-I)[B^{T}]\right]
    \end{align}
generated by time-depedent random unitaries $U, V^T$. 
In particular, assume $V^{T+1} = \exp(-iH^{T+1}\xi)$ is generated by a random Hamiltonian which is bounded $\lV H^{T+1} \rV \xi \le \eta$ and zero mean conditioned on the past $\BE_{T}[H^{T+1}]=0$. Then, 
\begin{align}
    \vertiii{A^{T+1}}_*^2 &\le  (1+\frac{\eta^2}{p})\vertiii{A^{T}}_{*}^2+ 8p\eta^2g(\eta)\vertiii{B^{T}}_{*}^2
\end{align}
for the expected p-th moments $\vertiii{ \cdot}_*=\vertiii{ \cdot}_{p}, \vertiii{ \cdot}_{D_P,p}$. Also, $g(\eta):= 1+\eta+\eta^2/2$.
\end{lem}
Note $g(\eta)$ approaches $1$ in the Brownian limit $\xi \rightarrow 0$. Remarkably, the interaction strengths come in as squares, strongly reflecting the stochastic nature. See Sec.~\ref{pf:induction_step} for the proof.
 
Our main theorem gives a general integral formula for moment bounds from recursively applying Lemma~\ref{lem:induction_step}. We first need a recursive form of Theorem~\ref{thm:det_sum_over_path} to higher order to take care of the Brownian limit.
\begin{thm}[Operator growth bounds for time-dependent random Hamiltonian.]\label{thm:sum_over_paths}
Consider a discrete time-dependent random circuit. Then\\ (A)Theorem~\ref{thm:det_sum_over_path} can be written as a system of recursive finite time-difference equations for any path $\gamma$ bactracking some self-avoiding path $\Gamma$ from target site $r$. Keeping the leading order (zero mean term at order $\xi$, non-zero mean term at $\xi^2$), 
\begin{align}
     O^{T+1}_\gamma &= \exp({\CL_{\gamma}\xi }) \left[[O^T_\gamma]+ \sum_{X\in \Delta(\gamma)} (I-e^{-\CL_{X}\xi})[O^T_{\gamma+X}] \right] +\CO(\xi^3)M +  \CO(\xi^2)Z.
\end{align}
where $\CL_\gamma=\CL^{(\Gamma)}_k$ is the intermediate unitary associated with $\gamma$, defined in Theorem~\ref{thm:det_sum_over_path}; $\Delta(\gamma)$ is the set of possible next steps of $\gamma$;  $Z$ is zero mean. \footnote{Both of which vanishes in the Brownian limit.}\\
(B) In the continuum (Brownian) limit $\xi \rightarrow 0, \tau:=T\xi^2 $, the commutator between local operators $O_0,A_r$ can be bounded by a weighted sum over self-avoiding paths  $\Gamma=\{X_\ell,\ldots,X_1\}\in \mathcal{S}_{r0}$
\begin{align}
\vertiii{\frac{1}{2}[A_r,O_0(\tau)]}_{*}^2
&\stackrel{\xi\rightarrow 0}{\le} \sum_{\Gamma} \displaystyle\int\limits_{\tau> \tau_\ell>\cdots> \tau_1\ge 0}\mathrm{d}\tau_\ell\ldots\mathrm{d}\tau_1 \prod_{1\le k\le \ell} \left(\exp{\left(\sum_{X \in  \Delta(k+1)}\frac{b_X^2(\tau_{k+1}-\tau_k)}{p}\right)} \cdot 8pb_{X_k}^2\right)  \vertiii{O_0}_{*}^2,\label{eq:brown_integral}
\end{align}
for the expected p-th moments $\vertiii{ \cdot}_*=\vertiii{ \cdot}_{p}, \vertiii{ \cdot}_{D_P,p}$. Also, $\tau_{\ell+1} =\tau,\ \Gamma(\ell+1)=\{r\}$, $\lV H_X\rV \le b_X$.
\end{thm}
In comparision with time-independent results, self-avoiding is not as crucial because we have more randomness at our disposal. See Sec.~\ref{sec:proof_brown_thm} for the proof.

\section{Evaluating the main theorem on systems}\label{sec:results_systems}
We will plug in our general formula (Theorem~\ref{thm:sum_over_paths}, Theorem~\ref{thm:time_indep_sum_over_path}) for various notable systems, and some of them require a moderate combinatorial effort. For generality, our OTOC bounds will be in terms of OTOC with low rank inputs. To recover bounds on OTOC with non-random states, the reader may set $D_P=1$ and do a word by word replacement in this section
\begin{align}
    \lV  [A_{r},O_0(\cdot)]P \rV \rightarrow \sqrt{\tr\bigg(\rho [A_r,O_0(\cdot)]^\dagger[A_r,O_0(\cdot)]\bigg)}, 
\end{align} 
since moments of OTOC with low-rank input bound the moments of OTOC with non-random state (Proposition~\ref{prop:lowrank_to_mixed}) and all concentration bounds below are derived from moment bounds.

For OTOC, our bounds work for most systems known to have deterministic bounds, expect for the $d>1$ long range interacting system where the nested-commutator expansion is ill-behaved~\cite{strictlylinear_KS}.
Unfortunately, the concentration of spectral norm requires more sophisticated control of the support of the operators, unlike OTOC. We only thus far obtain results for $1d$ short-ranged and $1d$ power-law systems where the control of support is more accessible.
\subsection{Short-ranged interacting systems}\label{sec:shortranged}
 Here we study the nearest neighbor interacting systems. Due to the simplicity of the interacting terms, they all have a  discrete bound very similar to the continuum, i.e. the finite time step effect is under control. 
\begin{cor}[Brownian $d=1$ nearest-neighbour system, OTOC]
Consider a $d=1$ nearest-neighbour Brownian Hamiltonian where each term is normalized $\lV H^{({T})}_{i,i+1}\rV\le 1$ and independent, zero mean conditioned on the past $\BE_{T-1}[H^{({T})}_{i,i+1}]=0$
\begin{equation}
     H^{({T})} :=\sum_{i} a H^{({T})}_{i,i+1}.
\end{equation}
Then the OTOC concentrates, for all projector to input subspace of dimension $dim(P)=D_P$,
\begin{align}
    \BP\bigg[ \frac{1}{2}\lV [A_{r},O_0(\tau)]P \rV \ge \epsilon \bigg] &\le\begin{cases}
     \displaystyle D_Pe^{a^2\tau/2}  \exp(-\frac{\epsilon^{2/r}}{16e^2\frac{a^2\tau}{r^2}}) & \text{if } (16e^2\frac{a^2\tau}{r})^{r} < \epsilon^2\\
     \displaystyle\frac{D_P}{\epsilon^2}e^{a^2\tau/2}(16e\frac{a^2\tau}{r})^{r} & \text{if } (16e^2\frac{a^2\tau}{r})^{r} \ge \epsilon^2.
     \end{cases}
\end{align}
or equivalently with failure probability $\delta$, $ \BP[ \frac{1}{2}\lV [A_{r},O_0(T)]P \rV \ge \epsilon(\delta) ]\le \delta$,
\begin{align}
    \epsilon(\delta):= 
    \begin{cases}
    \displaystyle\left[16e^2\frac{a^2\tau}{r^2}(\frac{a^2\tau}{2}+\ln(D_P/\delta))\right]^{r/2} &\text{if } \delta <D_Pe^{a^2\tau/2-r} \\
    \\
    \displaystyle\sqrt{ \frac{D_P}{\delta}e^{a^2\tau/2}(16e\frac{a^2\tau}{r})^{r}} &\text{if } \delta \ge D_Pe^{a^2\tau/2-r}.
    \end{cases} 
\end{align}
For any threshold $\epsilon >0$, points  outside ($\zeta$>0) the light cone has small OTOC almost surely
\begin{equation}
    \frac{r}{\tau(r)} = u = 16ee^{1/32e} a^2 +\zeta.\ \ \ \text{ ensures     }\ \ \ \limsup_{r\rightarrow \infty} \BP\bigg[\lV\frac{1}{2}[A_{r},O_{0}(\tau(r))]P\rV \ge \epsilon \bigg] = 0. 
\end{equation}

\end{cor}
The logarithmic dependence on failure probability $\delta$ captures the very strong tail bound that might be useful elsewhere. See Sec.~\ref{sec:1dnn_OTOC} for the proof as well as the discrete version. The operator norm bound depends on the local dimension $D$ and takes a more complicated form.
\begin{prop}[Brownian $d=1$ nearest-neighbour system, spectral norm]
Consider a $d=1$ nearest-neighbour Brownian Hamiltonian with local Hilbert space dimension $D$ and each term is normalized $\lV H^{({T})}_{i,i+1}\rV\le 1$ and independent zero mean conditioned on the past $\BE[H^{({T})}_{i,i+1}]=0$.
\begin{equation}
     H^{({T})} :=\sum_{i} a H^{({T})}_{i,i+1}.
\end{equation}
Then the operator norm concentrates for arbitrary failure probability $\delta_0/(1-\lambda)$
\begin{align}
\BP\left[\frac{1}{2}\lV [A_{r},O_0(\tau)] \rV\ge \frac{\left[\left( \ln(2D/\lambda) +\frac{1}{16e^2}- \frac{\ln \delta_{0}}{r}\right) \frac{16\mathrm{e}^2 a^2\tau}{r}\right]^{r/2}}{ 1- \sqrt{\left( \ln(2D/\lambda) +\frac{1}{16e^2}- \frac{\ln \delta_{0}}{r}\right) \frac{16\mathrm{e}^2 a^2\tau}{r} } }\right] \le \frac{\delta_0}{1-\lambda}.
\end{align} 
For any threshold $\epsilon >0$, commutator outside the light cone ($\zeta>0$) has small spectral norm almost surely
\begin{equation}
\frac{r}{\tau(r)}= 16\mathrm{e}^2a^2\left( \ln(2D) +\frac{1}{16e^2}\right)+\zeta  \ \ \ \text{ ensures     }\ \ \ \limsup_{r\rightarrow \infty} \BP\bigg[\lV\frac{1}{2}[A_{r},O_{0}(\tau(r))]P\rV \ge \epsilon \bigg] = 0. 
\end{equation}

\end{prop}
See Sec.~\ref{1d_nn_LR} for the proof as well as the discrete version. 
Note we can also simplify the expression by adding some overheads
\begin{align}
\BP\left[\frac{1}{2}\lV [A_{r},O_0(\tau)] \rV  \ge \frac{1}{ 1-1/\sqrt{2}}\left[( \ln (4D) + \frac{1}{16e^2}- \frac{\ln \delta_{0}}{r}) \frac{32\mathrm{e}^2a^2 \tau}{r }\right]^{r/2}\right] \le 2\delta_0.
\end{align} 
The dimensional factor $\CO(\ln(D^r))$is due to the operator supported on $\CO(r)$ sites of local dimension $D$. The complication is due to union bound over local support sizes of the operator. 
For $d>1$, unfortunately we only have a bound for OTOC.
\begin{cor}[Brownian $d>1$ nearest-neighbour system, OTOC]
Consider a $d>1$ nearest-neighbour Brownian Hamiltonian  where each term is normalized $\lV H^{({T})}_{x,x'}\rV\le 1$ and independent, zero mean conditioned on the past $\BE[H^{({T})}_{i,i+1}]=0$.
\begin{equation}
     H^{({T})} :=\sum_{<x,x'>} a H^{({T})}_{x,x'}
\end{equation}
Then the OTOC concentrates, for all projector to input subspace of dimension $dim(P)=D_P$,\begin{align}
     \BP[ \frac{1}{2}\lV [A_{r},O_0(\tau)]P \rV \ge \epsilon ] &\le\begin{cases}
     \displaystyle \frac{D_P}{1- 1/e}e^{da^2\tau} \exp(-\frac{\epsilon^{2/r}}{32e^2d\frac{a^2\tau}{r^2}}) & \text{if } (32e^2d\frac{a^2\tau}{r})^{r} < \epsilon^2\\
     \\
    \displaystyle \frac{D_P}{\epsilon^2}e^{da^2\tau}\frac{(32ed\frac{a^2\tau}{r})^{r}}{1-32ed\frac{a^2\tau}{r}} & \text{if }  (32e^2d\frac{a^2\tau}{r})^{r} \ge \epsilon^2.
     \end{cases}
\end{align}
or equivalently with failure probability $\delta$, $ \BP[ \frac{1}{2}\lV [A_{r},O_0(\tau)]P \rV \ge \epsilon(\delta) ]\le \delta$,
\begin{align}
    \epsilon(\delta):= 
    \begin{cases}
   \displaystyle\left[32e^2d\frac{a^2\tau}{r^2}(da^2\tau+\ln(\frac{D_P}{(1-1/e)\delta}))\right]^{r} &\text{if } \delta <\frac{D_P}{1-1/e}e^{a^2\tau d-r} \\
    \\
   \displaystyle \sqrt{\frac{D_Pe^{da^2\tau}}{\delta}\frac{(32ed\frac{a^2\tau}{r})^{r}}{1-32ed\frac{a^2\tau}{r}}}  &\text{if } \delta \ge \frac{D_P}{1-1/e}e^{a^2\tau d-r}
    \end{cases} 
\end{align}
For any threshold $\epsilon >0$, points  outside ($\zeta$>0) the light cone has small OTOC almost surely
\begin{equation}
    \frac{r}{\tau(r)} = u \ge 32ede^{1/32e} a^2+\zeta  \ \ \ \text{ ensures     }\ \ \ \limsup_{r\rightarrow \infty} \BP\bigg[\lV\frac{1}{2}[A_{r},O_{0}(\tau(r))]P\rV \ge \epsilon \bigg] = 0. 
\end{equation}

\end{cor}
See Section~\ref{sec:dnn_OTOC} for the proof as well as the discrete version.
\subsection{SYK-like k-local systems}\label{sec:k-local}
Random complete $k$-local system hosts most notably the SYK model and also spin liquids. From a quick application of Theorem~\ref{thm:sum_over_paths}, here we obtain a $\tau \ge \Omega(\log(N))$ upper bound from OTOC growth matching the scrambling time for both time-independent and Brownian Hamiltonians. Also, $k$ can be taken as large as $N^{1-\zeta}, \zeta >0$ without hindering the logarithm $\log(N)$.
\begin{cor}[time-independent $k$-local system]
Consider a complete $k$-body time-independent Hamiltonian where each term is normalized $\lV H^{}_{i_1\cdots i_k} \rV\le 1$, independent, and zero mean $\BE[H^{}_{i_1\cdots i_k}]=0$.
\begin{equation}
    H^{}_{k,N} :=\sum_{i_1< \ldots <i_k \le N}  \sqrt{\frac{J^2 (k-1)!}{kN^{k-1}}} H^{}_{i_1\cdots i_k}. 
\end{equation}
Then the OTOC concentrates, for all projector to input subspace of dimension $dim(P)=D_P$,
 \begin{align}
     \BP[\frac{1}{2}\lV  [A_{r},O_0(t)]P \rV \ge \epsilon ] &\le \begin{cases}
     D_P\exp \left( -(\frac{\ln(N\epsilon^2/(k-1))}{6})^3\frac{1}{J^2t^2}\right) &\text{if } (\frac{\ln(N\epsilon^2/(k-1))}{6Jt})^2>2\\
     \\
      D_P\left(e^{4Jt} \frac{(k-1)}{N\epsilon^2} \right) &\text{if } (\frac{\ln(N\epsilon^2/(k-1))}{6Jt})^2\le 2
     \end{cases} .
 \end{align}
Expressing in terms of $\delta$, $\BP\left[\frac{1}{2}\lV  [A_{r},O_0(t)]P \rV > \epsilon(\delta) \right]\le \delta$,
\begin{align}
    \epsilon(\delta):= 
    \begin{cases}
    \displaystyle \frac{k-1}{N}\exp \left(6\sqrt[3]{\ln(D_P/\delta)J^2t^2}\right)&\text{if } \delta < D_P e^{(4-6\sqrt{2})Jt} \\
    \\
    \displaystyle \sqrt{ D_P\left(e^{4Jt} \frac{(k-1)}{N\delta} \right)} &\text{if } \delta \ge D_P e^{(4-6\sqrt{2})Jt}
    \end{cases} .
\end{align}
For any threshold $\epsilon>0$, there is a scrambling time $t(N)$ below which ($\zeta>0$) the OTOC is small almost surely
\begin{align}
    D_P\left(e^{4Jt} \frac{(k-1)}{N} = \frac{1}{N^{\zeta}}\right) \ \ \ \text{ ensures     }\ \ \ \limsup_{N\rightarrow \infty} \BP\left(\lV\frac{1}{2}\left[A_{r},O_{0}(t(N))\right]P\rV \ge \epsilon \right) = 0, 
\end{align}
with a bound on asymptotic scrambling time in the large $N$ limit 
\begin{align}
    \displaystyle \frac{Jt_{scr}}{\ln(\frac{N}{k-1})} \stackrel{a.s.}{\ge} \frac{1}{4}.
\end{align}
\end{cor}
See Sec.~\ref{sec:proof_time-Indep_klocal} for the proof.
\begin{cor}[Brownian $k$-local system]
Consider a a complete $k$-body Brownian Hamiltonian where each term is normalized $\lV H^{(T)}_{i_1\cdots i_k} \rV\le 1$, independent, and zero mean conditioned on the past $\BE_{T-1}[H^{(T)}_{i_1\cdots i_k}]=0$
\begin{equation}
    H^{(\tau)}_{k,N} :=\sum_{i_1< \ldots <i_k \le N}  \sqrt{\frac{J^2 (k-1)!}{kN^{k-1}}} H^{(\tau)}_{i_1\cdots i_k}.
\end{equation}
Then the OTOC concentrates, for all projector to input subspace of dimension $dim(P)=D_P$,
 \begin{align}
     \BP[\frac{1}{2}\lV  [A_{r},O_0(\tau)]P \rV \ge \epsilon ] &\le \begin{cases}
     \displaystyle D_Pe^{J^2\tau/2} \exp({\frac{-[\ln(N\epsilon^2/(k-1))]^2}{32J^2\tau}}) &\text{if } \frac{\ln(N\epsilon^2/(k-1))}{16\tau J^2} > 2\\
    \displaystyle  D_Pe^{17J^2\tau/2}\frac{(k-1)}{N\epsilon^2} &\text{ if } \frac{\ln(N\epsilon^2/(k-1))}{16\tau J^2} \le 2
     \end{cases}.
 \end{align}
Expressing in terms of $\delta$, $\BP\left[\frac{1}{2}\lV  [A_{r},O_0(\tau)]P \rV \ge \epsilon(\delta) \right]\le \delta$,
\begin{align}
    \epsilon(\delta):= 
    \begin{cases}
    \displaystyle \sqrt{ \frac{(k-1)}{N}\exp(\sqrt{ 32J^2\tau (J^2\tau/2 -\ln(\delta/D_P) ) }) } &\text{if }  \delta < D_P e^{-23J^2\tau/2} \\
    \\
    \displaystyle \sqrt{ D_Pe^{17J^2\tau/2}\frac{k-1}{N\delta}} &\text{if } \delta \ge D_P e^{-23J^2\tau/2}
    \end{cases} .
\end{align}
For any threshold $\epsilon>0$, there is a scrambling time $\tau(N)$ below which ($\zeta>0$) the OTOC is small almost surely
\begin{align}
    D_P\left(e^{17J^2\tau/2} \frac{(k-1)}{N} = \frac{1}{N^{\zeta}}\right) \ \ \ \text{ ensures     }\ \ \ \limsup_{r\rightarrow \infty} \BP\left(\lV\frac{1}{2}\left[A_{r},O_{0}(\tau(N))\right]P\rV \ge \epsilon \right) = 0,
\end{align}
with a bound on asymptotic scrambling time in the large $N$ limit 
\begin{align}
   \displaystyle  \frac{J^2\tau_{scr}}{\ln(\frac{N}{k-1})} \stackrel{a.s.}{\ge} \frac{2}{17}.
\end{align}
\end{cor}
See Section~\ref{sec:klocal_OTOC} for the proof. 
A similar result in the time-independent random SYK was obtained using intense, specialized combinatorics~\cite{Lucas_2020}, here our proof is a consequence of the general formula (Theorem~\ref{thm:time_indep_sum_over_path}) and is shorter and more flexible. Our Brownian bound is new. 
\subsection{Power-law interaction systems in $d=1$ dimension }\label{sec:longrange}
Power-law interacting systems recently attracted attention due to their presence in almost every realistic system in lab. These models interpolate between an all-to-all interacting system ($\alpha=0$) and a local system ($\alpha=\infty$), and the intermediate regime has been thoroughly studied. Theoretically, not only a polynomial light cone exists~\cite{Tran_2019_polyLC}, there are also critical value of $\alpha$s beyond which a strictly linear light cone emerges~\cite{strictlylinear_KS,hierarachy,alpha_3_chenlucas}, i.e. the system is effectively local. In particular, the transition value also depends on the notion of scrambling, whether it is the Forbenius norm $(\alpha\le 5/2)$ or spectral norm ($\alpha =3$). However, numerical evidences and theoretical argument suggest for chaotic time-independent~\cite{Levy} (and Brownian~\cite{long_range_brownian}) Hamiltonians an even smaller critical value of $\alpha$ for which the system appears effectively local. In this section, we plug in our main theorem and obtain a upper bound matching the few proposals for the linear light cone cut-off at $\alpha = 1.5$~\cite{Levy,long_range_brownian}. Unfortunately we cannot prove for $\alpha =1$, proposed in a numerical study~\cite{alpha_1_numerics}.  
\begin{thm}[Brownian $d=1$ power-law interacting system, OTOC]\label{thm:brown_long_OTOC}
Consider a $d=1$ power law Brownian Hamiltonian where each term is normalized $\lV H^{({T})}_{ij}\rV\le 1$, independent, and zero mean conditioned on the past $\BE_{T-1}[H^{({T})}_{ij}]=0$
\begin{equation}
     H^{({T})} :=\sum_{i} \frac{1}{|i-j|^\alpha} H^{({T})}_{ij}.
\end{equation}
Let $\alpha >1$, define
\begin{align}
    \begin{cases}
    \beta = 1 \ &if \ \alpha >3/2\\
    \beta = 2\alpha-2 \ &if \ 3/2>\alpha >1\\
    \end{cases}.
\end{align}
Then the OTOC concentrates, for all projector to input subspace of dimension $dim(P)=D_P$,
\begin{align}
     \BP[ \frac{1}{2}\lV [A_{r},O_0(\tau)]P \rV \ge \epsilon ] &\le\begin{cases}
     \displaystyle D_P\exp(-\frac{r^\beta\epsilon^2}{c_\alpha\tau}) &\text{if } \epsilon^2\ge  c_\alpha\frac{\tau}{r^\beta}\\
     \\
     \displaystyle D_P\frac{c_\alpha\tau}{er^\beta\epsilon^2} &\text{if } \epsilon^2\le c_\alpha\frac{\tau}{r^\beta}
     \end{cases},
\end{align}
or equivalently with failure probability $\delta$, $ \BP[ \frac{1}{2}\lV [A_{r},O_0(\tau)]P \rV \ge \epsilon(\delta) ]\le \delta$,
     \begin{align}
    \epsilon(\delta):= 
    \begin{cases}
    \displaystyle \sqrt{c_\alpha\ln(D_P/\delta)\frac{\tau}{r^\beta}} &\text{if } \delta <\frac{D_P}{e} \\
    \\
    \displaystyle \sqrt{D_P\frac{c_\alpha}{e\delta}\frac{\tau}{r^\beta}} &\text{if } \delta \ge \frac{D_P}{e}
    \end{cases} .
\end{align}
For any threshold $\epsilon >0$, points  outside ($\zeta$>0) the light cone has small OTOC almost surely
\begin{align}
     \tau(r) = rf(r)  \ \ \ \text{ ensures     }\ \ \ \limsup_{r\rightarrow \infty} \BP\bigg[\lV\frac{1}{2}[A_{r},O_{0}(\tau(r))]P\rV \ge \epsilon \bigg] = 0,
\end{align}
with any function approaching infinity arbitrarily slowly, $\lim_{r\rightarrow\infty} f(r)=\infty$.
We also have a strictly linear light cone with finite failure probability 
\begin{align}
    \frac{r^\beta}{\tau} = \frac{\ln(D_P/\delta)}{c_\alpha\epsilon^2},
\end{align}
for $\epsilon >0$ and $1/e\ge \delta >0$. 
\end{thm}
See Section~\ref{sec:proof_brown_1d_long_OTOC} for proof.
In addition to OTOC, which was the central quantity in recent studies on chaotic power law system, here we append the calculation results for the operator norm, which yield yet another value of critical $\alpha$, in parallel with the hierarchy in deterministic case~\cite{hierarachy}. Interestingly, the empirical value of critical $\alpha$ in a numerical study of chaotic (non-random) time-independent spin chain~\cite{Levy} matched the prediction from OTOC yet not the Lieb-Robinson-like bound on spectral norm.
\begin{thm}[Brownian $d=1$ power-law interacting system, spectral norm]
Consider the same Brownian model as Thoerem~\ref{thm:brown_long_OTOC}. Let $\alpha >3/2$, define
\begin{align}
    \begin{cases}
    \beta = 1 \ &if \ \alpha >2\\
    \beta = 2\alpha-3 \ &if \ 2 >\alpha >3/2\\
    \end{cases}.
\end{align}
Then the spectral norm of commutator concentrates \begin{align}
    \BP(\frac{1}{2}\lV  [A_{r},O_0(\tau)] \rV \ge \epsilon )\le  \begin{cases}
     \displaystyle D^{a_\alpha}\exp\left(-c_\alpha\frac{r\epsilon^2}{\tau}\right) & \text{if } \alpha >2\\
    \displaystyle b_\alpha D^{a_\alpha r^{1-\alpha/2}} \exp\left(-{c_\alpha\frac{\epsilon^2r^{3\alpha/2 -2}}{\tau}}\right) &\text{if }2>\alpha >3/2
    \end{cases},
\end{align}
for some absolute constants $a_\alpha,b_\alpha,c_\alpha$ only depends on $\alpha$ and $D$ is the single site Hilbert space dimension. Or equivalently with failure probability $\delta$, $ \BP[ \frac{1}{2}\lV [A_{r},O_0(\tau)] \rV \ge \epsilon(\delta) ]\le \delta$,
     \begin{align}
    \epsilon(\delta):= 
    \begin{cases}
    \displaystyle\sqrt{\ln(D^{a_\alpha}/\delta)\frac{\tau}{c_\alpha r}} &\text{if } \alpha >2 \\
    \displaystyle \sqrt{
    \frac{\tau}{c_\alpha r^{3\alpha/2-2}}\left(\ln(D)a_\alpha r^{1-\alpha/2} - \ln(\delta/b_\alpha) \right)}  &\text{if }2>\alpha >3/2
    \end{cases} .
\end{align}
For any threshold $\epsilon >0$, points  outside the (algebraic) light cone has small OTOC almost surely
\begin{align}
    \begin{cases}
    \tau(r) = rf(r) &\text{if } \alpha >2\\
    \\
    \displaystyle \frac{r^{2\alpha-3}}{\tau} = \frac{\ln(D)a_\alpha}{c_\alpha\epsilon^2}+\zeta &\text{if }2>\alpha >3/2
    \end{cases} \ \ \ \text{ ensures     }\ \ \ \limsup_{r\rightarrow \infty} \BP\bigg[\lV\frac{1}{2}[A_{r},O_{0}(\tau(r))]\rV \ge \epsilon \bigg] = 0. 
\end{align}
 where $\zeta>0$, and $f(r)$ is any function approaching infinity arbitrarily slowly $\lim_{r\rightarrow\infty} f(r)=\infty$.
For $\alpha >2 $ we also have a strictly linear light cone but with finite failure probability 
\begin{align}
    \frac{r}{\tau} = \frac{\ln(D^{a_\alpha}/\delta)}{c_\alpha\epsilon^2}.
\end{align}
\end{thm}
See Sec.~\ref{sec:proof_brown_1d_long_LR} for the proof.

We also dervie bounds for the time-independent cases, with different critical values of $\alpha.$

\begin{thm}[Time-independent $d=1$ power-law interacting system, state dependent OTOC ]
Consider a $d=1$ power law Hamiltonian where each term is normalized $\lV H_{ij}\rV\le 1$, independent, and zero mean $\BE[H_{ij}]=0$
\begin{equation}
     H :=\sum_{i} \frac{1}{|i-j|^\alpha} H_{ij}.
\end{equation}
Let $\alpha >1$, define
\begin{align}
    \begin{cases}
    \beta = 1 \ &if \ \alpha >2\\
    \beta = \alpha- 1\ &if \ 2>\alpha >1\\
    \end{cases}.
\end{align}
Then the OTOC concentrates, for all projector to input subspace of dimension $dim(P)=D_P$,
\begin{align}
     \BP[ \frac{1}{2}\lV [A_{r},O_0(t)]P \rV \ge \epsilon ] &\le\begin{cases}
     \displaystyle D_P\exp(-\frac{r^{2\beta}\epsilon^2}{c_\beta t^2}) &\text{if } \epsilon^2\ge  c_\alpha\frac{t^2}{r^{2\beta}}\\
     \\
    \displaystyle D_P\frac{c_\alpha t^2}{er^{2\beta}\epsilon^2} &\text{if } \epsilon^2\le c_\alpha\frac{t^2}{r^{2\beta}},
     \end{cases}
\end{align}
or equivalently with failure probability $\delta$, $ \BP[ \frac{1}{2}\lV [A_{r},O_0(t)]P \rV \ge \epsilon(\delta) ]\le \delta$,
     \begin{align}
    \epsilon(\delta):= 
    \begin{cases}
    \displaystyle\sqrt{c_\alpha\ln(D_P/\delta)\frac{t^2}{r^{2\beta}}} &\text{if } \delta <\frac{D_P}{e} \\
    \\
    \displaystyle \sqrt{\frac{D_Pc_\beta}{e\delta}\frac{t^2}{r^{2\beta}}} &\text{if } \delta \ge \frac{D_P}{e}
    \end{cases} .
\end{align}
For any threshold $\epsilon >0$, points  outside the (algebraic) light cone has small OTOC almost surely
\begin{align}
     t(r) = r^\beta f(r) \ \ \ \text{ ensures     }\ \ \ \limsup_{r\rightarrow \infty} \BP(\lV\frac{1}{2}[A_{r},O_{0}(t(r))]P\rV \ge \epsilon ) = 0,
\end{align}
where $f(r)$ is any function approaching infinity arbitrarily slowly, $\lim_{r\rightarrow\infty} f(r)=\infty$.
Or with finite failure probability,
\begin{align}
    \frac{r^\beta}{t} = \sqrt{\frac{\ln(D_P/\delta)}{c_\alpha\epsilon^2}}
\end{align}
for $\epsilon >0$ and $1/e\ge \delta >0$. 
\end{thm}
See Section~\ref{sec:proof_time_indep_1d_long_OTOC} for proof.
\begin{thm}[Time-independent $d=1$ power-law interacting system, spectral norm]
Let $\alpha >3/2$, define
\begin{align}
    \begin{cases}
    \beta = 1 \ &if \ \alpha >5/2\\
    \beta =\alpha-3/2 \ &if \ 5/2 >\alpha >3/2\\
    \end{cases}.
\end{align}
Then the spectral norm of commutator concentrates 
\begin{align}
    \BP(\frac{1}{2}\lV  [A_{r},O_0(t)] \rV \ge \epsilon )\le  \begin{cases}
     \displaystyle D^{a_\alpha}\exp\left(-c_\alpha\frac{r^2\epsilon^2}{t^2}\right) & \text{if } \alpha >5/2\\
   \displaystyle b_\alpha D^{a_\alpha r^{1-2\alpha/5}} \exp\left(-{c_\alpha\frac{\epsilon^2r^{8\alpha/5 -2}}{t^2}}\right) &\text{if }5/2>\alpha >3/2
    \end{cases},
\end{align}
for some numerical constants $a_\alpha,b_\alpha,c_\alpha$ only depends on $\alpha$ and $D$ is the single site Hilbert space dimension. Or equivalently with failure probability $\delta$, $ \BP[ \frac{1}{2}\lV [A_{r},O_0(t)] \rV \ge \epsilon(\delta) ]\le \delta$,
     \begin{align}
    \epsilon(\delta):= 
    \begin{cases}
   \displaystyle \frac{t}{r}\sqrt{\frac{\ln(D^{a_\alpha}/\delta)}{c_\alpha}} &\text{if } \alpha >5/2 \\
     \displaystyle\frac{t}{r^{\alpha-3/2}}\sqrt{
    \frac{1}{c_\alpha}\left(\ln(D)a_\alpha - \frac{\ln(\delta/b_\alpha)}{r^{1-2\alpha/5}} \right)}  &\text{if }5/2>\alpha >3/2
    \end{cases} .
\end{align}
For any threshold $\epsilon >0$, commutator  outside the (algebraic) light cone has small spectral norm almost surely
\begin{align}
    \begin{cases}
    t(r) = rf(r) &\text{if } \alpha >5/2\\
    \\
    \displaystyle\frac{r^{\alpha-3/2}}{t} = \sqrt{\frac{\ln(D)a_\alpha}{c_\alpha\epsilon^2}}+\zeta &\text{if }5/2>\alpha >3/2
    \end{cases} \ \ \ \text{ ensures     }\ \ \ \limsup_{r\rightarrow \infty} \BP(\lV\frac{1}{2}[A_{r},O_{0}(t(r))]P\rV \ge \epsilon ) = 0,
\end{align}
 where $\zeta>0$, and $f(r)$ is any function approaching infinity arbitrarily slowly $\lim_{r\rightarrow\infty} f(r)=\infty$.For $\alpha >5/2 $ we also have a strictly linear light cone but with finite failure probability 
\begin{align}
    \frac{r}{t} = \sqrt{\frac{\ln(D^{a_\alpha}/\delta)}{c_\alpha\epsilon^2}}.
\end{align}
\end{thm}
See Sec.~\ref{sec:proof_time_indep_1d_long_LR} for the proof.

\section{Proofs}\label{sec:proofs}
We first include a minimal introduction for the matrix concentration results we will be using. This is a vast field and in this work only we import the simple and illustrative ones. There are more refined tools, but they are also more complicated.
\subsection{Matrix-valued martingales}
The theory of statistics does not end at the central limit theorem for asymptotic sum of i.i.d random numbers. The phenomena in the wild are often by no mean identical, independent, a sum, and not in the infinite limit, yet nonetheless appear to robustly concentrate around the mean. Among the zoo of extensions, a (scalar-valued) \textit{martingale} describes a random process that the future has zero mean conditioning on the past. This martingale condition turns out enough for strong concentration results akin to an i.i.d. sum. In our case, the martingale would be \textit{matrix-valued} and we would need tools from non-commutative functional analysis. Historically, the earliest general results were established by Lust-Piquard~\cite{lust_piquard_86}, Lust-Piquard and Pisier~\cite{lust_Pisier_91}, and Pisier and Xu~\cite{Pisier_1997}. Some more recent works and applications include~\cite{tropp2011freedmans,oliveira2010concentration,HNTR20:Matrix-Product, jungeZenq_nc15}. 

In particular, the main martingale tool kit we will be using is call \textit{uniform smoothness}. It is not the tightest, but the simplest and most robust when matrices are bounded--which we assume throughout. Indeed, our goal is to demonstrate martingale techniques can be natural for operator growth in random Hamiltonian (and beyond). More delicate but complicated martingale inequalities exist but may blur the overall picture. Initially, uniform smoothness was a statement on the geometry of the underlying non-commutative spaces (e.g. the Schatten p-norms). The first uniform smoothness results for Schatten p-norm was obtained by Nicole Tomczak-Jaegermann~\cite{Tomczak1974}, related ideas obtained by Piesier~\cite{piesier75}, and sharpen by Ball, Carlen, and Lieb~\cite{uniformconvexLieb94}. Inequalities for martingale can be extracted as a consequence of the geometry~\cite{naor_2012,HNTR20:Matrix-Product}. See Huang et. al~\cite{HNTR20:Matrix-Product} for a user friendly review on uniform smoothness and Naor~\cite{naor_2012} for the comprehensive background.

For a minimal technical introduction (following Tropp~\cite{tropp2011freedmans} and Huang et. al~\cite{HNTR20:Matrix-Product}), consider a filtration of the master sigma algebra $\CF_0\subset \CF_1 \subset \CF_2 \cdots \subset \CF_t \subset \cdots \CF$, where for each $\CF_j$ we denote the conditional expectation $\BE_j$. A martingale is a sequence of random variable $Y_t$ adapted to the filtration $\CF_t$ such that
\begin{align}
    \sigma(Y_t) &\subset \CF_t &\textrm{(causality)},\\
    \BE_{t-1} Y_t &= Y_{t-1} &\textrm{(status quo)}.
\end{align}
Intuitively, we can think of $t$ as a 'time' index and $\CF_t$ hosts possible events happening before $t$. The present at most depends on the past ('causality') and the tomorrow is in expectation the same as today ('status quo'). For simplicity, we often subtract the mean to obtain a \textit{martingale difference} sequence $D_t:=Y_t-Y_{t-1}$ such that
\begin{align}
    \BE_{t-1} D_t = 0.
\end{align}

We will get tail bounds via Markov's inequality. Denote the appropriate expected moments (or formally called the $L_q(S_p)$-norm),
\begin{equation}
    \vertiii{X}_{p,q} := \left(\BE[\lV X\rV_p^q]\right)^{1/q},
\end{equation}
where the Schatten p-norm as the sum over p-th power of singular values $\nu_i$ is denoted by
\begin{equation}
    \lV X\rV_p :=\tr [(X^\dagger X)^{p/2} ] = (\sum_i \nu_i^p)^{1/p}.
\end{equation}
The main input from uniform smoothness is replacing the deterministic triangle inequality, which is linear, with the following martingale moment bounds, which are quadratic. The inequalities for different underlying spaces are largely in the same format except for constants.\\

\textbf{(i)The operator norm.}
We will use a more general fact than we need,
\begin{prop}[{Subquadratic Averages.~\cite{ricardXu16},\cite[Proposition~4.3]{HNTR20:Matrix-Product}}] \label{prop:sub_average_pq}
Consider random matrices $X, Y$ of the same size that satisfy
$\BE[Y|X] = 0$. When $2 \le q \le p$,
\begin{equation}
\vertiii{X+Y}_{p,q}^2 \le \vertiii{X}_{p,q}^2  + C_p\vertiii{Y}_{p,q}^2 
\end{equation}
The constant $C_p = p - 1$ is the best possible.
\end{prop}
We will simply set $p=q$ to estimate the operator norm, 
\begin{align}
    (\BE \lV O\rV^p )^{1/p}\le (\BE \lV O\rV^p_p)^{1/p}=:\vertiii{O}_p .
\end{align}

\textbf{(ii)Low rank inputs.} For our purpose in physics, we often care about an input state/subspace on which the commutator is evaluated. We therefore extend uniform smoothness to the following:
\begin{cor}[{Subquadratic Averages with low rank input}] \label{prop:sub_average_2q_DP}
Consider random matrices $X, Y$ of the same size that satisfy
$\BE[Y|X] = 0$.
Define 
\begin{align}
\vertiii{X}_{{D_P},p}:=\sup_{rank(P)=D_P}\left(\BE[\lV XP\rV_p^p]\right)^{1/p} ,
\end{align}
where the supremum is taken over projectors $P$ with rank $D_P$.
When $2 \le p$,
\begin{align}
\vertiii{X+Y}_{{D_P},p}^2 &= \sup_{rank(P)=D_P}\vertiii{XP+YP}_{p}^2\\
&\le\sup_{rank(P)=D_P}( \vertiii{XP}_{p}^2  + C_p\vertiii{YP}_{p}^2)\\
&\le \sup_{rank(P)=D_P}\vertiii{XP}_{p}^2 + C_p\sup_{rank(P)=D_P}\vertiii{YP}_{p}^2\\
&= \vertiii{X}_{{D_P},p}^2+C_p\vertiii{Y}_{D_P,p}^2
\end{align}
with constant $C_p = p - 1$. 
\end{cor} 
This in fact upper bounds OTOC for non-random (mixed) states. Take the input subspace to be 1-dimensional,
\begin{prop}\label{prop:lowrank_to_mixed}
\begin{align}
\BE[\tr(\rho X^\dagger X)^{p/2}]^{2/p} \le \sup_{\ket{\psi}}\BE[\bra{\psi} X^\dagger X\ket{\psi}^{p/2}]^{2/p} = \vertiii{X}_{{D_P=1},p}^2.
\end{align}
\end{prop}
The proof decomposes $\rho$ into convex combination of pure states and uses Minkowski inequality for scalar $\ell_{p/2}$ space. \\
\textbf{(iii) general Schatten p-norms.} More generally, uniform smoothness for $(\BE[\lV X\rV_p^q])^{1/q}$ holds\footnote{Thanks Joel Tropp for pointing this out to me.} at any $p,q\ge 2$. It implies concentration of Schatten norms for arbitrary $p$, although the constants are not yet sharp. Their usage is identical except for different constants in Lemma~\ref{lem:induction_step} and Lemma~\ref{lem:time_indep+induction_step}, so we do not propagate these numerical values explicitly.
\begin{prop}[Uniform smoothness of $L_q(S_p)$] \label{prop:general_pq}
Consider random matrices $X, Y$ of the same size that satisfy
$\BE[Y|X] = 0$. When $2 \le q, p$,
\begin{equation}
\vertiii{X+Y}_{p,q}^2 \le \vertiii{X}_{p,q}^2  + C_{p,q}\vertiii{Y}_{p,q}^2 
\end{equation}
with (non-sharp) constant $C_{p,q} = \CO(p+q)$.
\end{prop}
Most of the proof is due to Noar~\cite{naor_2012}, plus some standard manipulation (See~\cite{HNTR20:Matrix-Product}). For a technical note, at $p=2$ there are better constants due to being a \textit{vector} martingale in the Hilbert-Schmidt inner product. We will not use it for OTOCs as it is upper-bounded by with low rank input (Proposition~\ref{prop:lowrank_to_mixed}). We include it for potential applications in other settings\footnote{Such as when the low-rank input bound is not available.}.
\begin{prop}[{Uniform smoothness of $L_q(S_2)$\cite[Fact 3]{chen2020quantum}}] \label{prop:sub_average_2q}
Consider random matrices $X, Y$ of the same size that satisfy
$\BE[Y|X] = 0$. When $2 \le q $,
\begin{equation}
\vertiii{X+Y}_{2,q}^2 \le \vertiii{X}_{2,q}^2  + C_q\vertiii{Y}_{2,q}^2. 
\end{equation}
The constant $C_q = q - 1$ is the best possible.
\end{prop}
\subsection{Warm-up: sum of independent bounded matrices}
The calculation will get more involved for martingales (Sec.\ref{sec:detail_cal_OTOC}, Sec.\ref{sec:detail_cal_Op}), but the mechanism can be captured by the sum of independent bounded matrices. 
\begin{eg}[sum of bounded matrices]\label{ex:sum_matrices}
Consider independent, bounded random $D\times D$ matrices, $\lV X_i\rV \le b_i$. Then the sum $S_N:=\sum_{i=1}^N X_i$ has exponential tail bounds in terms of the variance $v:=\sum_{i}^N b_i^2$,
\begin{align}
    \BP(\lV \sum^N_i X_i \rV_\infty > \epsilon )&\le D\exp(-\frac{\epsilon^2}{e^2v})\\
    \BP( \frac{\lV\sum^N_i X_i \rV_2}{\lV I\rV_2} > \epsilon )&\le \exp(-\frac{\epsilon^2}{e^2v})\\
    \sup_{P}\BP(\lV \sum^N_i X_i P \rV_\infty > \epsilon )&\le D_P\exp(-\frac{\epsilon^2}{e^2v}),
\end{align}
for $\epsilon > 2e^2v $ and input subspace dimension $D_P$. Or in terms of expectation values,
\begin{align}
       \BE \lV \sum^N_i X_i \rV_\infty &\lesssim \sqrt{v\ln(D)}\\
       \frac{\BE \lV\sum^N_i X_i \rV_2}{\lV I\rV_2} &\lesssim \sqrt{v}\\
    \sup_P\BE \lV \sum^N_i X_i P \rV_\infty&\lesssim \sqrt{v\ln(D_P)}.
\end{align}
\end{eg}
\begin{proof}
Consider $\vertiii{\cdot }_{*}= \vertiii{\cdot }_{p},\vertiii{\cdot }_{2,p}, \vertiii{\cdot }_{D_P,p}$. We obtain moment bounds in term of the variance $v=\sum^N_i b_i^2$,
\begin{align}
\vertiii{\sum^N_i X_i}^2_{*} &\le \vertiii{\sum^{N-1}_i X_i}^2_{*}+C_p\vertiii{X_N}^2_{*}\\
&\le \sum_{i}^N C_p\vertiii{X_i}^2_{*} \le \sum_{i}^N C_p b_i^2\vertiii{I}^2_{*}=  C_p v \vertiii{I}^2_{*},
\end{align}
where we recursively call uniform smoothness for the respective spaces (Proposition~\ref{prop:sub_average_pq},Proposition~\ref{prop:sub_average_2q_DP}), with the same constant $C_p=p-1$. The rest is a standard Markov's inequality:\\
(i) for spectral norm, using $\vertiii{\cdot}_p$,
\begin{align}
    \BP[\lV S_N \rV \ge \epsilon ]  \le \frac{\BE[\lV S_N \rV^p ]}{\epsilon^p}&\le  \frac{\BE[\lV S_N \rV_p^p ]}{\epsilon^p}=\frac{ \vertiii{S_N}_p^p}{\epsilon^p}\\
    &\le \left(\vertiii{I}_{p}\frac{\sqrt{\sum_{i}^N C_p b_i^2}}{\epsilon}\right)^p\le D(\frac{\sqrt{ (p-1) v}}{\epsilon})^p\\
    &\le \begin{cases}
      \displaystyle D\exp(-\frac{\epsilon^2}{e^2 v}) & \text{if } 2e^2v < \epsilon^2\\
      \displaystyle D\frac{v}{\epsilon^2} & \text{if } 2e^2v \ge \epsilon^2,
     \end{cases}
\end{align}
where we set $p = \max(2,\frac{\epsilon^2}{e^2v})$ and for $p=2$ we use $C_p=1$ and restore the Chebychev bound.\\
(ii) for Hilbert-Schmidt norm, using $\vertiii{\cdot}_{2,p}$,
\begin{align}
    \BP[\lV S_N \rV_2 \ge \epsilon ]  &\le  \frac{\BE[\lV S_N \rV_2^p ]}{\epsilon^p}=\frac{ \vertiii{S_N}_{2,p}^p}{\epsilon^p}\\
    &\le \left(\vertiii{I}_{2,p}\frac{\sqrt{\sum_{i}^N C_p b_i^2}}{\epsilon}\right)^p\le (\frac{\sqrt{ (p-1) v}\lV I\rV_2}{\epsilon})^p\\
    &\le \begin{cases}
      \displaystyle \exp(-\frac{\epsilon^2}{e^2 v\lV I\rV^2_2}) & \text{if } 2e^2v\lV I\rV^2_2 < \epsilon^2\\
      \displaystyle \frac{v\lV I\rV^2_2}{\epsilon^2} & \text{if } 2e^2v\lV I\rV^2_2 \ge \epsilon^2,
     \end{cases}
\end{align}
where we used $\vertiii{I}_{2,p} = \lV I\rV_2$, and set $p = \max(2,\frac{\epsilon^2}{e^2v\lV I\rV^2_2})$.\\
(iii) for low rank inputs, using $\vertiii{\cdot}_{D_P,p}$, 
\begin{align}
    \BP[\lV S_N P\rV \ge \epsilon ]  \le \frac{\BE[\lV S_N P\rV^p ]}{\epsilon^p}&\le  \frac{\BE[\lV S_N P\rV_p^p ]}{\epsilon^p}=\frac{ \vertiii{S_NP}_p^p}{\epsilon^p}\\
    &\le \left(\vertiii{D_P}_{p}\frac{\sqrt{\sum_{i}^N C_p b_i^2}}{\epsilon}\right)^p\le D_P(\frac{\sqrt{ (p-1) v}}{\epsilon})^p\\
    &\le \begin{cases}
      \displaystyle D_P\exp(-\frac{\epsilon^2}{e^2 v}) & \text{if } 2e^2v < \epsilon^2\\
      \displaystyle D_P\frac{v}{\epsilon^2} & \text{if } 2e^2v \ge \epsilon^2,
     \end{cases}
\end{align}
where the only difference is replacing $D$ with $D_P$. In fact, $(iii)$ includes $(i)$ by setting the input subspace to be full rank $D_P=D$, but we present both for clarity. Lastly, integrating the tail bound yields the expectation bound by 
\begin{align}
    \BE[|x|] = \int_0^\infty \BP(|x|>\epsilon) d\epsilon,
\end{align}
which produces a logarithm of dimension $\ln(D)$ and $\ln(D_P)$.
\end{proof}
\subsection{Reminders of useful facts in non-commutative analysis }
Before we turn to the proof, let us remind ourselves the useful properties for the underlying norms $\vertiii{ \cdot}_{*}:=\vertiii{ \cdot}_{p,q}, \vertiii{ \cdot}_{D_P,p}$ for $p,q\ge 2$. They are largely inherited from the (non-random) Schatten p-norm. 
\begin{fact}[non-commutative Minkowski]\label{fact:non-commutative_mink}
Each of the expected moment satisfies the triangle inequality and thus a valid norm. For any random matrix $X, Y$
\begin{align}
\vertiii{X+Y}_{*} \le \vertiii{X}_{*}+\vertiii{Y}_{*}.
\end{align}
\end{fact}
\begin{fact}[operator ideal norms]\label{fact:operator ideal}
For operators $A$ deterministic and $O$ random
\begin{align}
\vertiii{AO}_{*}, \vertiii{OA}_{*} \le \lV A\rV \cdot \vertiii{O}_{*}.
\end{align}
\end{fact}
\begin{fact}[unitary invariant norms]
For $U, V$ deterministic unitaries and random operator $O$
\begin{align}
\vertiii{UOV}_{*} = \vertiii{O}_{*}.
\end{align}
\end{fact}
Being operator ideal already implies unitary invariance, but we state it regardless. As the norm $\vertiii{\cdot}_{D_P,p}$ defined for low rank input is somewhat non-standard, we include a proof as follows.
\begin{proof}[Proof of Fact~\ref{fact:operator ideal} for low rank inputs]
\begin{align}
\vertiii{XA}_{{D_P},p}&=\sup_{rank(P)=D_P}(\BE[\lV XAP\rV_p^p])^{1/p}\\
&=\sup_{rank(P)=D_P}(\BE[\lV XP'A'\rV_p^p])^{1/p}\\
&=\sup_{rank(P')=D_P}(\BE[\lV XP'\rV_p^p])^{1/p} \lV A\rV\\
&=\vertiii{X}_{{D_P},p} \lV A\rV.
\end{align}
In the second line we use the singular value decomposition 
\begin{align}
    AP = USV= US_{D_P}S_{A'}V = US_{D_P}U^\dagger \cdot US_{A'}V := P'A',
\end{align} where we rewrite the diagonal elements as product $S=S_{D_P}S_{A'}$, where $S_{D_p}$ is a rank $D_P$ projector and $\lV S_{A'}\rV \le \lV S\rV \le \lV A\rV$. This is possible because $S$ must have smaller rank $rank(s) \le rank(P)=D_P$ and bounded by $\lV S\rV \le \lV PA\rV \le \lV A\rV$, 
\end{proof}
\subsection{Time-independent Hamiltonians}
\subsubsection{Proof of Lemma~\ref{lem:time_indep+induction_step}:}\label{sec:proof_time_indep_induction}
\begin{proof}
By triangle inequality and unitary invariance,
\begin{align}
    \vertiii{ A(t+dt)}_* &= \vertiii{\hat{U}(dt)\left[ A(t) + \sum_j \CL_j[B_j(t)]dt \right]}_*\\
    &= \vertiii{\left[ A(t) + \sum_j \CL_j[B_j(t)]dt \right]}_*\\
    &\le \vertiii{A(t)}_* + \vertiii{\sum_j \CL_j[B_j(t)]dt}_*.
\end{align}
Next, notice that $H_j$ are by assumption mutually independent conditioning on $B_j(t)$. This allows us to cast $\sum_j \CL_j[B_j(t)]$ into a martingale difference sequence by constructing a filtration of sigma algebras
$\CF_0\subset \CF_1 \subset \CF_2 \cdots \subset \CF_k$:
\begin{align}
    \CF_0 &:= \sigma(B_1(t),\cdots, B_k(t))\\
    \CF_j&:= \sigma(B_1(t),\cdots, B_k(t),H_1,\cdots, H_j),
\end{align}
where for each $\CF_j$ we denote the conditional expectation $\BE_j$. Indeed our sum forms a martingale difference $\BE_{j-1}\CL_jB_j(t) =0$, which allows us to call matrix martingale concentration (Proposition~\ref{prop:sub_average_2q},~\ref{prop:sub_average_pq}, Corrolary~\ref{prop:sub_average_2q_DP})
\begin{align}
     \vertiii{\sum_{j=1}^k \CL_j[B_j(t)]dt}^2_* &\le \vertiii{\sum_{j=1}^{k-1} \CL_j[B_j(t)]dt}^2_*+(p-1)\vertiii{\CL_k[B_k](t)dt}^2_* \le \sum_{j=1}^{k}(p-1)\vertiii{\CL_j[B_j(t)]dt}^2_*\\
     &\le \sum_{j=1}^{k}4pb_j^2\vertiii{B_j(t)dt}^2_*.
\end{align}
where we used the operator ideal (Fact~\ref{fact:operator ideal}) property of the p-th moments $\vertiii{\CL_j[B_j(t)]}_*\le 2\lV H_j\rV \vertiii{B_j(t)}_*$ and then the uniform bound $\lV H_j\rV\le b_j$. Furthermore, to get the linear recursion, 
\begin{align}
    \frac{d}{dt}(\vertiii{ A(t)}^2_*) = 2\vertiii{ A(t)}_* \frac{d}{dt}\vertiii{ A(t)}_* &\le 2\vertiii{ A(t)}_* \sqrt{4p \sum_j b_j^2\vertiii{B(t)_j}_{*}^2}\\
    &\le \lambda \vertiii{A(t)}^2_{*}+ \frac{4p}{\lambda}\sum_j b_j^2 \vertiii{B(t)_j}_{*}^2
\end{align}
where we used the arithmetic-geometric inequality with a tuneable parameter $\lambda$. 
\end{proof}

\subsubsection{Proof of Theorem~\ref{thm:time_indep_sum_over_path}}\label{proof:time_indep}
The sum over self-avoiding paths can be alternatively written as a system of recursive (differential) operator equations. The idea is as simple as turning an integral to a derivative, and the rest is careful accounting. Let us start with Theorem~\ref{thm:det_sum_over_path}, denoting the sum over self-avoiding path to site $r$ as 
\begin{align}
O_{\{r\}}(t) &:= \sum_{\Gamma={X_\ell,\ldots,X_1} \in \mathcal{S}_{r0}} \ \  \int\limits_{t>t_\ell>\cdots >t_1} \mathrm{d}t_1\cdots \mathrm{d}t_\ell \mathrm{e}^{\mathcal{L}(t-t_\ell)} \mathcal{L}_{X_\ell}  \mathrm{e}^{\mathcal{L}_{\ell-1} (t_\ell-t_{\ell-1})} \mathcal{L}_{X_{\ell-1}}\cdots \mathrm{e}^{\mathcal{L}_1(t_2-t_1)}\mathcal{L}_{X_1}\mathrm{e}^{\mathcal{L}_0t_1} [O_0] \\
& =:\sum_{\Gamma={X_\ell,\ldots,X_1} \in \mathcal{S}_{r0}} \hat{W}^t(\Gamma). 
\end{align}
where we can recover the commutator by $\frac{1}{2}\vertiii{[A_r,O(t)_0]}_*\le \lV A_r\rV\cdot \vertiii{O_{\{r\}}(t)}_*$. We also need to define the intermediate auxiliaries operators.
\begin{lem}[Part(A) of Thoerem~\ref{thm:time_indep_sum_over_path}. recursive form of Theorem~\ref{thm:det_sum_over_path}]\label{lem:recursive_time-indep}
Define an operator $O^T_\gamma$, associating with any subpath $\gamma\subset \Gamma$ backtracking along $\Gamma$ from $r$, ending anywhere on $\Gamma$ 
\begin{align}
    O^t_\gamma &:= \sum_{\Gamma' \in \mathcal{S}_{r0}, \gamma \subset \Gamma' } \int\limits_{t>t_m>\cdots >t_1} \mathrm{d}t_\ell\cdots \mathrm{d} t_1  \mathrm{e}^{\mathcal{L}_m(t-t_m)} \mathcal{L}_{X_m}  \mathrm{e}^{\mathcal{L}_{m-1} (t_m-t_{m-1})} \mathcal{L}_{X_{m-1}}\cdots \mathrm{e}^{\mathcal{L}_1(t_2-t_1)}\mathcal{L}_{X_1}\mathrm{e}^{\mathcal{L}_0t_1} [O_0] \\
    &=:\sum_{\Gamma' \in \mathcal{S}_{r0}, \gamma \subset \Gamma' } \hat{W}^t (\Gamma'|\gamma).
\end{align}
where $m = |\Gamma'|-|\gamma| \le |\Gamma'| =\ell$ is the length of the remaining path unspecified by $\gamma$, and the intermediate unitaries and Linbaldians are associated with $\Gamma'$: $\CL_{k}=\CL_{k}(\Gamma')$, $\mathcal{L}_{X_k}=\mathcal{L}_{X_k}(\Gamma')$.
Then they satisfy the recursive equation
\begin{align}
    \displaystyle O^{t+dt}_\gamma = e^{\CL_{\gamma}dt } \left[O^t_\gamma+ \sum_{X\in \Delta(\gamma)} \CL_{X}[O^t_{\gamma+X}]dt \right] +\CO(dt^2),
\end{align}
where $\CL_\gamma=\CL^{(\Gamma')}_m$ is the intermediate unitary associated with $\gamma$, defined in Theorem~\ref{thm:det_sum_over_path}; $\Delta(\gamma)$ is the set of permissible next steps of $\gamma$, defined by $\Delta(\gamma):=\Delta_m:= V^{\Gamma'}_m/V^{\Gamma'}_{m+1}$. 
\end{lem}
Note the expressions $\CL^{(\Gamma')}_m, \Delta(m)$ which is identified with $m$ and $\Gamma$ are not ambiguous and only depend on $\gamma$.
\begin{proof}
Taking derivative of the time-ordered integral,
\begin{align}
\frac{d}{dt}O^{t}_\gamma 
&=\sum_{\Gamma' \in \mathcal{S}_{r0}, \gamma \subset \Gamma' }\big( \int\limits_{t>t_m>\cdots >t_1} \mathrm{d}t_\ell\cdots \mathrm{d} t_1  \mathcal{L}_m\mathrm{e}^{\mathcal{L}_m(t-t_m)} \mathcal{L}_{X_m}  \mathrm{e}^{\mathcal{L}_{m-1} (t_m-t_{m-1})} \mathcal{L}_{X_{m-1}}\cdots \mathrm{e}^{\mathcal{L}_1(t_2-t_1)}\mathcal{L}_{X_1}\mathrm{e}^{\mathcal{L}_0t_1} [O_0]\\
&+\int\limits_{t>t_m>\cdots >t_1} \mathrm{d}t_\ell\cdots \mathrm{d} t_1 \mathcal{L}_{X_m}  \mathrm{e}^{\mathcal{L}_{m-1} (t_m-t_{m-1})} \mathcal{L}_{X_{m-1}}\cdots \mathrm{e}^{\mathcal{L}_1(t_2-t_1)}\mathcal{L}_{X_1}\mathrm{e}^{\mathcal{L}_0t_1} [O_0]
\big)\\
& =\CL_{m} [O^{t}_\gamma ]+ \sum_{\Gamma' \in \mathcal{S}_{r0}, \gamma \subset \Gamma' }
\mathcal{L}_{X_{m}} [\hat{W}^t (\Gamma'|\gamma+X_m)]\\
& = \CL_{\gamma} [O^{t}_\gamma] + \sum_{X \in \Delta(\gamma)} \mathcal{L}_X\sum_{\Gamma' \in \mathcal{S}_{r0}, \gamma+X \subset \Gamma' }
 [\hat{W}^t (\Gamma'|\gamma+X) ]\\
& = \CL_{\gamma} [O^{t}_\gamma] + \sum_{X \in \Delta(\gamma)}  \mathcal{L}_{X}[O^t_{\gamma+X}].
\end{align}
where in the third inequality we factor out the first possible steps $\CL_X, X\in\Delta_m$, which depend only on the path $\Delta_m=\Delta(\gamma)$; in the last line we recall the definition of $O^t_{\gamma+X}$. This completes the proof by the uniqueness of ODE solution. Note the initial conditions for completed self-avoiding path $\Gamma \in \mathcal{S}_{r0}$ is the starting operator $(O^{t=0}_\Gamma= O^{t=0})$, and uncompleted self-avoiding paths $\gamma$ vanish $(O^{t=0}_{\gamma\notin \mathcal{S}_{r0}}=0)$.
\end{proof}
The above lemma establishes the first part of Theorem~\ref{thm:time_indep_sum_over_path}. 
Next, to call Lemma~\ref{lem:time_indep+induction_step}, we need to verify that $\CL_X, X\in \Delta(\gamma)$ are independent conditioned on the event $O^t_{\gamma+X}, X\in \Delta(\gamma)$. This is why the self-avoiding decomposition (Theorem~\ref{thm:det_sum_over_path}) is the key to this derivation (Fig.~\ref{fig:forbidden}): for any paths enumerated in $O^t_{\gamma+X}$, \textit{none} of the interactions $\CL_{k-1},\cdots,\CL_0$ contain \textit{any} term $X \in \Delta(\gamma)$. This can be seen from that $V_k$ contains $X_{k+1}$ and excludes any interaction intersecting with $X_{k+1}$, and in particular excludes any $X \in \Delta(\gamma)$. 
Therefore, by Lemma~\ref{lem:time_indep+induction_step}, we get a system of recursive ODE bounds
\begin{align}
    \frac{d}{dt}(\vertiii{ O_\gamma(t)}^2_*) & \le \lambda_\gamma \vertiii{O_\gamma(t)}^2_{*}+ \frac{4p}{\lambda_\gamma}\sum_{X \in \Delta_m} b_X^2 \vertiii{O^t_{\gamma+X}}_{*}^2.
\end{align}
The formal solution to this system of ODEs is exactly the integral form in Theorem~\ref{thm:time_indep_sum_over_path}, which can be verified by taking derivatives like above. Technically this is a system of ordinary differential inequality that is treated by Gronwall~\cite{Gronwall}'s Lemma.
\subsection{Time-dependent Hamiltonians}
\subsubsection{Proof of Lemma~\ref{lem:induction_step}: Inductive step for operator growth from matrix martingale}

\begin{proof}[Proof of Lemma~\ref{lem:induction_step}]\label{pf:induction_step}
The recursive equation captures how operator at a previous time $B^{t-1}$ contributes to the adjacent operator $A^{t}$, due to $V$. 

\begin{align}
    \vertiii{A^t}_{*}^2 
    &= \vertiii{\hat{U}\left[ \hat{V}[A^{t-1}]+    (\hat{V}-I)[B^{t-1}]\right]}_{*}^2\\
    & = \vertiii{ A^{t-1}+    (\hat{V}^\dagger-I)B^{t-1}}_{*}^2\\
    &= \vertiii{A^{t-1}+    (\BE_{t-1}\hat{V}^\dagger-I)[B^{t-1}]+(\hat{V}^\dagger-\BE_{t-1}\hat{V}^\dagger)[B^{t-1}]}_{*}^2\\
    &\le \vertiii{A^{t-1}+    (\BE_{t-1}\hat{V}^\dagger-I)[B^{t-1}]}_{*}^2 +(p-1)\vertiii{(\hat{V}^\dagger-\BE_{t-1}\hat{V}^\dagger)[B^{t-1}]}_{*}^2,
    \end{align}
where we called the matrix martingale inequalities (Proposition.~\ref{prop:sub_average_2q},Proposition.~\ref{prop:sub_average_2q_DP}) for each norm.

We bound each term: \\
(i) The first term is \textbf{deterministic} conditioning on the times before $t$. Taking expectation, we are left with subleading orders of $\eta$, and we can generously bound by brute force triangle inequality:
\begin{align}
    \vertiii{A^{t-1}+    (\BE_{t-1}\hat{V}^\dagger-I)[B^{t-1}]}_{*}^2&=( \vertiii{A^{t-1}}_{*}+\vertiii{    (\BE_{t-1}\hat{V}^\dagger-I)[B^{t-1}]}_{*})^2\\
    &\le( \vertiii{A^{t-1}}_{*}+2\eta^2\vertiii{B^{t-1}}_{*})^2,
\end{align}
where in the inequality we used that $\BE [H,O] = 0$ and Taylor expanded to second order, used Jensen's inequality to pull expectation outside of norm, and used Fact~\ref{fact:operator ideal} that $\vertiii{\cdot}_{*}$ is an operator ideal norm: $\lV\BE[ e^{-iH} O e^{iH} - O + i[H,O]]\rV_* = \lV \BE \int_0^1 e^{-iHs}[H,[H,O]]e^{iHs}(1-s)ds \rV_* \le 2 \lV H\rV^2 \lV O\rV_*$.\\
(ii) We isolate the second term as the {\textbf{random fluctuation}}, entirely coming from the adjacent operator and the unitary.
\begin{align}
    \vertiii{(\hat{V}^\dagger-\BE_{t-1}\hat{V}^\dagger)[B^{t-1}]}_{*}^2 &= \vertiii{(\hat{V}^\dagger-I)[B^{t-1}]-(\BE_{t-1}\hat{V}^\dagger-I)[B^{t-1}]}_{*}^2\\
    &\le (2\eta + 2\eta^2)^2\vertiii{B^{t-1}}_{*}^2 \le 4\eta^2(1+\eta)^2\vertiii{B^{t-1}}_{*}^2,
\end{align}
where for the first term $\hat{V}^\dagger-I$ we Taylor expanded to first order $ \lV e^{-iH} O e^{iH} - O\rV_* = \lV \int_0^1 e^{-iHs}[H,O]e^{iHs}ds \rV_* \le 2 \lV H\rV \lV O\rV_*$, and then we generously used $\eta\le 1$. The second term $(\BE_{t-1}\hat{V}^\dagger-I)$ we already calculated above. 

Putting (i) and (ii) together yields the inductive inequality
\begin{align}
    \vertiii{A^t}_{*}^2 &\le \left( \vertiii{A^{t-1}}_{*}^2+4\eta^2\vertiii{A^{t-1}}_{*}\vertiii{[B^{t-1}]}_{*}+4\eta^4\vertiii{B^{t-1}}_{*}^2 \right)+ 4\eta^2(1+\eta)^2(p-1)\vertiii{B^{t-1}}_{*}^2\\
    &\le  (1+\frac{\eta^2}{p})\vertiii{A^{t-1}}_{*}^2+ 8p\eta^2g(\eta)\vertiii{B^{t-1}}_{*}^2,
\end{align}
where $g(\eta):= 1+\eta+\eta^2/2 \le 1+3\eta/2$ and we used the arithmetic-geometric inequality to remove the bilinear term, with a carefully tuned coefficient
\begin{align}
4\eta^2\vertiii{A^{t-1} }_*\vertiii{O^{t-1}_j }_*  \le \frac{\eta^2}{p}\vertiii{O^{t-1}_{\gamma} }_*^2 + 4p\eta^2\vertiii{O^{t-1}_j }_*^2,
\end{align}
and also $\eta^4 < \eta^2$ for simplification.\\
\end{proof}

\subsubsection{Proof of Theorem~\ref{thm:sum_over_paths}}\label{sec:proof_brown_thm}
Recall our Brownian circuit is a product of $T$ independent unitary evolution 
\begin{equation}
    U(T) = \exp(\sum -iH^{T}_X \xi ) \cdots \exp(\sum -iH^1_X \xi ),
\end{equation}
where each of $H^{\tau}_X$ are independent for different times $\tau$ or different terms $X$.

\textbf{Part A: deriving the recursive equations}

This is a Brownian treatment of self-avoiding path (Theorem~\ref{thm:det_sum_over_path}). In spirit, this section closely follows the proof of the time-independent case (Theorem~\ref{proof:time_indep}), and the main difference is that we have to keep track of the next order term in the infinitesimal step $\xi$.
\begin{lem}[Self-avoiding paths in Brownian systems]\label{lem:self-avoiding_brown}
\begin{align}
O(T) &\stackrel{a.s.}{=} \sum_{\Gamma={X_\ell,\ldots,X_1}  \in \mathcal{S}_{r0}}  \sum_{T> T_\ell>\cdots> T_1\ge 0} \hat{U}^{T,T_\ell+1} (e^{\xi\mathcal{L}_{X_\ell}}-1) \hat{U}^{T_\ell,T_{\ell-1}+1}_{\ell-1}  (e^{\xi\mathcal{L}_{X_{\ell-1}}}-1) \cdots \hat{U}_1^{T_2,T_1+1}(e^{\xi\mathcal{L}_{X_1}}-1) \hat{U}_0^{T_1,0}[O_0] \\
& =:\sum_{\Gamma={X_\ell,\ldots,X_1} \in \mathcal{S}_{r0}} \hat{W}^T(\Gamma) 
\end{align}
where the product of random evolution of Hamiltonians between two time slices is denoted by 
\begin{equation}
    \displaystyle \hat{U}_k^{(T_2,T_1)} :=\prod_{T_2-1\ge T \ge T_1}e^{\CL^{(T)}_k\xi}= \prod_{T_2-1\ge T \ge T_1}\exp(\mathcal{L^{(T)}} - \sum_{Y\in \partial V^\Gamma_k} \mathcal{L}^{(T)}_Y).
\end{equation}
\end{lem}
In contrast to time-independent case (Lemma~\ref{lem:recursive_time-indep}), the key distinction in the Brownian case is that we have to keep some of the second order terms because $\xi^2 T$ is now finite in the limit. This amounts to replacing $\CL_{X_k}dt$ with $(e^{\xi\mathcal{L}_{X_{k}}}-1)$ in the derivation of Theorem~\ref{thm:det_sum_over_path} in \cite{chen2019operator}. This replacement is exact since $(e^{\xi\mathcal{L}_{X_{k}}}-1)$ becomes commuting with each other -- up to higher order terms ('Trotter error') that vanish in the Brownian Limit.
\begin{lem} \label{lem:first_order_suzuki} For arbitrary ordering of independent zero mean bounded operators $H_i$,
\begin{align}
e^{\sum H_i \xi} = \prod_i e^{H_i \xi} + \CO(\xi^3)M +  \CO(\xi^2)Z = 1 + \sum _i e^{H_i \xi}- 1 + \CO(\xi^3)M' +  \CO(\xi^2)Z',
\end{align}
where $Z$,$Z'$ are zero mean. 
\end{lem}
\begin{proof}
A direct second order trotter error analysis yields $e^{\sum H_i \xi} - \prod_i e^{H_i \xi} = \text{const.} \sum_{i>j} [H_i,H_j] \xi^2 +O(\xi^3)$. 
\end{proof}
Therefore the ordering within a short period of time $\xi$ can be arbitrary. Then, the rest is analogous to the time-independent case (Theorem~\ref{thm:time_indep_sum_over_path}). Again, rewrite sum over self-avoiding path as a system of recursive (differential) operator equations. 
\begin{lem}[Recursive form of Lemma~\ref{lem:self-avoiding_brown}]\label{lem:recursive_brown}
Define an operator $O^T_\gamma$, associating with any subpath $\gamma\subset \Gamma$ backtracking along $\Gamma$ from $r$, ending anywhere on $\Gamma$. 
\begin{align}
    O^T_\gamma &:= \sum_{\Gamma' \in \mathcal{S}_{r0}, \gamma \subset \Gamma' }  \sum_{T> T_{m}>\cdots> T_1\ge 0} \hat{U}^{T,T_{m}+1}_{m} (e^{\xi\mathcal{L}_{X_{m}}}-1) \hat{U}^{T_{m},T_{{m}-1}+1}_{{m}-1}  (e^{\xi\mathcal{L}_{X_{{m}-1}}}-1) \cdots \hat{U}_1^{T_2,T_1+1}(e^{\xi\mathcal{L}_{X_1}}-1) \hat{U}_0^{T_1,0}[O_0]  \\
    &=:\sum_{\Gamma' \in \mathcal{S}_{r0}, \gamma \subset \Gamma' } \hat{W}^T (\Gamma'|\gamma),
\end{align}
where $m = |\Gamma'|-|\gamma| \le |\Gamma'| =\ell$ is the length of the remaining path unspecified by $\gamma$, and the intermediate evolution and hops are associated with $\Gamma'$: $\hat{U}^{T,T_{m}+1}_{m}=\hat{U}^{T,T_{m}+1}_{m}(\Gamma')$, $\mathcal{L}_{X_k}=\mathcal{L}_{X_k}(\Gamma')$.
Then they satisfy the recursive equation
\begin{align}
     O^{T+1}_\gamma &= \exp({\CL_{\gamma}\xi }) \left[[O^T_\gamma]+ \sum_{X\in \Delta(\gamma)} (I-e^{-\CL_{X}\xi})[O^T_{\gamma+X}] \right] +\CO(\xi^3)M +  \CO(\xi^2)Z,
\end{align}
where $\Delta(\gamma)$ is the set of the permissible next hops the path $\gamma$ can take, defined by $\Delta(\gamma):=\Delta_m:= V^{\Gamma'}_m/V^{\Gamma'}_{m+1}$, and is only dependent on $\gamma\subset \Gamma'$. All unitary evolution are at time $T$ but suppressed $\CL=\CL^{(T)}$.
\end{lem}
\begin{proof}
This is analogous to proof of Lemma~\ref{lem:recursive_time-indep}.
\begin{align}
O^{T+1}_\gamma &= \sum_{\Gamma' \in \mathcal{S}_{r0}, \gamma \subset \Gamma' } \sum_{T+1> T_{m}>\cdots> T_1\ge 0} \hat{U}^{T+1,T_{m}+1}_{m} (e^{\xi\mathcal{L}_{X_{m}}}-1) \hat{U}^{T_{m},T_{{m}-1}+1}_{{m}-1}   \cdots (e^{\xi\mathcal{L}_{X_1}}-1) \hat{U}_0^{T_1,0}[O_0]\\ 
&= \sum_{\Gamma' \in \mathcal{S}_{r0}, \gamma \subset \Gamma' }[\sum_{T> T_{m}>\cdots> T_1\ge 0} \hat{U}^{T+1,T}_{m}\cdot\hat{U}^{T,T_{m}+1}_{m} (e^{\xi\mathcal{L}_{X_{m}}}-1) \hat{U}^{T_{m},T_{{m}-1}+1}_{{m}-1}   \cdots (e^{\xi\mathcal{L}_{X_1}}-1) \hat{U}_0^{T_1,0}[O_0]\\
&+ \sum_{T=T_{m}>\cdots> T_1\ge 0}  (e^{\xi\mathcal{L}_{X_{m}}}-1)\cdot  \hat{U}^{T_{m},T_{{m}-1}+1}_{{m}-1}   \cdots (e^{\xi\mathcal{L}_{X_1}}-1) \hat{U}_0^{T_1,0}[O_0]
]\\
& =\hat{U}^{T+1,T}_{m} [O^{T}_\gamma] + \sum_{\Gamma' \in \mathcal{S}_{r0}, \gamma \subset \Gamma' }
(e^{\xi\mathcal{L}_{X_{m}}}-1) [\hat{W}^T (\Gamma'|\gamma+X_m)]\\
& = e^{\xi\mathcal{L}_{\gamma}} [O^{T}_\gamma] + \sum_{X \in \Delta(\gamma)} (e^{\xi\mathcal{L}_{X}}-1)[\sum_{\Gamma' \in \mathcal{S}_{r0}, \gamma+X \subset \Gamma' }
 \hat{W}^T (\Gamma'|\gamma+X)] \\
& = e^{\xi\mathcal{L}_{\gamma}} [O^{T}_\gamma] + \sum_{X \in \Delta(\gamma)} (e^{\xi\mathcal{L}_{X}}-1)[O^T_{\gamma+X}],
\end{align}
where in the fourth equality we note that $\hat{U}^{T+1,T}_{m}$ depends only on $\gamma$ and similarly we rewrite $\Delta_m=\Delta(\gamma)$ for the first possible steps $\CL_X, X\in\Delta_m$; in the last line we used Lemma~\ref{lem:self-avoiding_brown}.

To put into the desired form, rearrange the product
\begin{align}
& =e^{\xi\mathcal{L}_{\gamma}} \left[O^{T}_\gamma + \sum_{X \in \Delta(\gamma)} e^{-\xi\mathcal{L}_{\gamma}}(e^{\xi\mathcal{L}_{X}}-1)[O^T_{\gamma+X}]\right]\\
& =e^{\xi\mathcal{L}_{\gamma}} \left[O^{T}_\gamma + \sum_{X \in \Delta(\gamma)} e^{-\xi\mathcal{L}_{\gamma}+\xi\mathcal{L}_{X}}(1-e^{-\xi\mathcal{L}_{X}})[O^T_{\gamma+X}]\right]\\
& =e^{\xi\mathcal{L}_{\gamma}} \left[O^{T}_\gamma + \sum_{X \in \Delta(\gamma)} (1-e^{-\xi\mathcal{L}_{X}})[O^T_{\gamma+X}]\right]+ \CO(\xi^3)M+\CO(\xi^2)Z,
\end{align}
where in last line we used that for independent terms in the Hamiltonian, $(e^{\xi\mathcal{L}_{X'}}-1) (1-e^{\xi\mathcal{L}_{X}})= \CO(\xi^3)M+\CO(\xi^2)Z$, which vanishes in the Brownian limit. Note that $\CL_\gamma$ contains terms $\CL_X$ for $X\in \Delta(\gamma)$.
\end{proof}

\textbf{Part B: call Lemma~\ref{lem:induction_step} and sum over self-avoiding paths}


\begin{proof}[Second part of Theorem~\ref{thm:sum_over_paths}]
Continue with
\begin{align}
    \displaystyle e^{\xi\mathcal{L}_{\gamma}} \left[O^{T}_\gamma + \sum_{X \in \Delta(\gamma)} (1-e^{-\xi\mathcal{L}_{X}})[O^T_{\gamma+X}]\right]+ \CO(\xi^2)M+\CO(\xi^3)Z.
\end{align}
During a time step $T$ to $T+1$, we peel off each $X \in \Delta(\gamma) $ by calling Lemma~\ref{lem:induction_step}.
\begin{align}
    \vertiii{O^{T+1}_\gamma}_{*}^2 &\simeq \vertiii{ O^T_\gamma+ \sum_{j} (I-e^{-\CL_{X_j}})O^T_{\gamma+X_j} }_{*}^2\\
    & \lesssim (1+\frac{\eta_{X_1}^2}{p})\vertiii{O^T_\gamma+ \sum_{j>1} (I-e^{-\CL_{X_j}})O^T_{\gamma+X_j} }_{*}^2+ 8p\eta_{X_1}^2\vertiii{O^{t-1}_{\gamma+X_1}}_{*}^2\\
    & \lesssim (1+ \sum_{X\in \Delta(\gamma)}\frac{\eta_{X}^2}{p}) \vertiii{O^{T}_\gamma}_{*}^2 + \sum_{X\in \Delta(\gamma)} 8p\eta_{X}^2\vertiii{O^{T}_{\gamma+X}}_{*}^2,
\end{align}
where we suppressed subleading terms $\CO(\xi^3)M +  \CO(\xi^2)Z$ by $\simeq$ and $\lesssim$. Note for zero mean $Z$ to contribute to the a $p$-th moment, it must come with another term $\xi^2 H_iH_j$, which makes it $\CO(\xi^4)$.  The formal solution to the scalar recursion is exactly the global form~\eqref{eq:brown_integral}. 
\end{proof}

\section{Conclusions and discussions}
We formulate the first operator growth bounds suitable for time-independent and Brownian zero mean random Hamiltonian, in spectral norm or for OTOC with some non-random state. The bounds are given by a general, plug-and-play formula as a weighted sum over paths, where spatial (and temporal) noises accumulate as sum-of-squares. This reflects the stochastic nature and fundamentally differs from the worst-case linear behavior from the triangle-inequality. Our bounds match existing evidences or low bounds in power-law interacting systems and k-local systems. Our Brownian bounds consolidate the foundation to analyze continuous time random circuits, especially when intuition from discrete random circuits does not apparently carry over.

An immediate follow-up work would be deriving concentration bounds for the spectral norm in general systems such as 2d short-ranged system. We only thus far derive bounds for 1d systems due to lack of control over operator support in higher dimension. 

Looking afar, we expect our spectral norm bounds to find application wherever traditional Lieb-Robinson bounds served as a sub-routine, and our OTOC bounds would give intuition for chaotic systems in the wild. Depending on the models at hand, one may need to upgrade the matrix martingale tools. We currently choose the most demonstrative ones but not the more refined one that may account for heavy-tailed distribution of matrices. In a general context, we expect the techniques from matrix concentration to be applied in other randomized settings in quantum physics.





\section{Acknowledgements}
I thank Andrew Lucas for inspiring comments in a precursor of this work~\cite{chen2019operator}. I also thank Joel Tropp for introducing the idea of uniform smoothness to me and related comments. Thank Hsin-Yuan (Robert) Huang, Fernando G.S.L. Brandao, De Huang, Minh Tran, Angelo Lucia for helpful discussion. CFC is support by Caltech TA  fellowship.

\appendix
\section{ Detailed calculations for concentration of the OTOC in several systems}\label{sec:detail_cal_OTOC}
Here we evaluate the sum over paths and then call Markov's inequality. Depending on the system, the combinatorics might require extra care.

\subsection{Brownian $d=1$ nearest-neighbour system, OTOC }\label{sec:1dnn_OTOC}
Consider a $d=1$ nearest-neighbour Brownian Hamiltonian
\begin{equation}
    H :=\sum_{i} a H_{i,i+1}
\end{equation}
where each term is normalized $\lV H_{i,i+1}\rV\le 1$ and zero mean $\BE[H_{i,i+1}]=0$. We will derive both discrete and continuous bound. First, take a discrete brick wall circuit $\prod \exp(-iaH_{i,i+1}\xi)$ with discrete time step\footnote{In terms of modeling time-independent Hamiltonian, this is a coherent time.} $\xi$ and depth $2\cdot T$. 
Here we prepare ourselves with some numerical values, which simplifies as there is only one self-avoiding path $\Gamma$. There are at most $|\Delta(\gamma)|\le 1$ possible next steps amid any self-avoiding path. Since all terms $H_X$ have the same bound, $\eta_X \le \eta = a \xi$. Plug in Theorem ~\ref{thm:sum_over_paths}:
\begin{align}
\frac{1}{2}\vertiii{[A_r,O(T)_{0}]}_{D_P,p}^2&\le (1+\frac{\eta^2}{p})^T \vertiii{O^0}_{D_P,p}^2  (8p\eta^2g(\eta))^\ell \binom{T+\ell-1}{\ell}\\
&\le D_P^{2/p}e^{a^2\xi^2T/p}  (8pe\frac{\eta^2g(\eta)(T+r-1)}{r})^{r}
\end{align}
where $\binom{T+\ell-1}{\ell}$ is the number of possible time ordering, \footnote{Since there are even and odd terms, consecutive hops may occur at same T. This why we bound by $\binom{T+\ell-1}{\ell}$ instead of $\binom{T}{\ell}$ This discretization issue disappears in the Brownian limit.} and we defined a temporary value $\lambda:=8e(\frac{\eta^2g(\eta)(T+r-1)}{r})$. 
By Markov inequality with tunable parameter $p = \max( \frac{1}{e\lambda}\epsilon^{2/r},2)$ \footnote{p cannot be smaller than $2$ in Proposition~\ref{prop:sub_average_pq}},
 \begin{align}
     \sup_P \BP[\lV O^{T}_{\{r\}} P\rV \ge \epsilon ] & \le   \frac{\sup_P\BE[\lV O^{T}_{\{r\}}P \rV_p^p ]}{\epsilon^p}\le \frac{ \vertiii{O^{T}_{\{r\}}}_{D_P,p}^p}{\epsilon^p}\\
     & = \begin{cases}
     D_Pe^{\eta^2T/2}  \exp(-\frac{\epsilon^{2/r}}{16e^2(\frac{\eta^2g(\eta)(T+r-1)}{r^2})}) & \text{if } (16e^2\frac{\eta^2g(\eta)(T+r-1)}{r})^{r} < \epsilon^2\\
     \frac{D_Pe^{\eta^2T/2}}{\epsilon^2}(16e\frac{\eta^2g(\eta)(T+r-1)}{r})^{r} & \text{if } (16e^2\frac{\eta^2g(\eta)(T+r-1)}{r})^{r} \ge \epsilon^2.
     \end{cases}
 \end{align}
 Or expressed in terms of $\delta$:
 \begin{align}
     \BP\left[\lV O^{T}_{\{r\}} P\rV > \epsilon(\delta) \right]\le \delta
 \end{align}
\begin{align}
    \epsilon(\delta):= 
    \begin{cases}
    \left[16e^2(\frac{\eta^2g(\eta)(T+r-1)}{r^2})(\frac{\eta^2T}{2}+\ln(D_P/\delta))\right]^{r/2} &\text{if } \delta <D_Pe^{\eta^2T/2-r} \\
    \\
    \sqrt{ \frac{D_Pe^{\eta^2T/2}}{\delta}(16e\frac{\eta^2g(\eta)(T+r-1)}{r})^{r}} &\text{if } \delta \ge D_Pe^{\eta^2T/2-r}.
    \end{cases} 
\end{align}
The logarithmic dependence on $\delta$ captures the very strong tail bound that might be useful elsewhere. 
Our next goal is extracting the asymptotic velocity, which only requires the second moment bound and Chebyshev. Take the order of limit: pick $\epsilon >0$, then for any $\zeta>0$, define $T(r)$ such that
\begin{align}
     e^{\eta^2T/2r}16e\frac{\eta^2g(\eta)(T+r-1)}{r} = 1- \zeta.
\end{align}
Then we obtain a Lieb-Robinson velocity almost surely
\begin{align}
    \lim_{r\rightarrow \infty} \BP(\lV\frac{1}{2}[A_{r},O_{0}(T(r))]P\rV \ge \epsilon ) =0. 
\end{align}
The expression of $\frac{r}{T(r)}$ can be simplified in the Brownian limit $\xi \rightarrow 0, \tau = \xi^2T$, or physically when the 'coherence time' is short compared to the hopping term $\xi \ll 1/a $,
\begin{equation}
    \frac{r}{\tau(r)} = u \le 16ee^{1/32e} a^2
\end{equation}
where we used $16e\frac{\eta^2(T+r-1)}{r}\le 1$ to estimate the exponential term.

\subsection{Brownian $d>1$ nearest-neighbour system, OTOC}\label{sec:dnn_OTOC}
Consider a $d>1$ nearest-neighbour Brownian Hamiltonian
\begin{equation}
    H^(T) := \sum_{<xx'>} a H^{(T)}_{x,x'} ,
\end{equation}
where each term is normalized $\lV H^{(T)}_{x,x'}\rV\le 1$ and independent zero mean conditioned on the past $\BE_{T-1}[H^{(T)}_{x,x'}]=0$. We will derive both discrete and continuous bound. First, take a discrete brick wall circuit $\prod \exp(-iaH_{x,x'}\xi)$ with discrete time step $\xi$ (``the coherent time") and depth $2d\cdot T$ (i.e. $T$ rounds).
Here we prepare ourselves with some numerical values. There are at most $|\Delta(\gamma)|\le R=2d$ possible next steps at any self-avoiding path. Since all terms $H_X$ have the same bound, $\eta_X \le \eta = a \xi$. Plug in the sum over self-avoiding paths (Theorem ~\ref{thm:sum_over_paths})
\begin{align}
\vertiii{O^{T}_{\{r\}}}_{D_P,p}^2&\le (1+\frac{\eta^2}{p})^{RT} \vertiii{O^0}_{D_P,p}^2 \sum_{\Gamma} (8p\eta^2g(\eta))^\ell \binom{T+\ell-1}{\ell}\\
&\le D_Pe^{\eta^2RT/p} \sum_{\ell \ge r} (2d8e\frac{p\eta^2g(\eta)(T+\ell-1)}{\ell})^\ell\\
&\le D_Pe^{\eta^2RT/p}  \frac{(p\lambda)^{r}}{1- p\lambda}
\end{align}
with the temporary variable $\lambda:=16ed(\frac{\eta^2g(\eta)(T+r-1)}{r})$. 

By Markov inequality with tunable parameter $p =\max( \frac{1}{e\lambda}\epsilon^{1/r},2)$,
 \begin{align}
     \sup_P \BP[\lV O^{T}_{\{r\}} P\rV \ge \epsilon ] & \le   \frac{\sup_P\BE[\lV O^{T}_{\{r\}}P \rV_p^p ]}{\epsilon^p}\le \frac{ \vertiii{O^{T}_{\{r\}}}_{D_P,p}^p}{\epsilon^p}\\
     & = \begin{cases}
     \displaystyle D_P\frac{e^{\eta^2RT/2}}{1- 1/e} \exp(-\frac{\epsilon^{2/r}}{32e^2d(\frac{\eta^2g(\eta)(T+r-1)}{r^2})}) & \text{if } (32e^2d\frac{\eta^2g(\eta)(T+r-1)}{r})^{r} < \epsilon^2\\
     \\
    \displaystyle D_P\frac{e^{\eta^2RT/2}}{\epsilon^2}\frac{(32ed\frac{\eta^2g(\eta)(T+r-1)}{r})^{r}}{1-32ed\frac{\eta^2g(\eta)(T+r-1)}{r}} & \text{if }  (32e^2d\frac{\eta^2g(\eta)(T+r-1)}{r})^{r} \ge \epsilon^2.
     \end{cases}
 \end{align}

 Or expressed in terms of $\delta$:
 \begin{align}
     \BP\left[\lV O^{T}_{\{r\}} P\rV > \epsilon(\delta) \right]\le \delta,
 \end{align}
\begin{align}
    \epsilon(\delta):= 
    \begin{cases}
   \left[32e^2d(\frac{\eta^2g(\eta)(T+r-1)}{r^2})(\frac{\eta^2RT}{2}+\ln(D_P/(1-1/e)\delta))\right]^{r} &\text{if } \delta <\frac{D_P}{1-1/e}e^{\eta^2Td-r} \\
    \\
   \displaystyle \sqrt{\frac{D_Pe^{\eta^2RT/2}}{\delta}\frac{(32ed\frac{\eta^2g(\eta)(T+r-1)}{r})^{r}}{1-32ed\frac{\eta^2g(\eta)(T+r-1)}{r}}}  &\text{if } \delta \ge \frac{D_P}{1-1/e}e^{\eta^2Td-r}
    \end{cases} 
\end{align}
The logarithmic dependence on $\delta$ captures the very strong tail bound that might be useful elsewhere. 
Our next goal is extracting the asymptotic velocity, which only requires the second moment bound and Chebyshev. Take the order of limit: pick $\epsilon >0$, then for any $\zeta>0$, define $T(r)$ such that
\begin{align}
     e^{\eta^2Td/r}32ed\frac{\eta^2g(\eta)(T+r-1)}{r} = 1- \zeta.
\end{align}
Then we obtain a Lieb-Robinson velocity almost surely
\begin{align}
    \lim_{r\rightarrow \infty} \BP(\lV\frac{1}{2}[A_{r},O_{0}(T(r))]P\rV \ge \epsilon ) =0. 
\end{align}
The expression of $\frac{r}{T(r)}$ can be simplified in the Brownian limit $\xi \rightarrow 0, \tau = \xi^2T$ or physically when the 'coherence time' is short compared to the hopping term $\xi \ll 1/a $,
\begin{equation}
    \frac{r}{\tau(r)\xi^2} = u \le 32ede^{1/32e} a^2
\end{equation}
where we used $32ed\frac{\eta^2(T+r-1)}{r}\le 1$ to estimate the exponential term.

\subsection{Time-independent k-local systems}
\label{sec:proof_time-Indep_klocal}
Consider a complete $k$-local random Hamiltonian
\begin{equation}
    H^{}_{k,N} :=\sum_{i_1< \ldots <i_k \le N}  \sqrt{\frac{J^2 (k-1)!}{kN^{k-1}} }H^{}_{i_1\cdots i_k} 
\end{equation}
where each term is normalized $\lV H_{i_1\cdots i_k}\rV\le 1$ and zero mean $\BE[H_{i_1\cdots i_k}]=0$. 
Here we prepare ourselves with some numerical values: there are at most $|\Delta(\gamma)|\le R =  (k-1)\frac{\binom{N}{k}}{N/k} \le\frac{ N^{k-1}}{(k-2)!}$ possible next steps at any self-avoiding path. Since all k-local terms $H_X$ have the same bound, $b^2_X \le b^2 = \frac{J^2 (k-1)!}{kN^{k-1}}$ and hence $Rb^2\le J^2$. Plug in the sum over self-avoiding paths (Theorem ~\ref{thm:sum_over_paths})
\begin{align}
\vertiii{\frac{1}{2}[A_r,O_0(t)]}_{D_P,p}^2
&\le \sum_{\Gamma}  \int\limits_{t>t_\ell>\cdots >t_1} \mathrm{d}t_1\cdots \mathrm{d}t_\ell \prod_{1\le k\le \ell} \left(e^{\beta_k(t_{k+1}-t_k)} \frac{4(p-1) b_{X_j}^2}{\beta_k} \right) \vertiii{O^0}_{D_P,p}^2\\
&\le e^{\sqrt{4(p-1) Rb^2}t} \vertiii{O^0}_{D_P,p}^2 \sum_{\text{self-avoiding paths } \Gamma: r \rightarrow 0 } (\sqrt{\frac{4(p-1)b^2}{R}})^\ell \frac{t^\ell}{\ell!}\\
&\le e^{\sqrt{4(p-1) Rb^2}t} \sum_{\ell=1}^\infty \frac{(k-1)}{N} (\sqrt{4(p-1)Rb^2})^\ell \frac{t^\ell}{\ell!} \vertiii{O^0}_{D_P,p}^2\\
&\le D^{2/p}_Pe^{4Jt\sqrt{(p-1)}} \frac{(k-1)}{N}
\end{align}
where in the second inequality we set a convenient parameter $\beta_k = \sqrt{4(p-1)\sum_{\Delta(\gamma)} b_j^2}= \sqrt{4(p-1) Rb^2}$; in the third the number of path is overestimated to be $R^{\ell-1}\cdot R(k-1)/N$; in the last we used $\vertiii{O^0}_{D_P,p}^2 \le D_P^{2/p}$ for $\lV O\rV \le 1$. By Markov inequality with tuneable parameter $p = \max\left((\frac{\ln(N\epsilon^2/(k-1))}{6Jt})^2,2\right)$,
 \begin{align}
     \sup_P \BP[\lV O^{t}_{\{r\}} P\rV \ge \epsilon ] &\le D_P\left(e^{4Jt\sqrt{(p-1)}} \frac{(k-1)}{N\epsilon^2} \right)^{p/2}\\
     &\le \begin{cases}
     D_P\exp \left( -(\frac{\ln(N\epsilon^2/(k-1))}{6})^3\frac{1}{J^2t^2}\right) &\text{if } (\frac{\ln(N\epsilon^2/(k-1))}{6Jt})^2>2\\
     \\
      D_P\left(e^{4Jt} \frac{(k-1)}{N\epsilon^2} \right) &\text{if } (\frac{\ln(N\epsilon^2/(k-1))}{6Jt})^2\le 2
     \end{cases} ,
 \end{align}
where in the Markov's inequality for $p>2$ we conveniently use $p \ge p-1$ while keep $p-1$ at $p=2$. Expressing in terms of $\delta$:
 \begin{align}
     \BP\left[\lV O^{t}_{\{r\}} P\rV > \epsilon(\delta) \right]\le \delta,
 \end{align}
\begin{align}
    \epsilon(\delta):= 
    \begin{cases}
    \frac{k-1}{N}\exp \left(6\sqrt[3]{\ln(D_P/\delta)J^2t^2}\right)&\text{if } \delta < D_P e^{(4-6\sqrt{2})Jt} \\
    \\
    \sqrt{ D_P\left(e^{4Jt} \frac{(k-1)}{N\delta} \right)} &\text{if } \delta \ge D_P e^{(4-6\sqrt{2})Jt}
    \end{cases} 
\end{align}

For the asymptotic bound on scrambling time we only use the second moment bound. For any $\epsilon>0$, $\zeta>0$, set $t(N)$ such that
\begin{align}
    D_P\left(e^{4Jt} \frac{(k-1)}{N} = \frac{1}{N^{\zeta}}\right),
\end{align}
then 
\begin{align}
    \lim_{N\rightarrow \infty} \BP(\lV\frac{1}{2}[A_{r},O_{0}(t(N))]P\rV \ge \epsilon ) =0. 
\end{align}
with a bound on asymptotic scrambling time
\begin{align}
    Jt_{scr} \stackrel{a.s.}{\ge} \frac{1}{4}\ln(\frac{N\epsilon^2}{k-1})
\end{align}
where we keep the dependence on $k$ in case $k$ scales with $N$.

\subsection{Brownian $k$-local systems}\label{sec:klocal_OTOC}
Consider a complete $k$-body Brownian Hamiltonian 
\begin{equation}
    H^{(T)}_{k,N} :=\sum_{i_1< \ldots <i_k \le N}  \sqrt{\frac{J^2 (k-1)!}{kN^{k-1}}} H^{(T)}_{i_1\cdots i_k}, 
\end{equation}
where each term is normalized $\lV H_{i_1\cdots i_k}\rV\le 1$ and independent zero mean conditioned on the past $\BE_{T-1}[H^{(T)}_{i_1\cdots i_k}]=0$. 
Consider the Brownian limit that $T\xi^2=\tau$ is held fixed. 
Here we prepare ourselves with some numerical values: there are at most $|\Delta(\gamma)|\le R =  (k-1)\frac{\binom{N}{k}}{N/k} \le\frac{ N^{k-1}}{(k-2)!}$ possible next steps at any self-avoiding path. Since all k-local terms $H_X$ have the same bound, $\eta^2_X \le \eta^2 = \frac{J^2 (k-1)!}{kN^{k-1}} \xi^2$. Plug in the sum over self-avoiding paths (Theorem ~\ref{thm:sum_over_paths})
\begin{align}
\vertiii{O^T_{\{r\}}}_{D_P,p}^2&\le e^{\eta^2RT/p} \vertiii{O^0}_{D_P,p}^2 \sum_{\text{self-avoiding paths } \gamma: r \rightarrow 0 } \frac{(8p\eta^2T)^\ell}{\ell!}\\
&\le D^{2/p}_Pe^{\eta^2RT/p} \sum_{\ell=1}^\infty \frac{(k-1)}{N} \frac{(R8p\eta^2T)^\ell}{\ell!}\\
&\le D^{2/p}_Pe^{\eta^2RT/p}\frac{(k-1)e^{p\lambda T}}{N}.
\end{align}
with a temporary value $\lambda:=8R\eta^2 \le 8\xi^2J^2$.
By Markov inequality with tunable parameter $p=\max( \frac{\ln(N\epsilon^2/(k-1))}{16\xi^2TJ^2},2)$,
 \begin{align}
     \BP[\lV O^T_{\{r\}}P \rV \ge \epsilon ] &\le  \frac{\sup_P\BE[\lV O^T_{\{r\}}P \rV^p ]}{\epsilon^p} \le \frac{ \vertiii{O^T_{\{r\}}}_{D_P,p}^p}{\epsilon^p}\\
     &=\begin{cases}
     D_Pe^{J^2\tau/2} \exp({\frac{-[\ln(N\epsilon^2/(k-1))]^2}{32J\tau }}). &\text{if } \frac{\ln(N\epsilon^2/(k-1))}{16\tau J^2} > 2\\
     D_Pe^{17J^2\tau /2}\frac{(k-1)}{N\epsilon^2} &\text{if } \frac{\ln(N\epsilon^2/(k-1))}{16\tau J^2} \le 2
     \end{cases}.
 \end{align}
For failure probability $\delta$, 
Expressing in terms of $\delta$:
 \begin{align}
     \BP\left[\lV O^{\tau}_{\{r\}} P\rV > \epsilon(\delta) \right]\le \delta,
 \end{align}
\begin{align}
    \epsilon(\delta):= 
    \begin{cases}
    \sqrt{ \frac{(k-1)}{N}\exp(\sqrt{ 32J^2\tau  (J^2\tau /2 -\ln(\delta/D_P) ) }) } &\text{if }  \delta < D_P e^{-23J^2\tau /2} \\
    \\
    \sqrt{ D_Pe^{17J^2\tau /2}\frac{k-1}{N\delta}} &\text{if } \delta \ge D_P e^{-23J^2\tau /2}
    \end{cases} .
\end{align}
For the asymptotic bound on scrambling time we only use the second moment bound. For any $\epsilon>0$, $\zeta>0$, set $T(N)$ such that
\begin{align}
    D_P\left(e^{17J^2\tau/2} \frac{(k-1)}{N} = \frac{1}{N^{\zeta}}\right).
\end{align}
Then 
\begin{align}
    \lim_{N\rightarrow \infty} \BP(\lV\frac{1}{2}[A_{r},O_{0}(\tau(N))]P\rV \ge \epsilon ) =0, 
\end{align}
with a bound on asymptotic scrambling time
\begin{align}
    J^2\tau_{scr} \stackrel{a.s.}{\ge} \frac{2}{17}\ln(\frac{N\epsilon^2}{k-1}),
\end{align}
where we keep the dependence on $k$ in case $k$ scales with $N$. Note the $17/2$ exponent can be slightly improved if we instead tune the arithmetic-geometric inequality in the derivaiton of inductive Lemma (Lemma~\ref{lem:induction_step}).  This is another short proof for the $\tau \ge \ln(N)$ bound on fast scrambling of OTOC, very flexible on whichever $k$-body term being used. 

\subsection{Time-independent, $d=1$ power-law interacting system, OTOC}\label{sec:proof_time_indep_1d_long_OTOC}
Consider a $d=1$ power-law time-independent Hamiltonian where each term is normalized $\lV H^{}_{ij}\rV\le 1$ and zero mean $\BE[H^{}_{ij}]=0$
\begin{equation}
     H :=\sum_{i} \frac{1}{|i-j|^\alpha} H_{ij}.
\end{equation}

The proof largely follows~\cite{alpha_3_chenlucas}, which is an interesting result but some of the detailed calculations are rather specialized, and may not be enlightening to the reader. Therefore, we do not re-introduce the subject and simply inherited some of the definitions, and recommend the enthusiastic reader to parse both papers together. For readers seeking for applications other than power-law interacting systems, the takeaway here is that we can use the concentration bounds as a \textit{subroutine} after we pre-process the more complicated evolution into pieces that resembles a sum over paths. 

\begin{figure}[t]
    \centering
    \includegraphics[width=0.9\textwidth]{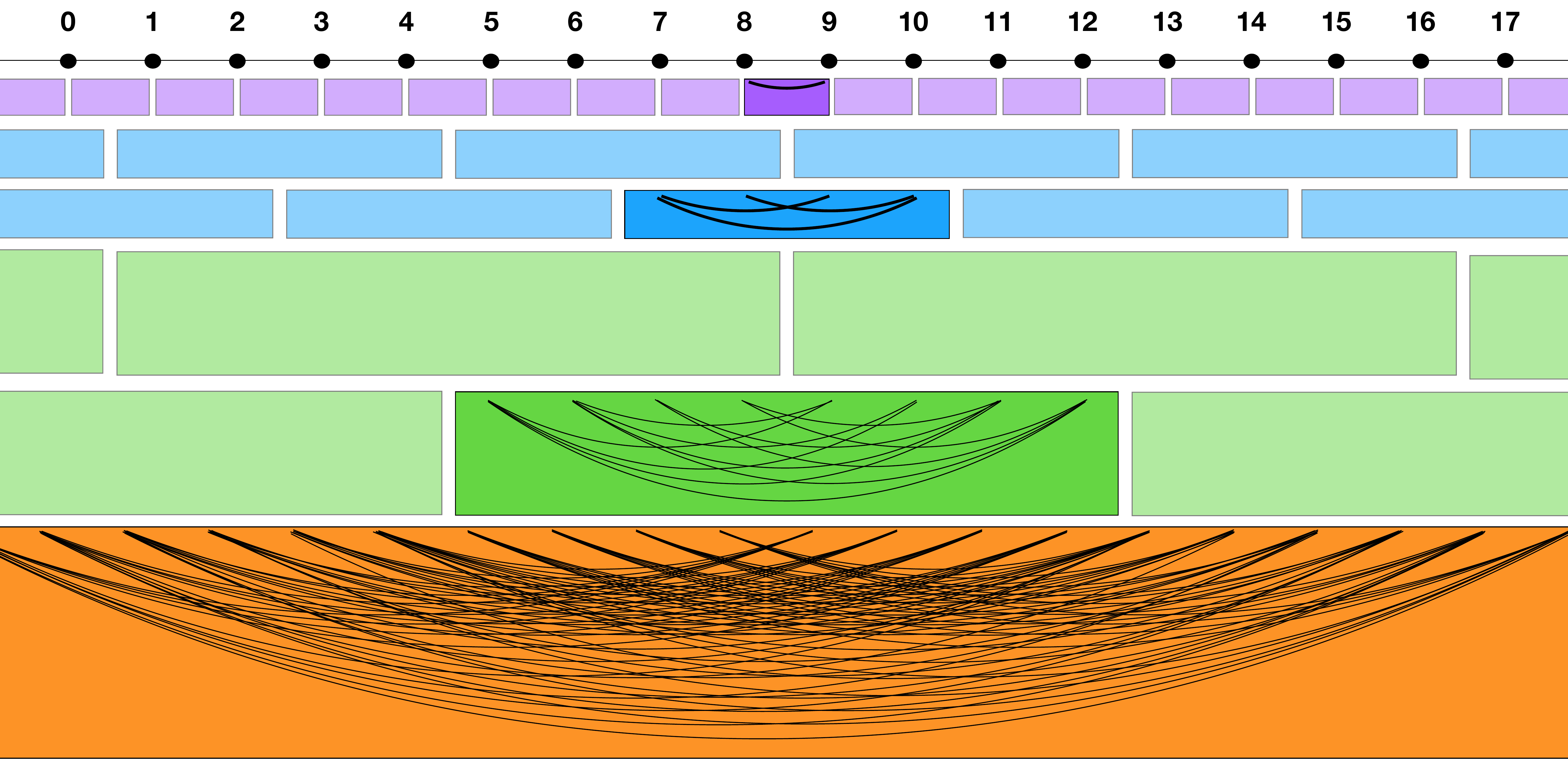}
    \caption{ Regrouping the Hamiltonian into scales alike as in~\cite{alpha_3_chenlucas}. For any operator growing past $r$ there must be a scale with sufficiently many advancing hops. The difference for Brownian system is that, after plugging in our formula (Theorem~\ref{thm:sum_over_paths}), essentially the terms in a block sum by squares. Each block of scale $\ell$ has strength $\CO(\ell^{1-\alpha})$ in the random case instead of $\CO(\ell^{2-\alpha})$ in the deterministic case, shifting the critical $\alpha$.
    }
    \label{fig:scale by scale}
\end{figure}

\textit{proof.} Immediately apply Theorem~\ref{thm:sum_over_paths} would yield incorrect result. Instead, we have to first decompose the unitary into scales (Fig.\ref{fig:scale by scale}):
\begin{align}
\vertiii{\frac{1}{2}[A_r,O_0(t)]}_{D_P,p}& \le \vertiii{[1-\prod^{}_k(1-\chi_k)]O^t}_{D_P,p}.
\end{align}.
The differences are (1) changing some parameter in $N_k$:
\begin{equation}
N_k :=\begin{cases}
\left\lceil \frac{1}{2} \frac{2^{-k(\alpha-2)/2}}{ M}  \frac{r}{2^{k}} \right\rceil &\text{if}\ k\le n^*\\
1 &\text{if}\ k > n_*
\end{cases}\label{eq:Nq},
\end{equation}
where $M:=\sum_{k^\prime=1}^{n_*} 2^{-k^\prime(\alpha-2)/2}$ is a normalization factor.\\

(2) within each equivalence class $\chi_k$, we further regroup by the different forward paths, labeled by their steps $m_{N_k},\cdots,m_1$. 
\begin{align}
    \chi_kO^{t} = \sum_{forward\ paths:\ m_{N_k}>\cdots>m_1} \chi_kO^{t}(m_{N_k},\cdots,m_1).
\end{align}
Each of the forward path has exactly the same expression as a $N_{k}+1$ sites nearest neighbour spin chain with $2^{2(k-1)}$ terms between the $S_i, S_{i+1}$ 'super'sites (Fig.~\ref{fig:multi-scale}). From here we can apply Theorem~\ref{thm:time_indep_sum_over_path} on each forward path $(m_{N_k},\cdots,m_1)$, with the number of available hops $\Delta(\gamma) = 2^{2(k-1)}$.
\begin{align}
\vertiii{\chi_kO^{t}(m_{N_k},\cdots>m_1)}_{D_P,p}\le (\sqrt{4pR_k\eta_k^2})^{N_k} \frac{t^{N_k}}{N_k!}\vertiii{O^0}_{D_P,p}
\end{align}
We simply use the triangle inequality between these different forward paths, which only adds constant overheads without hindering the critical $\alpha$.
\begin{align}
\vertiii{\chi_kO^{t}}_{D_P,p}\le \binom{2\lceil \frac{r}{2^k}\rceil+1}{N_k} (\sqrt{4pR_k\eta_k^2})^{N_k} \frac{t^{N_k}}{N_k!}\vertiii{O^0}_{D_P,p}=: C_k
\end{align}
where $\eta_k \le \xi \frac{1}{2^{(k-1)\alpha}}$ is the maximum strength, and the sum over block yields $[\sum_{k - \text{block}}\eta_k^2]=\eta_k^2R_k \le \xi^2 2^{-(k-1)2(\alpha-1)}$, featuring a sum of squares; The above function has transition at (a) starting value $k_*$ at which a long $k$-forward path has a single coupling: $N_k=1$ for $k\ge k_*$. (b)$\frac{r}{2^k} \le 1 $. (a) occurs when 
\begin{equation}
\frac{M}{R}\ge \frac{1}{2^{1+k_*(\alpha/2)}},
\end{equation}
Hence 
\begin{align}
    C_k \le \begin{cases}
    (e^2 2^{4+\alpha}\sqrt{p} \frac{M^2t}{r})^{N_k} &\ \text{if } k< k_*\\
    4e^2\sqrt{p} \frac{rt}{2^{\alpha(k-1)}} &\ \text{if } n_*\ge k\ge k_*\\
    8e^2\sqrt{p} \frac{t}{2^{(k-1)(\alpha-1)}} &\ \text{if } k > n_*
    \end{cases}
\end{align}
And for cross terms, the different scales do not affect each other, and hence it becomes a simple product. Formally this becomes a n-dimensional lattice of nearest neighbor spin chain (Fig.~\ref{fig:multi-scale}), and applying Theorem~\ref{thm:sum_over_paths} on it yields the product form.
\begin{align}
\vertiii{\chi_{k_1}\cdots \chi_{k_n}O^t}_{D_P,p}\le [\prod_k \binom{2\lceil \frac{r}{2^k}\rceil+1}{N_k}][\prod_k (\sqrt{4pR_k\eta_k^2})^{N_k} \frac{t^{N_k}}{N_k!}]\vertiii{O^0}_{D_P,p} \le C_{k_1}\cdots C_{k_n} 
\end{align}
The rest will be a matter of accounting. \begin{align}
\vertiii{O^t}_{D_P,p}& = \vertiii{[1-\prod_k(1-\chi_k)]O^t}_{D_P,p}\\
&\le \prod_k(1+C_{k})-1\\
&\le \exp(\sum_k C_{k}) - 1\\
&\le 2 \sum_k C_{k}
\end{align}
where we used the elementary $e^x-1 \le 2 x$ for $e^x \le 2$. 
where 
\begin{align}
    \sum_k C_k = \sum_{k=1}^{k_*-1} (e^2 2^{4+\alpha}\sqrt{p} \frac{M^2t}{r})^{N_k}+ \sum_{k=k_*}^{n_*} 4e^2\sqrt{p} \frac{rt}{2^{\alpha(k-1)}}+ 
    \sum_{k=n_*}^{\infty} 8e^2\sqrt{p} \frac{t}{2^{(k-1)(\alpha-1)}}
\end{align}

\textbf{Case $\alpha > 2$:}  
Calculate value $M < \sum_{q=1}^\infty 2^{-q(\alpha-2)/2} = \frac{1}{1-2^{-(\alpha-2)/2}}:=M'.$ and plug in each summand:
\begin{align}
 \sum_{k=1}^{k_*-1} (e^2 2^{4+\alpha}\sqrt{p} \frac{M^2t}{r})^{N_k}  &:= \sum_{k=1}^{k_*-1} \left( \sqrt{p}\frac{b_1t}{r}\right)^{N_k} \le \frac{\sqrt{p}\frac{b_1t}{r}}{1-\sqrt{p}\frac{b_1t}{r} }
\end{align} 
where in the bound by geometric series we used that 
\begin{equation}
N_1 > N_2 > \cdots > N_{k_*-1}.  \label{eq:Nqinequality}
\end{equation}
To derive this, note that the argument of the ceiling function in (\ref{eq:Nq}) changes by a factor of $2^{\alpha^\prime/2}$ each time $q$ changes by 1.  When $\alpha^\prime>2$, this factor is larger than 2, so once the argument is larger than 1, it changes by at least 1: $N_q \le N_{q-1}-1$. For the second term for those $k \ge k_*$, they decay at first by a factor of $2^\alpha$ and then $2^{\alpha-1}$ above $n_*$. 
\begin{align}
\sum_{k=k_*}^{n_*} 4e^2\sqrt{p} \frac{rt}{2^{\alpha(k-1)}}+ 
    \sum_{k=n_*}^{\infty} 8e^2\sqrt{p} \frac{t}{2^{(k-1)(\alpha-1)}} &\le 4e^2\sqrt{p} \frac{rt}{2^{\alpha(k_*-1)}}\frac{1}{1-2^{\alpha - 1}}\\
&\le \frac{4e^22^{2+\alpha}M'^2}{1-2^{\alpha - 1}}\sqrt{p} \frac{t}{r}\\
&:= \sqrt{p}\frac{b_2t}{r}
\end{align} We arrive at concentration by setting $p =\max( \frac{r^2\epsilon^2}{c^2 t^2 e},2)$
\begin{align}
     \sup_P\BP[\lV O^t_{\{r\}}P \rV \ge \epsilon ] &\le  \frac{\BE[\lV O^t_{\{r\}} \rV_2^p ]}{\epsilon^p} \le \frac{ \vertiii{O^t_{\{r\}}}_{D_P,p}^p}{\epsilon^p}\\
     &\le D_P(\frac{2}{\epsilon}\frac{\sqrt{p}\frac{b_1t}{r}}{1-\sqrt{p}\frac{b_1t}{r} } +  \frac{2}{\epsilon}\sqrt{p}\frac{b_2t}{r}) ^{p}\\
     &\le D_P (\frac{2}{\epsilon}\frac{\sqrt{p}\frac{b_1t}{r}}{1-\frac{b_1}{2(b_1+b_2)} } +  \frac{2}{\epsilon} \sqrt{p}\frac{b_2t}{r} ) ^{p} := (\sqrt{p}\frac{ct}{\epsilon r} ) ^{p} \\
     &\le \begin{cases}
     D_P\exp(-\frac{r^2\epsilon^2}{2ec^2t^2}) &\text{if } \epsilon^2 > 2e\frac{c^2 t^2}{ r^2}\\
     \\
     D_P\frac{2c^2t^2}{r^2\epsilon^2} &\text{if } \epsilon^2\le  2e\frac{c^2 t^2}{ r^2}
     \end{cases}
 \end{align}
where in the third line we used the otherwise vacuous constraint $\epsilon \le 1$ and defined $c=\frac{2b_1}{1-b_1/2(b_1+b_2)}+2b_2$. Expressing in terms of $\delta$:
 \begin{align}
     \BP\left[\lV O^{t}_{\{r\}} P\rV > \epsilon(\delta) \right]\le \delta,
 \end{align}
\begin{align}
    \epsilon(\delta):= 
    \begin{cases}
    \sqrt{2e\ln(D_P/\delta)\frac{c^2t^2}{r^2}} &\text{if } \delta <\frac{D_P}{e} \\
    \\
    \sqrt{D_P\frac{2c^2t^2}{r^2\delta}} &\text{if } \delta \ge \frac{D_P}{e}
    \end{cases} .
\end{align}
The logarithmic dependence on $\delta$ captures the very strong tail bound that might be useful elsewhere. 
Our next goal is extracting the asymptotic velocity, which only requires the second moment bound and Chebyshev. Take the order of limit: pick $\epsilon >0$, then for any function approaching infinity abrbitrarily slowly, $\lim_{r\rightarrow\infty} f(r)=\infty$, define $T(r)$ such that
\begin{align}
     t = rf(r),
\end{align}
we obtain a linear light cone almost surely
\begin{align}
    \lim_{r\rightarrow \infty} \BP(\lV\frac{1}{2}[A_{r},O_{0}(t(r))]P\rV \ge \epsilon )= 0. 
\end{align}
Note we can also get a velocity for a fixed $\epsilon >0, D_P/e >\delta >0$:
\begin{align}
    u = c\frac{\sqrt{2e\ln(D/\delta)}}{\epsilon^2}.
\end{align}
\textbf{Case $1<\alpha<2$.}
Moreover, we now find \begin{equation}
M = \sum_{q=1}^{n_*} 2^{q(2-\alpha)/2} < r^{(2-\alpha)/2} \sum_{k^\prime=0}^\infty 2^{-k^\prime((2-\alpha)/2} = \frac{r^{(2-\alpha)/2}}{1-2^{-(2-\alpha)/2}}=: M'
\end{equation}
and that 
\begin{equation}
\frac{1}{2^{k_*(\alpha)/2}} \le \frac{2M}{r} \le  \frac{2M'}{r} \le \frac{2r^{-\alpha/2}}{1-2^{-(2-\alpha)/2}} 
\end{equation}
\begin{align}
 \sum_{k=1}^{k_*-1} (e^2 2^{4+\alpha}\sqrt{p} \frac{M^2t}{r})^{N_k}  &= \sum_{k=1}^{k_*-1} \left( \sqrt{p}\frac{b_1t}{r^{\alpha-1}}\right)^{N_k} \le 2\frac{\sqrt{p}\frac{b_1t}{r^{\alpha-1}}}{1-\sqrt{p}\frac{b_1t}{r^{\alpha-1}} }
\end{align} 
In the last line, we get a factor 2 due to $N_k$ growing by $2^{\alpha/2} \ge \sqrt{2}$ each time $k$ varies by 1: $N_1 > N_3 > N_5\cdots > N_{k_*-1}$.

\begin{align}
\sum_{k=k_*}^{n_*} 4e^2\sqrt{p} \frac{rt}{2^{\alpha(k-1)}}+ 
    \sum_{k=n_*}^{\infty} 8e^2\sqrt{p} \frac{t}{2^{(k-1)(\alpha-1)}} &\le 4e^2\sqrt{p} \frac{rt}{2^{\alpha(k_*-1)}}\frac{1}{1-2^{\alpha - 1}}\\
&\le \frac{4e^22^{2+\alpha}M'^2}{1-2^{\alpha - 1}}\sqrt{p} \frac{t}{r}\\
&:= \sqrt{p}\frac{b_2t}{r^{\alpha-1}}
\end{align}
The rest is a close analog of the $\alpha>2$ case except for a few constants. We arrive at concentration by setting $p =\max( \frac{r^{2\alpha-2}\epsilon^2}{t^2 ec},2)$
\begin{align}
     \sup_P\BP[\lV O^t_{\{r\}}P \rV \ge \epsilon ] &\le  \frac{\BE[\lV O^t_{\{r\}} \rV_2^p ]}{\epsilon^p} \le \frac{ \vertiii{O^t_{\{r\}}}_{D_P,p}^p}{\epsilon^p}\\
     &\le D_P(\frac{4}{\epsilon}\frac{b_1 \sqrt{p} \frac{t}{r^{\alpha-1}}}{1-b_1 \sqrt{p} \frac{t}{r^{\alpha-1}} } +  \frac{2}{\epsilon}b_2 \sqrt{p} \frac{t}{r^{\alpha-1}} ) ^{p}\\
     &\le D_P (\frac{4}{\epsilon}\frac{b_1 \sqrt{p} \frac{t}{r^{\alpha-1}}}{1-\frac{b_1}{2(2b_1+b_2)} } +  \frac{2}{\epsilon}b_2 \sqrt{p} \frac{t}{r^{\alpha-1}} ) ^{p} := (c\sqrt{p}\frac{t}{r^{\alpha-1}} ) ^{p} \\
     &\le \begin{cases}
     D_P\exp(-\frac{r^{2\alpha-2}\epsilon^2}{2ec^2 t^2}) &\text{if } \epsilon^2>  2e\frac{c^2t^2}{r^{2\alpha-2}}\\
     \\
     D_P\frac{2c^2t^2}{r^{2\alpha-2}\epsilon^2} &\text{if } \epsilon^2\le  2e\frac{c^2t^2}{r^{2\alpha-2}}
     \end{cases}
 \end{align}
where in the third line we used the otherwise vacuous constraint $\epsilon \le 1$ and defined $c=\frac{4b_1}{1-b_1/2(2b_1+b_2)}+2b_2$. Expressing in terms of $\delta$:
 \begin{align}
     \BP\left[\lV O^{t}_{\{r\}} P\rV > \epsilon(\delta) \right]\le \delta,
 \end{align}
\begin{align}
    \epsilon(\delta):= 
    \begin{cases}
    \sqrt{2e\ln(D_P/\delta)\frac{c^2t^2}{r^{2\alpha-2}}} &\text{if } \delta <\frac{D_P}{e} \\
    \\
    \sqrt{D_P\frac{2c^2t^2}{r^{2\alpha-2}\delta}} &\text{if } \delta \ge \frac{D_P}{e}
    \end{cases} .
\end{align}
The logarithmic dependence on $\delta$ captures the very strong tail bound that might be useful elsewhere. 
Our next goal is extracting the asymptotic velocity, which only requires the second moment bound and Chebyshev. Take the order of limit: pick $\epsilon >0$, then for any function approaching infinity arbitrarily slowly, $\lim_{r\rightarrow\infty} f(r)=\infty$, define $t(r)$ such that
\begin{align}
     t = r^{\alpha-1}f(r),
\end{align}
we obtain a linear light cone almost surely
\begin{align}
    \lim_{r\rightarrow \infty} \BP(\lV\frac{1}{2}[A_{r},O_{0}(t(r))]P\rV \ge \epsilon )= 0. 
\end{align}
Note we can also get a algebraic velocity for a fixed $\epsilon >0, D_P/e >\delta >0$:
\begin{align}
    \frac{r^{\alpha-1}}{t} = \sqrt{2ec^2\frac{\ln(D/\delta)}{\epsilon^2}}.
\end{align}

\subsection{Brownian $d=1$ power-law interacting system}\label{sec:proof_brown_1d_long_OTOC}
Consider a $d=1$ power law Brownian Hamiltonian where each term is normalized $\lV H^{(T)}_{ij}\rV\le 1$ and independent zero mean conditioned on the past $\BE_{T-1}[H^{(T)}_{ij}]=0$
\begin{equation}
     H^{(T)} :=\sum_{i} \frac{1}{|i-j|^\alpha} H^{(T)}_{ij}.
\end{equation} 
Consider the Brownian limit that $T\xi^2:=\tau$ is held fixed. 

The proof large follows~\cite{alpha_3_chenlucas}. We cannot immediately apply Theorem~\ref{thm:sum_over_paths}, but instead have to first decompose the unitary into $\log(r)$ scales:
\begin{align}
\vertiii{\frac{1}{2}[A_r,O(T)_0]}_{D_P,p}& \le \vertiii{[1-\prod_k(1-\chi_k)]O^T}_{D_P,p}.
\end{align}
The differences are (1) changing some parameter in $N_k$:
\begin{equation}
N_k :=\begin{cases}
\left\lceil \frac{1}{2} \frac{2^{-k(2\alpha-3)/3}}{ M}  \frac{r}{2^{k}} \right\rceil &\text{if}\ k\le n^*\\
1 &\text{if}\ k > n_*
\end{cases}\label{eq:Nq_br_OTOC}
\end{equation}
where $M:=\sum_{k^\prime=1}^{n_*} 2^{-k^\prime(2\alpha-3)/3}$ is a normalization factor.\\

\begin{figure}[t]
    \centering
    \includegraphics[width=0.9\textwidth]{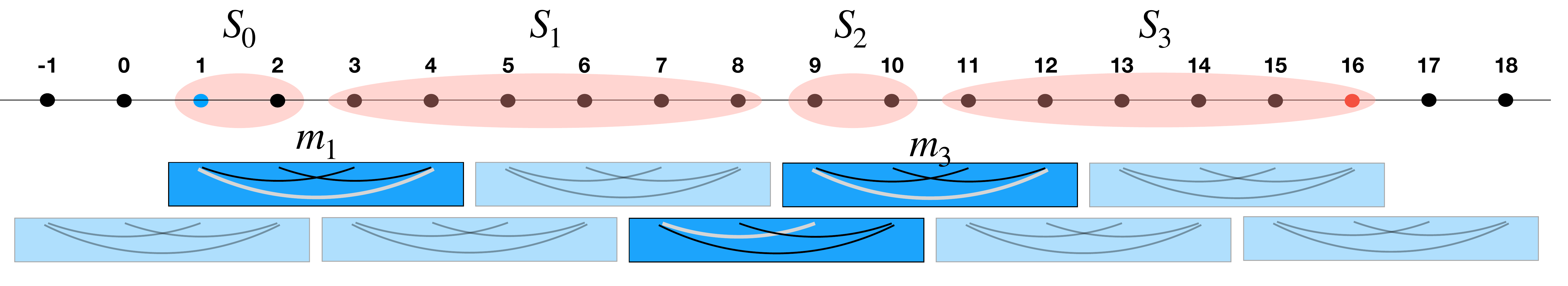}
    \caption{
    The effective nearest neighbor chain on which we apply the formula (Theorem~\ref{thm:sum_over_paths}). Within each scale, there are different forward paths labeled by $(m_1,\cdots,m_{N_k})$. For each forward paths, the evolved operator $\chi_kO^{T}(m_{N_k},\cdots,m_1)$ is identical to a nearest neighbor chain with 'super'sites $S_i$ and multiple interacting terms in between. An self-avoiding path in this system would be taking a term in each $m_i$, colored white in the figure. Different paths sum as squares by calling Theorem~\ref{thm:sum_over_paths}.
    }
    \label{fig:forward paths}
\end{figure}
\begin{figure}[t]
    \centering
    \includegraphics[width=0.9\textwidth]{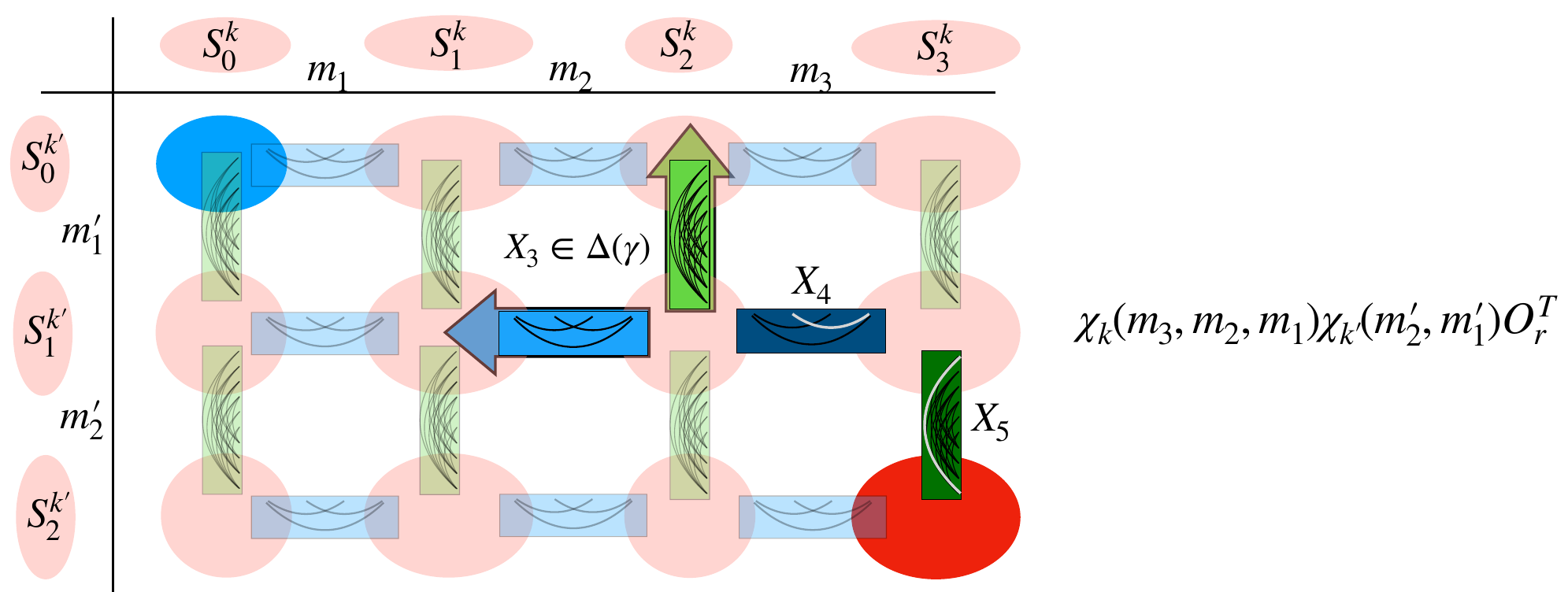}
    \caption{ The effective nearest-neighbor chain on which we apply the formula (Theorem~\ref{thm:sum_over_paths}). Note that we can only move forward at each scale (the green and blue arrows), so it should be thought of as a product of many spin chain. The dimension of the lattice depends on the number of scales $\chi_k$. An self-avoiding path would be taking a term in each $m_i$, colored white. 
    }
    \label{fig:multi-scale}
\end{figure}

(2) within each equivalence class $\chi_k$, we further regroup by the different forward paths, labeled by their steps $m_{N_k},\cdots,m_1$. 
\begin{align}
    \chi_kO^{T} = \sum_{forward\ paths:\ m_{N_k}>\cdots>m_1} \chi_kO^{T}(m_{N_k},\cdots,m_1)
\end{align}
Each of the forward path has exactly the same expression as a $N_{k}+1$ sites nearest neighbour spin chain with $2^{2(k-1)}$ terms between the $S_i, S_{i+1}$ 'super'sites (Fig.~\ref{fig:multi-scale}). From here we can apply~\ref{thm:sum_over_paths} on each forward path $(m_{N_k},\cdots,m_1)$, with the number of available hops $\Delta(\gamma) = 2^{2(k-1)}$.
\begin{align}
\vertiii{\chi_kO^{T}(m_{N_k},\cdots>m_1)}_{D_P,p}^2\le e^{\frac{\eta_k^2R_kT}{p}}\frac{(8pR_k\eta_k^2T)^{N_k}}{N_k!} \vertiii{O^0}_{D_P,p}^2
\end{align}
We simply use the triangle inequality between these different forward paths, which only adds constant overheads without hindering the critical $\alpha$.
\begin{align}
\vertiii{\chi_kO^{T}}_{D_P,p}^2
\le \binom{2\lceil \frac{r}{2^k}\rceil+1}{N_k}^2 e^{\frac{\eta_k^2R_kT}{p}}\frac{(8pR_k\eta_k^2T)^{N_k}}{N_k!} \vertiii{O^0}_{D_P,p}^2=: C_k
\end{align}
where $\eta_k \le \xi \frac{1}{2^{(k-1)\alpha}}$ is the maximum strength, and the sum over block yields $[\sum_{k - \text{block}}\eta_k^2]=\eta_k^2R_k \le \xi^2 2^{-(k-1)2(\alpha-1)}$, featuring a sum of squares; The above function has transition at (a) starting value $k_*$ at which a long $k$-forward path has a single coupling: $N_k=1$ for $k\ge k_*$. (b)$\frac{r}{2^k} \le 1 $. (a) occurs when 
\begin{equation}
\frac{M}{R}\ge \frac{1}{2^{1+k_*(2\alpha/3)}},
\end{equation}
Hence 
\begin{align}
    C_k \le e^{\frac{\eta_k^2R_kT}{p}} \cdot \begin{cases}
    (e^3 2^{8+2\alpha}p \frac{M^3\tau}{r})^{N_k} &\ \text{if } k< k_*\\
    16e^3p \frac{r^2\tau}{2^{2\alpha(k-1)}} &\ \text{if } n_*\ge k\ge k_*\\
    2^7e^3p \frac{\tau}{2^{(k-1)(2\alpha-2)}} &\ \text{if } k > n_*
    \end{cases}
\end{align}
And for cross terms, the different scales do not affect each other and hence it becomes a simple product. Formally this becomes a n-dimensional lattice of nearest neighbor spin chain (Fig.~\ref{fig:multi-scale}), and applying Theorem~\ref{thm:sum_over_paths} on it yields the product form.
\begin{align}
\vertiii{\chi_{k_1}\cdots \chi_{k_n}O^T}_{D_P,p}^2\le [\prod_k \binom{2\lceil \frac{r}{2^k}\rceil+1}{N_k}^2] [\prod_k e^{\frac{\eta_k^2R_kT}{p}}][\prod_k \frac{(8pR_k\eta_k^2T)^{N_k}}{N_k!}]\vertiii{O^0}_{D_P,p}^2 \le C_{k_1}\cdots C_{k_n} 
\end{align}
The rest will be a matter of accounting. \begin{align}
\vertiii{O^T}_{D_P,p}& = \vertiii{[1-\prod_k(1-\chi_k)]O^T}_{D_P,p}\\
&\le \prod_k(1+\sqrt{C_{k}})-1\\
&\le \exp(\sum_k \sqrt{C_{k}}) - 1\\
&\le 2 \sum_k \sqrt{C_{k}}
\end{align}
where we used the elementary $e^x-1 \le 2 x$ for $e^x \le 2$. 
where 
\begin{align}
    \sum_k \sqrt{C_k} = \sum_{k=1}^{k_*-1} e^{\frac{\eta_k^2R_kT}{2p}}(\sqrt{e^32^{9+2\alpha}M^3}\sqrt{p\frac{\tau}{r}})^{N_k}+ \sum_{k=k_*}^{n_*} e^{\frac{\eta_k^2R_kT}{2p}}4\sqrt{2e}e \sqrt{p\frac{r^2\tau}{2^{(k-1)2\alpha}}}+ 
    \sum_{k=n_*}^{\infty} e^{\frac{\eta_k^2R_kT}{2p}} 16e\sqrt{e}\sqrt{p \frac{\tau}{2^{(k-1)(2\alpha-2)}}}
\end{align}

\textbf{Case $\alpha > 3/2$:}  
Calculate values $\eta_k^2R_k\le \xi^2 2^{-(2\alpha-2)(k-1)}$, and $M < \sum_{k=1}^\infty 2^{-k(2\alpha-3)/3} = \frac{1}{1-2^{-(2\alpha-3)/3}}:=M'.$ and plug in each summand:
\begin{align}
 \sum_{k=1}^{k_*-1} e^{\frac{\eta_k^2R_kT}{2p}}(\sqrt{e^32^{8+2\alpha}M^3} \sqrt{p\frac{\tau}{r}})^{N_k} &\le \sum_{k=1}^{k_*-1} (e^{\frac{\tau}{r}\frac{2^{2\alpha-2}M'}{p2^{k(4\alpha/3-1)}}}\sqrt{e^32^{8+2\alpha}M^3} \sqrt{p\frac{\tau}{r}})^{N_k} \\
&\le \sum_{k=1}^{k_*-1} \left(\exp(\frac{1}{ e^32^{10}e^3M'^2})\sqrt{e^32^{8+2\alpha}M'^3} \sqrt{p\frac{\tau}{r}}\right)^{N_k} \\
&:= \sum_{k=1}^{k_*-1} \left(b_1 \sqrt{p\frac{\tau}{r}}\right)^{N_k} \le \frac{b_1 \sqrt{p\frac{\tau}{r}}}{1-b_1 \sqrt{p\frac{\tau}{r}} }
\end{align} 
where in the bound by geometric series we used that 
\begin{equation}
N_1 > N_2 > \cdots > N_{k_*-1}.  \label{eq:Nqinequality_br_OTOC}
\end{equation}
To derive this, note that the argument of the ceiling function in (\ref{eq:Nq_br_OTOC}) changes by a factor of $2^{\alpha^\prime/2}$ each time $q$ changes by 1.  When $\alpha^\prime>2$, this factor is larger than 2, so once the argument is larger than 1, it changes by at least 1: $N_q \le N_{q-1}-1$. In the second inequality we used the $(e^32^{8+2\alpha}M'^3\tau/r \le 1 $ as for later $\tau$ the bound already become vacuous, $M\le M'$, and a generous bound on the exponent $2^{k(4\alpha/3-1)}p^2 > 1 $. 
For the second term for those $k \ge k_*$, they decay at first by a factor of $2^\alpha$ and then $2^{\alpha-1}$ above $n_*$. 
\begin{align}
\sum_{k=k_*}^{n_*} e^{\frac{\eta_k^2R_kT}{2p}}4\sqrt{e}e \sqrt{p\frac{r^2\tau}{2^{(k-1)(2\alpha)}}}+ \sum_{k=n_*}^{\infty} e^{\frac{\eta_k^2R_kT}{2p}} 8e\sqrt{2e}\sqrt{p \frac{\tau}{2^{(k-1)(2\alpha-2)}}} &\le e^{\frac{\eta_{k_*}^2R_{k_*}T}{2p}}4\sqrt{e}e \sqrt{p\frac{r^2\tau}{2^{({k_*}-1)2\alpha}}}\frac{1}{1-2^{\alpha - 1}}\\
&\le \exp(\frac{1}{p^22^7e^3M'^2})4\sqrt{e}e \sqrt{p2^{2\alpha+2}\frac{M'^3\tau}{r}}\frac{1}{1-2^{\alpha - 1}}\\
&:= b_2 \sqrt{p\frac{\tau}{r}}
\end{align} We arrive at concentration by setting $p =\max( \frac{r\epsilon^2}{\tau ec},2)$
\begin{align}
     \sup_P\BP[\lV O^T_{\{r\}}P \rV \ge \epsilon ] &\le  \frac{\BE[\lV O^T_{\{r\}} \rV_2^p ]}{\epsilon^p} \le \frac{ \vertiii{O^T_{\{r\}}}_{D_P,p}^p}{\epsilon^p}\\
     &\le D_P(\frac{2}{\epsilon}\frac{b_1 \sqrt{p\frac{\tau}{r}}}{1-b_1 \sqrt{p\frac{\tau}{r}} } +  \frac{2}{\epsilon}b_2 \sqrt{p\frac{\tau}{r}} ) ^{p}\\
     &\le D_P (\frac{2}{\epsilon}\frac{b_1 \sqrt{p\frac{\tau}{r}}}{1-\frac{b_1}{2(b_1+b_2)} } +  \frac{2}{\epsilon}b_2 \sqrt{p\frac{\tau}{r}} ) ^{p} := (\sqrt{cp\frac{\tau}{r\epsilon^2}} ) ^{p} \\
     &\le \begin{cases}
     D_P\exp(-\frac{r\epsilon^2}{2ec\tau}) &\text{if } \epsilon^2\ge  2ec\frac{\tau}{r}\\
     \\
     D_P\frac{2c\tau}{r\epsilon^2} &\text{if } \epsilon^2\le 2ec\frac{\tau}{r}
     \end{cases}
 \end{align}
where in the third line we used the otherwise vacuous constraint $\epsilon \le 1$ and defined $c^2=\frac{2b_1}{1-b_1/2(b_1+b_2)}+2b_2$. Expressing in terms of $\delta$:
 \begin{align}
     \BP\left[\lV O^{\tau}_{\{r\}} P\rV > \epsilon(\delta) \right]\le \delta
 \end{align}
\begin{align}
    \epsilon(\delta):= 
    \begin{cases}
    \sqrt{2ec\ln(D_P/\delta)\frac{\tau}{r}} &\text{if } \delta <\frac{D_P}{e} \\
    \\
    \sqrt{D_P\frac{2c\tau}{r\delta}} &\text{if } \delta \ge \frac{D_P}{e}
    \end{cases} 
\end{align}
The logarithmic dependence on $\delta$ captures the very strong tail bound that might be useful elsewhere. 
Our next goal is extracting the asymptotic velocity, which only requires the second moment bound and Chebyshev. Take the order of limit: pick $\epsilon >0$, then for any function approaching infinity abrbitrarily slowly, $\lim_{r\rightarrow\infty} f(r)=\infty$, define $T(\ell_0)$ such that
\begin{align}
     \tau = rf(r),
\end{align}
we obtain a linear light cone almost surely
\begin{align}
    \lim_{r\rightarrow \infty} \BP(\lV\frac{1}{2}[A_{r},O_{0}(\tau(r))]P\rV \ge \epsilon )= 0. 
\end{align}
Note we can also get a velocity for a fixed $\epsilon >0, D_P/e >\delta >0$:
\begin{align}
    u = 2ec\frac{\ln(D/\delta)}{\epsilon^2}.
\end{align}
\textbf{Case $1<\alpha<3/2$.}
Moreover, we now find \begin{equation}
M = \sum_{q=1}^{n_*} 2^{q(3-2\alpha)/3} < r^{(3-2\alpha)/3} \sum_{k^\prime=0}^\infty 2^{-k^\prime((3-2\alpha)/3} = \frac{r^{(3-2\alpha)/3}}{1-2^{-(3-2\alpha)/3}}=: M'
\end{equation}
and that 
\begin{equation}
\frac{1}{2^{k_*(2\alpha)/3}} \le \frac{2M}{r} \le  \frac{2M'}{r} \le \frac{2r^{-2\alpha/3}}{1-2^{-(3-2\alpha)/3}} 
\end{equation}

\begin{align}
 \sum_{k=1}^{k_*-1} e^{\frac{\eta_k^2R_kT}{2p}}(\sqrt{e^32^{8+2\alpha}M^3p\frac{\tau}{r}})^{N_k} & \le \sum_{k=1}^{k_*-1} (e^{\frac{\tau}{r}\frac{2^{2\alpha-2}M'}{p2^{k(4\alpha/3-1)}}}\sqrt{e^32^{8+2\alpha}M^3p\frac{\tau}{r}})^{N_k} \\
 & \le \sum_{k=1}^{k_*-1} (e^{\frac{\tau}{r}\frac{2^{2\alpha-2}M'}{p2^{k(4\alpha/3-1)}}}\sqrt{\frac{e^32^{8+2\alpha}}{(1-2^{-(3-2\alpha)/3})^3}p\frac{\tau}{r^{2\alpha-2}}})^{N_k} \\
 & \le \sum_{k=1}^{k_*-1} (\exp({\frac{\tau}{r^{2\alpha/3}}\frac{2^{2\alpha-2}}{p(1-2^{-(3-2\alpha)/3})2^{k(4\alpha/3-1)}}})\sqrt{\frac{e^32^{8+2\alpha}}{(1-2^{-(3-2\alpha)/3})^3}p\frac{\tau}{r^{2\alpha-2}}})^{N_k} \\
  & \le \sum_{k=1}^{k_*-1} \exp({\frac{(1-2^{-(3-2\alpha)/3})^2}{e^32^{10}2^{k(4\alpha/3-1)}}}\sqrt{\frac{e^32^{8+2\alpha}}{(1-2^{-(3-2\alpha)/3})^3}p\frac{\tau}{r^{2\alpha-2}}})^{N_k} \\
  &:= \sum_{k=1}^{k_*-1} \left(b_1 \sqrt{p\frac{\tau}{r^{2\alpha -2}}}\right)^{N_k}\le 2 \sum_{n=1}^{\infty} \left(b_1 \sqrt{p\frac{\tau}{r^{2\alpha -2}}}\right)^{n} \le 2\frac{b_1 \sqrt{p\frac{\tau}{r^{2\alpha -2}}}}{1-b_1 \sqrt{p\frac{\tau}{r^{2\alpha -2}}} }
\end{align} 
where in defining $b_1$ we loosely set $p =1$ in the exponent. In the last line, we get a factor 2 due to $N_k$ growing by $2^{\alpha^\prime/2} \ge \sqrt{2}$ each time $k$ varies by 1: $N_1 > N_3 > N_5\cdots > N_{k_*-1}$. We used the $\frac{e^32^{8+2\alpha}}{(1-2^{-(3-2\alpha)/3})^3}\frac{\tau}{r^{2\alpha/3}}\le \frac{e^32^{8+2\alpha}}{(1-2^{-(3-2\alpha)/3})^3}\frac{\tau}{r^{2\alpha-2}} \le1 $ as for large $\tau$ the bound already become vacuous and $M\le M'$
\begin{align}
\sum_{k=k_*}^{n_*} e^{\frac{\eta_k^2R_kT}{2p}}4\sqrt{e}e \sqrt{\frac{r^2\tau}{2^{(k-1)(2\alpha)}}}&+ \sum_{k=n_*}^{\infty} e^{\frac{\eta_k^2R_kT}{2p}} 4\sqrt{2}e\sqrt{p \frac{\tau}{2^{(k-1)(2\alpha-2)}}} \le e^{\frac{\eta_{k_*}^2R_{k_*}T}{2p}} \frac{4\sqrt{e}e}{1-2^{\alpha - 1}} \sqrt{p\frac{r\tau}{2^{({k_*}-1)(2\alpha)}}}\\
&\le \exp(\frac{\tau 2^{2\alpha -2}8^{1-1/\alpha}}{r^{2\alpha -2} 2p(1-2^{-(3-2\alpha)/2})^{3(1-1/\alpha)}})\frac{4\sqrt{2e}e}{1-2^{\alpha - 1}} \sqrt{4p\frac{\tau}{r^{2\alpha-2}(1-2^{-(3-2\alpha)/3})^3}}\\
&\le \exp(\frac{(1-2^{-(3-2\alpha)/3})^{3/\alpha}}{2^7e^38^{1/\alpha}})\frac{4\sqrt{2e}e}{1-2^{\alpha - 1}} \sqrt{4p\frac{\tau}{r^{2\alpha-2}(1-2^{-(3-2\alpha)/3})^3}}\\
&:= b_2 \sqrt{q\frac{\tau}{r^{2\alpha -2}}}
\end{align}
where in defining $b_2$ we loosely set $p =1$ in the exponent. we used that $\frac{e^32^{8+2\alpha}}{(1-2^{-(3-2\alpha)/3})^3}\frac{\tau}{r^{2\alpha/3}}\le \frac{e^32^{8+2\alpha}}{(1-2^{-(3-2\alpha)/3})^3}\frac{\tau}{r^{2\alpha-2}} \le1 $ as for large $\tau$ the bound already become vacuous and $M\le M'$.
The rest is a close analog of the $\alpha>3/2$ case except for a few constants. We arrive at concentration by setting $p =\max( \frac{r^{2\alpha-2}\epsilon^2}{\tau ec},2)$
\begin{align}
     \sup_P\BP[\lV O^T_{\{r\}}P \rV \ge \epsilon ] &\le  \frac{\BE[\lV O^T_{\{r\}} \rV_2^p ]}{\epsilon^p} \le \frac{ \vertiii{O^T_{\{r\}}}_{D_P,p}^p}{\epsilon^p}\\
     &\le D_P(\frac{4}{\epsilon}\frac{b_1 \sqrt{p\frac{\tau}{r^{2\alpha-2}}}}{1-b_1 \sqrt{p\frac{\tau}{r^{2\alpha-2}}} } +  \frac{2}{\epsilon}b_2 \sqrt{p\frac{\tau}{r^{2\alpha-2}}} ) ^{p}\\
     &\le D_P (\frac{4}{\epsilon}\frac{b_1 \sqrt{p\frac{\tau}{r^{2\alpha-2}}}}{1-\frac{b_1}{2(2b_1+b_2)} } +  \frac{2}{\epsilon}b_2 \sqrt{p\frac{\tau}{r^{2\alpha-2}}} ) ^{p} := (\sqrt{cp\frac{\tau}{r^{2\alpha-2}\epsilon^2}} ) ^{p} \\
     &\le \begin{cases}
     D_P\exp(-\frac{r^{2\alpha-2}\epsilon^2}{2ec\tau}) &\text{if } \epsilon^2\ge  2ec\frac{\tau}{r^{2\alpha-2}}\\
     \\
     D_P\frac{2c\tau}{r^{2\alpha-2}\epsilon^2} &\text{if } \epsilon^2\le 2ec\frac{\tau}{r^{2\alpha-2}}
     \end{cases}
 \end{align}
where in the third line we used the otherwise vacuous constraint $\epsilon \le 1$ and defined $c^2=\frac{4b_1}{1-b_1/2(2b_1+b_2)}+2b_2$. Expressing in terms of $\delta$:
 \begin{align}
     \BP\left[\lV O^{\tau}_{\{r\}} P\rV > \epsilon(\delta) \right]\le \delta
 \end{align}
\begin{align}
    \epsilon(\delta):= 
    \begin{cases}
    \sqrt{2ec\ln(D_P/\delta)\frac{\tau}{r^{2\alpha-2}}} &\text{if } \delta <\frac{D_P}{e} \\
    \\
    \sqrt{D_P\frac{2c\tau}{r^{2\alpha-2}\delta}} &\text{if } \delta \ge \frac{D_P}{e}
    \end{cases} 
\end{align}
The logarithmic dependence on $\delta$ captures the very strong tail bound that might be useful elsewhere. 
Our next goal is extracting the asymptotic velocity, which only requires the second moment bound and Chebyshev. Take the order of limit: pick $\epsilon >0$, then for any function approaching infinity abrbitrarily slowly, $\lim_{r\rightarrow\infty} f(r)=\infty$, define $T(\ell_0)$ such that
\begin{align}
     \tau = r^{2\alpha-2}f(r),
\end{align}
we obtain a linear light cone almost surely
\begin{align}
    \lim_{r\rightarrow \infty} \BP(\lV\frac{1}{2}[A_{r},O_{0}(\tau(r))]P\rV \ge \epsilon )= 0. 
\end{align}
Note we can also get a velocity for a fixed $\epsilon >0, D_P/e >\delta >0$:
\begin{align}
    \frac{r^{2\alpha-2}}{\tau} = 2ec\frac{\ln(D/\delta)}{\epsilon^2}.
\end{align}

\section{Concentration of Lieb-Robinson bounds}\label{sec:detail_cal_Op}
Unlike OTOC, the concentration of Lieb-Robinson bounds (spectral norm) requires more sophisticated control of the support of the operators. This is due to a fundamental feature of matrix concentration: the maximal eigenvalue has a extra $\ln(d)$ dependence on the dimension. 

The proof relied on combining the detailed decomposition of operator from traditional Lieb-Robinson bounds and the matrix martingale techniques for bounding the higher moments. 
\subsection{$d=1$ nearest neighbor interacting system }\label{1d_nn_LR}
\textbf{Decomposing the unitary by locality.}
The time evolved operator can be decomposed by how far the operator has reached
\begin{align}
O(T)= U(T)O(0)U^\dagger(T): = O_0^T +  \ldots + O_\ell^T + \ldots
\end{align}
where 
\begin{align}
    O_\ell^T &:= \nexists_{>\ell}\exists_{\ell}O(T)\\
    &=e^{-iH_{<\ell}T}Oe^{-iH_{<\ell}T}-e^{-iH_{<\ell-1}T}Oe^{-iH_{<\ell-1}T} \label{eq:exists}
\end{align}
Throughout times, the decomposition satisfies the desired properties: 
\begin{enumerate}
\item Consistency: 
\begin{equation}
U \sum_k\left[ O^{T-1}_k\right] U^\dagger = \sum_k O_k^{T}.
\end{equation}
\item Operator $O^t_k$ is trivial on sites $s>k$
\begin{equation}
O^T_k = Tr_{>k}(O^t_k) \otimes \tau_{>k}. 
\end{equation}
\end{enumerate}
\textbf{Sum over paths and then Markov.} 

\begin{align}
     \vertiii{(\nexists_{>\ell}\exists_{\ell}O(T))_{\le\ell}}_p^2  &\le (1+2\frac{\eta^2}{p})^T\sum_{\Gamma} (8p\eta^2 )^\ell \binom{T+\ell-1}{\ell}\vertiii{O^{0}_{0}}_p^2 \\
     &\le 2 (1+2\frac{\eta^2}{p})^T \binom{T+\ell-1}{\ell} (8p\eta^2g(\eta) )^\ell D^{4\ell/p}  \le (8\mathrm{e} \frac{pT\eta^2g(\eta)}{\ell})^\ell D^{4\ell/p} e^{\eta^22T/p},
\end{align}
 where by $(\nexists_{>\ell}\exists_{\ell}O(T))_{\le\ell}$ we consider only the non-trivial component of the operator $I_{>\ell}\otimes (\nexists_{>\ell}\exists_{\ell}O(T))_{\le\ell}=\nexists_{>\ell}\exists_{\ell}O(T)$, the dimension of which comes in at $D^{2\ell}$.
 By Markov inequality with tunable parameter $p$ and $\lV O^{T}_{2k} \rV \le \lV O^{T}_{2k} \rV_p$,
\begin{align}
\BP[\lV O^{T}_{\ell} \rV \ge \epsilon ] \le \BP[ \lV (O^{T}_{\ell})_{\le\ell} \rV_p\ge \epsilon ]&\le \frac{\BE[\lV (O^{T}_{\ell} )_{\le\ell}\rV_p^p ]}{\epsilon^p}\\
\le (8\mathrm{e} \frac{pT\eta^2g(\eta)}{\ell})^{\ell p/2} D^{2\ell} e^{\eta^2T}\epsilon^{-p}.
\end{align}
Setting the optimal parameter $p = \ell \epsilon^{2/\ell} /8 \mathrm{e}^2T\eta^2 $, depending on each $\ell, T$
\begin{align}\label{eq:concentration bound for decomposed}
\BP[\lV O^{T}_{2k} \rV \ge \epsilon ] \le \exp(-\frac{\ell^2\epsilon^{2/\ell}}{16\mathrm{e}^2 T\eta^2g(\eta)}) D^{2\ell} e^{\eta^2T}.
\end{align}
This equation captures the essential mathematics. \\ 
\textbf{Union bound finishing.} The rest is to do a quick and dirty conversion to the final concentration bound for the sum of decomposed operators
\begin{align}
    O^T_{\ge\ell_0}  := O^{T}_{\ell_0} + O^{T}_{\ell_0+1} \cdots. 
\end{align}
The norm of this operator is tied to the desired commutator norm
\begin{align}
    \frac{1}{2}\lV[O_0(T), X_{\ell_0} ]\rV &= \frac{1}{2}\lV [O^T_{>\ell_0}, X_{\ell_0} ]\rV \\
    &\le \lV O^T_{\ge\ell_0} \rV \lV X_{\ell_0} \rV. 
\end{align}
The strategy is to do a union bound by putting together concentration bounds for each $O^T_{\ell_0 + m}$, tuning an error $\epsilon_{\ell_0 + m}$ corresponding to a failure probability $\delta_{\ell_0 + m}$.
\begin{align}
    \BP[\lV O^{T}_{\ell_0+m} \rV \ge \epsilon_{\ell_0 + m} ] &\le \delta_{\ell_0 + m}\\
    \epsilon_{\ell_0 + m} &= \left[(\ell \ln (2D) + \eta^2T- \ln \delta_{\ell_0 + m} ) \frac{16\mathrm{e}^2 T\eta^2g(\eta)}{\ell^2 }\right]^{\ell/2}
\end{align}
where we simply invert Eq.(\ref{eq:concentration bound for decomposed}) to get $\epsilon$ in terms of $\delta$. In order that the union bound works, set the failure probabilities to decay with rate $\lambda <1$
\begin{align}
\delta_{\ell_0+m} = \delta_0 \lambda^m.     
\end{align}
Then 
\begin{align}
\epsilon_{\ell_0 + m} &= \left[(\ell \ln (2D) + \eta^2T- \ln \delta_{0} - m \ln \lambda  ) \frac{16\mathrm{e}^2 T\eta^2g(\eta)}{\ell^2 }\right]^{\ell/2} \\
&\le     \left[(\ell \ln (2D/\lambda) + \eta^2T- \ln \delta_{0}) \frac{16\mathrm{e}^2 T\eta^2g(\eta)}{\ell^2 }\right]^{\ell/2}\\
&\le \left[\left( \ln(2D/\lambda) +\frac{1}{16e^2}- \frac{\ln \delta_{0}}{\ell}\right) \frac{16\mathrm{e}^2 T\eta^2g(\eta)}{\ell}\right]^{\ell/2},
\end{align}
In the second line we used $\ell = \ell_0 +m \ge m$ to simplify the expression; in the last line we used $\ln(2D/\lambda) \frac{16\mathrm{e}^2 T\eta^2g(\eta)}{\ell}\le 1$ to get $\eta^2T \le \ell /\ln(2D/\lambda)16e^2\le 1/16e^2$. The sum of error will be important later for plugging in the union bound.
\begin{align}
    \sum^{\infty}_{m = 0} \epsilon_{\ell_0 + m}  &= \sum^{\infty}_{m = 0} \left[\left( \ln(2D/\lambda) +\frac{1}{16e^2}- \frac{\ln \delta_{0}}{\ell}\right) \frac{16\mathrm{e}^2 T\eta^2g(\eta)}{\ell}\right]^{\ell/2}\\
    &\le \sum^{\infty}_{m = 0} \left[\left( \ln(2D/\lambda) +\frac{1}{16e^2}- \frac{\ln \delta_{0}}{\ell_0}\right) \frac{16\mathrm{e}^2 T\eta^2g(\eta)}{\ell_0}\right]^{(\ell_0+m)/2}\\
    &\le \frac{\left[\left( \ln(2D/\lambda) +\frac{1}{16e^2}- \frac{\ln \delta_{0}}{\ell_0}\right) \frac{16\mathrm{e}^2 T\eta^2g(\eta)}{\ell_0}\right]^{\ell_0/2}}{ 1- \sqrt{\left[\left( \ln(2D/\lambda) +\frac{1}{16e^2}- \frac{\ln \delta_{0}}{\ell_0}\right) \frac{16\mathrm{e}^2 T\eta^2g(\eta)}{\ell_0}\right]^{} } } =: \epsilon'_{\ell_0}\\
\end{align}
where in the first inequality we used that the expression $ \left[(\ell_0 \ln (2D/\lambda) + \eta^2T- \ln \delta_{0}) \frac{16\mathrm{e}^2 t\eta^2g(\eta)}{\ell_0^2 }\right]$ is decreasing with $m$; in the second we used the geometric sum.
We now use the union bound
\begin{align}
    \BP[\lV O^T_{\ge \ell_0} \rV \ge \epsilon'_{\ell_0}] \le \BP[\lV O^T_{\ge \ell_0} \rV \ge \sum^{\infty}_{m = 0} \epsilon_{\ell_0 + m} ] &\le \BP[\sum_{m=0}^\infty\lV O^T_{\ell_0+m} \rV \ge \sum^{\infty}_{m = 0} \epsilon_{\ell_0 + m} ] \\
    &\le \BP[\bigcup_{m=0}^\infty \lV O^T_{\ell_0+m} \rV \ge \epsilon_{\ell_0 + m} ] \\
    & \le \sum_{m=0}^\infty \BP[\lV O^T_{\ell_0+m} \rV \ge \epsilon_{\ell_0 + m} ] \\
    & \le \sum_{m=0}^\infty \delta_{\ell_0+m} = \frac{\delta_0}{1-\lambda},
\end{align}
showing the concentration bound with failure probabilty $\delta_0$ and arbitrary parameter $\lambda<1$. We can also simplify the expression with some constant overheads
\begin{align}
\BP\left[\frac{1}{2}\lV [A_{r},O_0(T)] \rV  \ge \frac{1}{ 1-1/\sqrt{2}}\left[( \ln (4D) + \frac{1}{16e^2}- \frac{\ln \delta_{0}}{r}) \frac{32\mathrm{e}^2 T\eta^2g(\eta)}{r }\right]^{r/2}\right] \le 2\delta_0
\end{align} 
where we set $\lambda = 1/2,$ weaken $16\mathrm{e}^2 T\eta^2g(\eta) $ in the nominator by a factor of 2 so that the denominator is bounded by $1-1/\sqrt{2}$ when the bound is not vacuous.\\
\textbf{Extracting the asymptotic velocity.}
For arbitrary $\zeta>0$, let $T(\ell_0)$ such that 
\begin{equation}
\left( \ln(2D) +\frac{1}{16e^2}\right) \frac{16\mathrm{e}^2 T(\ell_0)\eta^2g(\eta)}{\ell_0} = 1-\zeta.
\end{equation}
Then almost surely,
\begin{align}
    \limsup_{\ell_0\rightarrow \infty} \BP\left(\lV\frac{1}{2}[A_{\ell_0},O_{0}(T(\ell_0))]\rV \ge \epsilon \right)= 0. 
\end{align}
Taking $\lambda,\zeta$ aribitrarily small implies a Lieb-Robinson velocity bound independent of threshold $\epsilon$.
\begin{align}
   \frac{\ell_0}{ T(\ell_0)\xi^2} \le g(\eta)a^2(16\mathrm{e}^2\ln(2D) +1)\stackrel{\eta \rightarrow 0}{=} a^2(16\mathrm{e}^2\ln(2D) +1).
\end{align}
where $g(\eta)=1+\eta+\eta^2/2$ is a mild finite-size effect which disappears in the continuum.

\subsection{Time-independent $d=1$ power-law interacting system }\label{sec:proof_time_indep_1d_long_LR}
Consider a $d=1$ power-law time-independent Hamiltonian where each term is normalized $\lV H^{}_{ij}\rV\le 1$ and zero mean $\BE[H^{}_{ij}]=0$
\begin{equation}
     H :=\sum_{i} \frac{1}{|i-j|^\alpha} H_{ij}.
\end{equation} 
The distinction from OTOC derivation(Sec.~\ref{sec:proof_time_indep_1d_long_OTOC}) is that we have to control the support of operators. Decompose the operator by how far it has reached, labeled by $2^mr_0$
\begin{align}
{[A_{r_0},O^t]}& = {[A_{r_0},\exists_{\pm r_0}O^t]} = {[A_{r_0},(\exists_{\pm r_0}\nexists_{\pm 2r_0}+ \exists_{\pm 2r_0})O^t]}\\
& = [A_{r_0},{\sum_{m=0}^\infty \exists_{\pm 2^m r_0}\nexists_{\pm 2^{m+1}r_0}O^t}].
\end{align}
where we used $\exists_{\pm 2^mr_0}$ to select the terms in $O^t$ that has a path traversing beyond $\pm 2^mr_0$; on the other hand $\nexists_{\pm 2^{m+1}r_0}O^t$ restricts the interaction to be within $[-2^{m+1}r_0,2^{m+1}r_0]$. This is the same usage as Eq.~\eqref{eq:exists}. For each $r=2^mr_0$,
\begin{align}
\exists_{\pm r}\nexists_{\pm 2r}O^t=\exists_{\pm 2r}\left( U^{\dagger t}_{\pm 2r}OU_{\pm 2r}^t\right)= [1-\prod_k(1-\chi^{r}_k)]O^t.
\end{align}
The following calculation will be analogous to Sec.~\ref{sec:proof_brown_1d_long_LR}. For each operator supported on $[-2r,2r]=[-2^{m+1}r_0,2^{m+1}r_0]$, let $N_k$ associated with each scale $\chi_k$ be
\begin{equation}
N_k :=\begin{cases}
\left\lceil \frac{1}{2} \frac{2^{-k(2\alpha-5)/5}}{ M}  \frac{r}{2^{k}} \right\rceil &\text{if}\ k\le n^*\\
1 &\text{if}\ k > n_*,
\end{cases}\label{eq:Nq_ti_spectral}
\end{equation}
where $M:=\sum_{k^\prime=1}^{n_*} 2^{-k^\prime(2\alpha-5)/5}$ is a normalization factor. Now, 
\begin{align}
    \chi_kO^t = 2 \sum_{forward\ paths:\ m_{N_k}>\cdots>m_1} \chi_kO^t(m_{N_k},\cdots>m_1)
\end{align}
Each of the forward path has exactly the same expression as a $N_{k}+1$ sites nearest neighbour spin chain from and there are multiple interacting terms between $i, i+1$. The factor 2 accounts for the path reach $+r$ or $-r$. For each spin chain, $\Delta(\gamma) = 2^{2(k-1)}$.
\begin{align}
\vertiii{\chi_kO^t(m_{N_k},\cdots>m_1)}_{p}\le (\sqrt{4pR_k\eta_k^2})^{N_k} \frac{t^{N_k}}{N_k!}\vertiii{O^0}_{p}^2.
\end{align}
By Minkowski(triangle inequality), we can sum over choices of $m_{N_k}$ to get
\begin{align}
\vertiii{\chi_kO^t}_{p}
\le 2\binom{2\lceil \frac{r}{2^k}\rceil+1}{N_k} (\sqrt{4pR_k\eta_k^2})^{N_k} \frac{t^{N_k}}{N_k!}\vertiii{O^0}_{p}
\end{align}
where $\eta_k \le \xi \frac{1}{2^{(k-1)\alpha}}$ is the maximum strength, and the sum over block yields $[\sum_{k - \text{block}}\eta_k^2]=\eta_k^2R_k \le \xi^2 2^{-(k-1)2(\alpha-1)}$, featuring a sum of squares; The above function has transition at (a) starting value $k_*$ at which a long $k$-forward path has a single coupling: $N_k=1$ for $k\ge k_*$. (b)$\frac{r}{2^k} \le 1 $. (a) occurs when 
\begin{equation}
\frac{M}{R}\ge \frac{1}{2^{1+k_*2\alpha/5}},
\end{equation}
Hence 
\begin{align}
    \vertiii{\chi_kO^t}_p \le 2D^{2r/p} C_{k,p}:= 2D^{2r/p}  \cdot \begin{cases}
    (e^2 2^{4+\alpha}\sqrt{p} \frac{M^2t}{r2^{k\alpha/5}})^{N_k} &\ \text{if } k< k_*\\
    4e^2\sqrt{p} \frac{rt}{2^{\alpha(k-1)}} &\ \text{if } n_*\ge k\ge k_*\\
    8e^2\sqrt{p} \frac{t}{2^{(k-1)(\alpha-1)}} &\ \text{if } k > n_*
    \end{cases}\\
\end{align}
And for the sub-leading terms they form a product(Fig.~\ref{fig:multi-scale})
\begin{align}
\vertiii{\chi_{k_1}\cdots \chi_{k_n}O^t}_{p}\le 2 \vertiii{O^0}_{p} \prod_k \left[ \binom{2\lceil \frac{r}{2^k}\rceil+1}{N_k} (\sqrt{4pR_k\eta_k^2})^{N_k} \frac{t^{N_k}}{N_k!}\right] \le 2D^{2r/p}C_{k_1,p}\cdots C_{k_n,p} 
\end{align}
For each $k$, Let $p(k):=p_12^{(k-1)2\alpha/5}>p_1$:
\begin{align}
    \vertiii{\chi_k\prod_{k'>k}(1-\chi_{k'})O^t}_{p_1} &\le \vertiii{\chi_k\prod_{k'>k}(1-\chi_{k'})O^t}_{p(k)} \\
    &\le {2D^{2r/p(k)}C_{k,p(k)}}\prod_{k'>k}(1+{C_{k',p(k)}}) \\
    &\le {2D^{2r/p(k)}C_{k,p(k)}}\exp(\sum_{k'>k}{C_{k',p(k)}}) \\
    &\le {2D^{2r/p(k)}C_{k,p(k)}}\exp(\frac{1}{1-2^{-\alpha/5}}) \\
\end{align}
where in the first inequality is a Lyapunov's inequality for $p(k)>p_1$, and in the third inequality we called the following proposition and implicitly assumed the condition $C_{k,p(k)}\le 1$ as otherwise the bound is vacuous. 
\begin{prop}
For $k' <k$, $C_{k,p(k)}\le 1 \implies C_{k',p(k)} \le 2^{-(k-k')\alpha/5}$.
\end{prop}
\begin{proof}
As $\eta^kR_k$ is decreasing in $k$, and there is a factor of $2^{(-k'\alpha/5)N_{k'}}, 2^{-\alpha k'}, 2^{-k'(\alpha-1)}$. as $k'$ increases.
\end{proof}

\textbf{Case $\alpha > 5/2$:}
For $k\le k_*$,
\begin{align}
D^{2r/p(k)}C_{k,p(k)} &= \left(D^{2r/p(k)N_k}e^2 2^{4+\alpha}\sqrt{p} \frac{M^2t}{r2^{k\alpha/5}}\right)^{N_k}\\
&\le \left( D^{\frac{4M'2^{2\alpha/5}}{p_1}}e^2 2^{4+4\alpha/5}M^2 \frac{\sqrt{p_1}t}{r}\right)^{N_k}\\
&\le \left( D^{\frac{a}{p_1}} b\sqrt{p_1} \frac{t}{r}\right)^{N_k}
\end{align}
Hence 
\begin{align}
\vertiii{ \exists_{\pm r}\nexists_{\pm 2r}O^t}_{p_1}& = \vertiii{[1-\prod_k(1-\chi_k)]O^t}_{p_1}\\
&\le \sum_k \vertiii{\chi_k\prod_{k'>k}(1-\chi_{k'})O^t}_{p_1}\\
&\le \sum_k 2\exp(\frac{1}{1-2^{-\alpha/5}}) D^{2r/p(k)}C_{k,p(k)}\\
&\le \exp(\frac{1}{1-2^{-\alpha/5}})\frac{2}{1-2^{-\alpha+1}} \frac{D^{\frac{a}{p_1}} b\sqrt{p_1} \frac{t}{r_0}}{1-D^{\frac{a}{p_1}} b\sqrt{p_1} \frac{t}{r_0}}\\
&:= c_2\frac{D^{\frac{a}{p_1}} b\sqrt{p_1} \frac{t}{r_0}}{1-D^{\frac{a}{p_1}} b\sqrt{p_1} \frac{t}{r_0}}
\end{align}
where $(1-2^{-\alpha+1})$ came from the geometric series for $k'\le k_*$. Summing up tails 
\begin{align}
\vertiii{\sum_{m=0}^\infty \exists_{\pm 2^m r_0}\nexists_{\pm 2^{m+1}r_0}O^t }_{p_1} &\le \sum_{m=0}^\infty\vertiii{ \exists_{\pm 2^m r_0}\nexists_{\pm 2^{m+1}r_0}O^t }_{p_1}\\
&\le \sum_{m=0}^\infty c_2\frac{D^{\frac{a}{p_1}} b\sqrt{p_1} \frac{t}{r_0}}{1-D^{\frac{a}{p_1}} b\sqrt{p_1} \frac{t}{r_0}}\\
&\le \frac{ c_2}{1-1/2}\frac{D^{\frac{a}{p_1}} b\sqrt{p_1} \frac{t}{r_0}}{1-D^{\frac{a}{p_1}} b\sqrt{p_1} \frac{t}{r_0}}:=c'_2\frac{D^{\frac{a}{p_1}} b\sqrt{p_1} \frac{t}{r_0}}{1-D^{\frac{a}{p_1}} b\sqrt{p_1} \frac{t}{r_0}}
\end{align}
where $1-1/\sqrt{2}$ came from the geometric series for $m \ge 0$. Set $p =\max( \frac{r^2\epsilon^2}{t^2ec},2)=\frac{r^2\epsilon^2}{t^2ec}$ to get \footnote{$D^a > e$, hence when the bound is not vacuous, we have $\frac{r^2\epsilon^2}{t^2ec}>2$}
\begin{align}
     \BP[\lV \sum_{m=0}^\infty \exists_{\pm 2^m r_0}\nexists_{\pm 2^{m+1}r_0}O^t \rV \ge \epsilon ] \le \BP[ \sum_{m=0}^\infty \lV (\exists_{\pm 2^m r_0}\nexists_{\pm 2^{m+1}r_0}O^t)_{\le 2^{m+1}r_0} \rV_p\ge \epsilon ]
     & \le ( \frac{c'_2}{\epsilon}  \frac{D^{\frac{a}{p_1}} b\sqrt{p_1} \frac{t}{r_0}}{1-D^{\frac{a}{p_1}} b\sqrt{p_1} \frac{t}{r_0}})^{p_1}\\
     &\le ( \frac{c'_2}{\epsilon}\frac{D^{\frac{a}{p_1}} b\sqrt{p_1} \frac{t}{r_0}}{1-\frac{1}{c'_2}})^{p_1}:= D^{a} ( \sqrt{cp_1 \frac{t^2}{r_0^2\epsilon^2}})^{p_1}\\
     &\le D^{a}\exp(-\frac{r_0^2\epsilon^2}{2ect^2}) &\text{if } \epsilon^2\ge  2ec\frac{t^2}{r_0^2},
\end{align}
where in the fourth inequality we used $c'D^{\frac{a}{p_1}} b\sqrt{p_1} \frac{t}{r_0} \le 1 $ otherwise the bound becomes vacuous. Expressing in terms of $\delta$:
 \begin{align}
     \BP\left[\lV \sum_{m=0}^\infty \exists_{\pm 2^m r_0}\nexists_{\pm 2^{m+1}r_0}O^t\rV > \epsilon(\delta) \right]\le \delta,
 \end{align}
\begin{align}
    \epsilon(\delta):= \sqrt{2ec\ln(D^{a}/\delta)\frac{t^2}{r_0^2}}. 
\end{align}
The logarithmic dependence on $\delta$ captures the very strong tail bound that might be useful elsewhere. 
Our next goal is extracting the asymptotic velocity. Take the order of limit: pick $\epsilon >0$, then for any function approaching infinity arbitrarily slowly, $\lim_{r\rightarrow\infty} f(r)=\infty$, define $t(r_0)$ such that
\begin{align}
     t = r_0f(r_0),
\end{align}
we obtain a linear light cone almost surely
\begin{align}
    \lim_{r_0\rightarrow \infty} \BP(\lV\frac{1}{2}[A_{r_0},O_{0}(t(r_0))]P\rV \ge \epsilon )= 0. 
\end{align}
Note we can also get a velocity for a fixed $\epsilon >0, \delta >0$:
\begin{align}
    \frac{r_0}{t(r_0)} = \frac{\sqrt{2ec\ln(D^{a}/\delta)}}{\epsilon}.
\end{align}

\textbf{Case $3/2<\alpha < 5/2$:}
Moreover, we now find 
\begin{equation}
M = \sum_{q=1}^{n_*} 2^{q(5-2\alpha)/5} < r^{1-2\alpha/5} \sum_{k^\prime=0}^\infty 2^{-k^\prime(5-2\alpha)/5} = \frac{r^{1-2\alpha/5}}{1-2^{-(5-2\alpha)/5}}=: M'
\end{equation}
and that 
\begin{equation}
\frac{1}{2^{k_*2\alpha/5}} \le \frac{2M}{r} \le  \frac{2M'}{r} \le \frac{2r^{-2\alpha/5}}{1-2^{-(5-2\alpha)/5}}. 
\end{equation}

For $k\le k_*$,
\begin{align}
D^{2r/p(k)}C_{k,p(k)} &= \left(D^{2r/p(k)N_k}e^2 2^{4+\alpha}\sqrt{p} \frac{M^2t}{r2^{k\alpha/5}}\right)^{N_k}\\
&\le \left( D^{\frac{4M'2^{2\alpha/5}}{p_1}}e^2 \frac{2^{4+4\alpha/5}}{(1-2^{-(5-2\alpha)/5})^2} \frac{\sqrt{p_1}t}{r^{4\alpha/5-1}}\right)^{N_k}\\
&\le \left( D^{\frac{a}{p_1}} b\sqrt{p_1} \frac{t}{r^{4\alpha/5-1}}\right)^{N_k}.
\end{align}
Hence 
\begin{align}
\vertiii{ \exists_{\pm r}\nexists_{\pm 2r}O^t}_{p_1}& = \vertiii{[1-\prod_k(1-\chi_k)]O^t}_{p_1}\\
&\le \sum_k \vertiii{\chi_k\prod_{k'>k}(1-\chi_{k'})O^t}_{p_1}\\
&\le \sum_k 2\exp(\frac{1}{1-2^{-\alpha/5}}) D^{2r/p(k)}C_{k,p(k)}\\
&\le \exp(\frac{1}{1-2^{-\alpha/5}})\frac{2}{1-2^{-\alpha+1}} \frac{D^{\frac{a}{p_1}} b\sqrt{p_1} \frac{t}{r^{4\alpha/5-1}}}{1-D^{\frac{a}{p_1}} b\sqrt{p_1} \frac{t}{r^{4\alpha/5-1}}}\\
&:= c_2 \frac{D^{\frac{a}{p_1}} b\sqrt{p_1} \frac{t}{r^{4\alpha/5-1}}}{1-D^{\frac{a}{p_1}} b\sqrt{p_1} \frac{t}{r^{4\alpha/5-1}}},
\end{align}
where $(1-2^{-\alpha+1})$ came from the geometric series for $k'\le k_*$.
\begin{align}
     \BP[\lV \exists_{\pm r}\nexists_{\pm 2r}O^t \rV \ge \epsilon ] \le \BP[ \lV (\exists_{\pm r}\nexists_{\pm 2r}O^t)_{\le 2r} \rV_p\ge \epsilon ]&\le \frac{\BE[\lV (\exists_{\pm r}\nexists_{\pm 2r}O^t )_{\le\ell}\rV_p^p ]}{\epsilon^p}\\
     & \le ( \frac{c_2}{\epsilon}  \frac{D^{\frac{a}{p_1}} b\sqrt{p_1} \frac{t}{r^{4\alpha/5-1}}}{1-D^{\frac{a}{p_1}} b\sqrt{p_1} \frac{t}{r^{4\alpha/5-1}}})^{p_1}\\
     & \le ( \frac{c_2}{\epsilon}  \frac{D^{\frac{a}{p_1}} b\sqrt{p_1} \frac{t}{r^{4\alpha/5-1}}}{1-1/c_2})^{p_1}:=( D^{\frac{a}{p_1}} \sqrt{p_1\frac{c't^2}{\epsilon^2r^{8\alpha/5-2}}} )^{p_1}\\   
     &\le D^{a'r^{1-2\alpha/5}} \exp(-c{\frac{\epsilon^2r^{8\alpha/5-2}}{t^2}})\label{eq:long_range_deviation_decomposed_ti_spectral}
\end{align}
where we set\footnote{Here we do not have to worry about the constraint $p \ge 2$, because when that happen, $p=2$, $D^{a'r^{1-\alpha/5}} \exp(-1)\ge 1$, i.e. the bound is vacuous anyway.} $\displaystyle p_1 =\max( \frac{1}{e}\frac{\epsilon^2r^{8\alpha/5-2}}{c't^2},2)=\frac{1}{e}\frac{\epsilon^2r^{8\alpha/5-2}}{c't^2}$, and in the third inequality we used $c_2D^{\frac{a}{p_1}} b\sqrt{p_1} \frac{t}{r^{4\alpha/5-1}} \le 1 $ otherwise the bound becomes vacuous. In the last inequality we set $a':=\frac{2^{2+2\alpha/5}}{1-2^{-(5-2\alpha)/5}}$ to emphasize the $r$ dependence and $c=1/2ec'$.

The strategy is to do a union bound by putting together concentration bounds for each $\lV \exists_{\pm r}\nexists_{\pm 2r}O^t \rV := f(2^mr_0)$, tuning an error $\epsilon_{2^mr_0}$ corresponding to a failure probability $\delta_{2^mr_0}$.
\begin{align}
    \epsilon_{2^mr_0} &= \sqrt{
    \frac{t^2}{cr^{8\alpha/5-2}}(\ln(D)a'r^{1-2\alpha/5} - \ln(\delta_{2^mr_0}) )
    }
    \\ \textrm{ensures\ } &\BP[f(2^mr_0) \ge \epsilon_{2^mr_0} ] \le \delta_{2^mr_0}
\end{align}
where we simply invert Eq.(\ref{eq:long_range_deviation_decomposed_ti_spectral}) to get $\epsilon$ in terms of $\delta$. In order that the union bound works, set the failure probabilities to decay with
\begin{align}
\delta_{2^mr_0} = \delta_{r_0}\exp(-2^{m(1-2\alpha/5)}).     
\end{align}
Then 
\begin{align}
\epsilon_{2^mr_0} &= \sqrt{
    \frac{t^2}{cr^{8\alpha/5-2}}(\ln(D)a'r^{1-2\alpha/5} - \ln(\delta_{r_0})+2^{m(1-2\alpha/5)} )
    }
\end{align}
The sum of error will be important later for plugging in the union bound.
\begin{align}
    \sum^{\infty}_{m = 0} \epsilon_{2^mr_0}  &= \sum^{\infty}_{m = 0} \sqrt{
    \frac{t^2}{cr^{8\alpha/5-2}}\left(\ln(D)a'r^{1-2\alpha/5} - \ln(\delta_{r_0})+2^{m(1-2\alpha/5)}\right)}\\   
    &\le\sqrt{
    \frac{t^2}{c_0r_0^{8\alpha/5-2}}\left(\ln(D)a'r_0^{1-2\alpha/5} - \ln(\delta_{r_0})+1 \right)} 
     =: \epsilon'_{r_0}
\end{align}
where we used the geometric since the summand is decaying by $\sqrt{2^{m(3-2\alpha)}}<1$ and $c_0=c(1-2^{3/2-\alpha})^2$.
We now use the union bound
\begin{align}
    \BP[\lV O^t_{\ge r_0} \rV \ge \epsilon'_{r_0}] \le \BP[\lV O^t_{\ge r_0} \rV \ge \sum^{\infty}_{m = 0} \epsilon_{r_0 + m}  ]&\le \BP[\sum_{m=0}^\infty \lV \exists_{\pm r}\nexists_{\pm 2r}O^t \rV \ge \sum^{\infty}_{m = 0} \epsilon_{2^mr_0} ] \\
    &\le \BP[\bigcup_{m=0}^\infty \lV O^t_{2^mr_0} \rV \ge \epsilon_{2^mr_0} ] \\
    & \le \sum_{m=0}^\infty \BP[\lV O^t_{2^mr_0} \rV \ge \epsilon_{2^mr_0} ] \\
    & \le \sum_{m=0}^\infty \delta_{2^mr_0} = \sum_{m=0}^\infty \delta_{r_0}\exp(-2^{m(1-2\alpha/5)}).\\
    &\le \frac{\delta_{r_0}}{e(1-1/e)(1-2\alpha/5)} = \delta_{r_0}c_3
\end{align}
showing the concentration bound that
\begin{align}
    \BP(\lV O^t_{\ge r_0} \rV \ge \epsilon )\le  c_3eD^{a'r_0^{1-2\alpha/5}} \exp(-{c_0\frac{\epsilon^2r_0^{8\alpha/5 -2}}{t^2}})
\end{align}
and with failure probability $\delta = \delta_{r_0}c_3$,
\begin{align}
    \lV O^t_{\ge r_0} \rV\le \epsilon(\delta) = \sqrt{
    \frac{t^2}{c_0r_0^{8\alpha/5-2}}\left(\ln(D)a'r_0^{1-2\alpha/5} - \ln(\delta/c_3e) \right)} 
\end{align}
In particular we can obtain asymptotic velocity. For any $\epsilon>0$, $\zeta>0$, let $r_0(t)$ 
\begin{align}
    \frac{r_0(t)^{2\alpha-3}}{t^2} = \frac{\ln(D)a'}{c_0\epsilon^2} +\zeta
\end{align}
Then the commutator norm vanishes outside the algebraic light cone $t(r_0)$ almost surely
\begin{align}
    \lim_{r_0\rightarrow \infty} \BP(\lV\frac{1}{2}[A_{r_0},O_{0}(t(r_0))]\rV \ge \epsilon )= \lim_{r_0\rightarrow \infty}c_3e\exp(-\zeta c_0\epsilon^2r_0^{1-2\alpha/5})=0. 
\end{align}

\subsection{Brownian $d=1$ power law interacting system }\label{sec:proof_brown_1d_long_LR}
Consider a $d=1$ power law Brownian Hamiltonian where each term is normalized $\lV H^{(T)}_{ij}\rV\le 1$ and zero mean independent conditioned on the past $\BE_{T-1}[H^{(T)}_{ij}]=0$.
\begin{equation}
     H^{(T)} :=\sum_{i} \frac{1}{|i-j|^\alpha} H^{(T)}_{ij}
\end{equation} 
Consider the Brownian limit that $T\xi^2=\tau$ is held fixed.
The distinction from OTOC derivation (Sec.~\ref{sec:proof_brown_1d_long_OTOC}) is that we have to control the support of operators. Decompose the operator by how far it has reached, labeled by $2^mr_0$
\begin{align}
{[A_{r_0},O^{T} ]}& = {[A_{r_0},\exists_{\pm r_0}O^{T} ]} = {[A_{r_0},(\exists_{\pm r_0}\nexists_{\pm 2r_0}+ \exists_{\pm 2r_0})O^{T} ]}\\
& = [A_{r_0},{\sum_{m=0}^\infty \exists_{\pm 2^m r_0}\nexists_{\pm 2^{m+1}r_0}O^{T} }].
\end{align}
where we used $\exists_{\pm 2^mr_0}$ to select the terms in $O^{T} $ that has a path traversing beyond $\pm 2^mr_0$; on the other hand $\nexists_{\pm 2^{m+1}r_0}O^{T} $ restricts the interaction to be within $[-2^{m+1}r_0,2^{m+1}r_0]$. This is the same usage as Eq.~\eqref{eq:exists}. For each $r=2^mr_0$,
\begin{align}
\exists_{\pm r}\nexists_{\pm 2r}O^{T} =\exists_{\pm 2r}\left( U^{\dagger T}_{\pm 2r}OU_{\pm 2r}^T\right)= [1-\prod_k(1-\chi^{r}_k)]O^{T} .
\end{align}
Once $r$ is fixed, the following calculation will be analogous to Sec.~\ref{sec:proof_time_indep_1d_long_LR}. For each operator supported on $[-2r,2r]=[-2^{m+1}r_0,2^{m+1}r_0]$, let $N_k$ associated with each scale $\chi_k$ be
\begin{equation}
N_k :=\begin{cases}
\left\lceil \frac{1}{2} \frac{2^{-k(2\alpha-4)/4}}{ M}  \frac{r}{2^{k}} \right\rceil &\text{if}\ k\le n^*\\
1 &\text{if}\ k > n_*,
\end{cases}\label{eq:Nq_br_spectral}
\end{equation}
where $M:=\sum_{k^\prime=1}^{n_*} 2^{-k^\prime(2\alpha-4)/4}$ is a normalization factor.
\begin{align}
    \chi_kO^{T}  = 2 \sum_{forward\ paths:\ m_{N_k}>\cdots>m_1} \chi_kO^{T} (m_{N_k},\cdots>m_1).
\end{align}
Each of the forward path has exactly the same expression as a $N_{k}+1$ sites nearest neighbour spin chain from and there are multiple interacting terms between $i, i+1$. The factor 2 accounts for the path reach $+r$ or $-r$. For each spin chain, $\Delta(\gamma) = 2^{2(k-1)}$.
\begin{align}
\vertiii{\chi_kO^{T} (m_{N_k},\cdots>m_1)}_{D_P,p}^2\le e^{\frac{\eta_k^2R_kT}{p}}\frac{(8pR_k\eta_k^2T)^{N_k}}{N_k!} \vertiii{O^0}_{p}^2
\end{align}
By Minkowski,
\begin{align}
\vertiii{\chi_kO^{T}}_{D_P,p}^2
\le 4\binom{2\lceil \frac{r}{2^k}\rceil+1}{N_k}^2 e^{\frac{\eta_k^2R_kT}{p}}\frac{(8pR_k\eta_k^2T)^{N_k}}{N_k!}\vertiii{O^0}_{p}^2
\end{align}
where $\eta_k \le \xi \frac{1}{2^{(k-1)\alpha}}$ is the maximum strength, and the sum over block yields $[\sum_{k - \text{block}}\eta_k^2]=\eta_k^2R_k \le \xi^2 2^{-(k-1)2(\alpha-1)}$, featuring a sum of squares; The above function has transition at (a) starting value $k_*$ at which a long $k$-forward path has a single coupling: $N_k=1$ for $k\ge k_*$. (b)$\frac{r}{2^k} \le 1 $. (a) occurs when 
\begin{equation}
\frac{M}{R}\ge \frac{1}{2^{1+k_*\alpha/2}},
\end{equation}
Hence 
\begin{align}
    \vertiii{\chi_kO^{T} }^2_p \le 4D^{4r/p} C_{k,p}:= 4D^{4r/p} e^{\frac{\eta_k^2R_kT}{p}} \cdot \begin{cases}
    (e^3 2^{8+2\alpha}M^3p \frac{\tau}{r2^{k\alpha/2}})^{N_k} &\ \text{if } k< k_*\\
    16e^3p \frac{r^2\tau}{2^{2\alpha(k-1)}} &\ \text{if } n_*\ge k\ge k_*\\
    2^7e^3p \frac{\tau}{2^{(k-1)(2\alpha-2)}} &\ \text{if } k > n_*
    \end{cases}
\end{align}

And for the sub-leading terms 
\begin{align}
\vertiii{\chi_{k_1}\cdots \chi_{k_n}O^{T} }_{p}^2\le 4 \vertiii{O^0}_{p}^2 \prod_k \left[ \binom{2\lceil \frac{r}{2^k}\rceil+1}{N_k}^2 e^{\frac{\eta_k^2R_kT}{p}}\frac{(8pR_k\eta_k^2T)^{N_k}}{N_k!}\right] \le 4D^{4r/p}C_{k_1,p}\cdots C_{k_n,p} 
\end{align}
For each $k$, Let $p(k):=p_12^{(k-1)\alpha/2}>p_1$:
\begin{align}
    \vertiii{\chi_k\prod_{k'>k}(1-\chi_{k'})O^{T} }_{p_1} &\le \vertiii{\chi_k\prod_{k'>k}(1-\chi_{k'})O^{T} }_{p(k)} \\
    &\le \sqrt{4D^{4r/p(k)}C_{k,p(k)}}\prod_{k'>k}(1+\sqrt{C_{k',p(k)}}) \\
    &\le \sqrt{4D^{4r/p(k)}C_{k,p(k)}}\exp(\sum_{k'>k}\sqrt{C_{k',p(k)}}) \\
    &\le \sqrt{4D^{4r/p(k)}C_{k,p(k)}}\exp(\frac{1}{1-2^{-\alpha/4}}) 
\end{align}
where in the first inequality is a standard moment inequality for $p(k)>p_1$, and in the third inequality we called the following proposition and implicitly assumed the condition $C_{k,p(k)}\le 1$ as otherwise the bound is vacuous. 
\begin{prop}
For $k' <k$, $C_{k,p(k)}\le 1 \implies C_{k',p(k)} \le 2^{-(k-k')\alpha/2}$.
\end{prop}
\begin{proof}
As $\eta^kR_k$ is decreasing in $k$, and there is a factor of $2^{(-k'\alpha/2)N_{k'}}, 2^{-2\alpha k'}, 2^{-k'(2\alpha-2)}$. as $k'$ increases.
\end{proof}

\textbf{Case $\alpha > 2$:}
For $k\le k_*$,
\begin{align}
D^{4r/p(k)}C_{k,p(k)} &= \left( D^{4r/p(k)N_k} e^{\frac{\eta_k^2R_kT}{pN_k}}e^3 2^{8+2\alpha}M'^3p \frac{\tau}{r2^{k\alpha/2}}\right)^{N_k}\\
&\le \left( D^{\frac{8M'2^{\alpha/2}}{p_1}} c_1 e^3 2^{8+3\alpha/2}M'^3p_1 \frac{\tau}{r}\right)^{N_k}\\
&:= \left( D^{\frac{a}{p_1}} bp_1 \frac{\tau}{r}\right)^{N_k}
\end{align}
where 
\begin{prop}
$e^{\frac{\eta_k^2R_kT}{pN_k}} \le c_1$
\end{prop}
\begin{proof}
\begin{align}
\frac{\eta_k^2R_kT}{pN_k} &\le \frac{\tau2^{2\alpha}M'}{p2^{k(3\alpha/2-2)}r} \\
&\le \frac{1}{p_1^2e^3 2^{8-\alpha/2}M'^3 2^{k(3\alpha/2-2)}}\\
&\le \frac{1}{p_1^2e^3 2^{8-\alpha/2}M'^3}:=c_1
\end{align}
where we used that $e^3 2^{8+3\alpha/2}M'^3p_1 \frac{\tau}{r}\le 1$ otherwise the bound is vacuous.
\end{proof}
Hence 
\begin{align}
\vertiii{ \exists_{\pm r}\nexists_{\pm 2r}O^{T} }_{p_1}& = \vertiii{[1-\prod_k(1-\chi_k)]O^{T} }_{p_1}\\
&\le \sum_k \vertiii{\chi_k\prod_{k'>k}(1-\chi_{k'})O^{T} }_{p_1}\\
&\le \sum_k 2\exp(\frac{1}{1-2^{-\alpha/4}}) D^{2r/p(k)}\sqrt{C_{k,p(k)}}\\
&\le \exp(\frac{1}{1-2^{-\alpha/4}})\frac{2}{1-2^{-2\alpha+2}} \frac{D^{\frac{a}{p_1}} bp_1 \frac{\tau}{r}}{1-D^{\frac{a}{p_1}} bp_1 \frac{\tau}{r}}\\
&:= c_2 \frac{\sqrt{D^{\frac{a}{p_1}} bp_1 \frac{\tau}{r}}}{1-\sqrt{D^{\frac{a}{p_1}} bp_1 \frac{\tau}{r}}}
\end{align}
where $(1-2^{-2\alpha+2})$ came from the geometric series for $k'\le k_*$. Summing up tails for different $r$,
\begin{align}
 \sum_{m=0}^\infty\vertiii{ \exists_{\pm 2^m r}\nexists_{\pm 2^{m+1}r}O^{\tau} }_{p_1}&\le \sum_{m=0}^\infty c_2 \frac{\sqrt{D^{\frac{a}{p_1}} bp_1 \frac{\tau}{2^mr}}}{1-\sqrt{D^{\frac{a}{p_1}} bp_1 \frac{\tau}{2^mr}}}\\
&\le \frac{ c_2}{1-1/\sqrt{2}} \frac{\sqrt{D^{\frac{a}{p_1}} bp_1 \frac{\tau}{r}}}{1-\sqrt{D^{\frac{a}{p_1}} bp_1 \frac{\tau}{r}}}:=c'_2\frac{\sqrt{D^{\frac{a}{p_1}} bp_1 \frac{\tau}{r}}}{1-\sqrt{D^{\frac{a}{p_1}} bp_1 \frac{\tau}{r}}}
\end{align}
where $1-1/\sqrt{2}$ came from the geometric series for $\ge 0$. Set $p =\max( \frac{r\epsilon^2}{\tau ec},2)=\frac{r\epsilon^2}{\tau ec}$ to get
\begin{align}
     \BP[\lV O^{\tau}_{\{r\}} P\rV > \epsilon ] \le \BP[ \sum_m \lV \exists_{\pm 2^mr}\nexists_{\pm 2^{m+1}r}O^{\tau} \rV\ge \epsilon ] &\le \BP[ \sum_m \lV (\exists_{\pm 2^mr}\nexists_{\pm 2^{m+1}r}O^{\tau})_{\le 2^{m+1}r} \rV_p\ge \epsilon ]\\
     & \le ( \frac{c'_2}{\epsilon}  \frac{\sqrt{D^{\frac{a}{p_1}} bp_1 \frac{\tau}{r}}}{1-\sqrt{D^{\frac{a}{p_1}} bp_1 \frac{\tau}{r}}})^{p_1}\\
     &\le ( \frac{c'_2}{\epsilon}\frac{\sqrt{D^{\frac{a}{2p_1}} bp_1 \frac{\tau}{r}}}{1-\frac{1}{c'_2}})^{p_1}:= D^{a2} ( \sqrt{cp_1 \frac{\tau}{r\epsilon^2}})^{p_1}\\
     &\le D^{a/2}\exp(-\frac{r\epsilon^2}{2ec\tau}) 
\end{align}
where in the fourth inequality we used $c'\sqrt{D^{\frac{a}{p_1}} bp_1 \frac{\tau}{r}} \le 1 $ otherwise the bound becomes vacuous. Expressing in terms of $\delta$:
 \begin{align}
     \BP\left[\lV O^{\tau}_{\{r\}} P\rV > \epsilon(\delta):=\sqrt{2ec\ln(D^{a/2}/\delta)\frac{\tau}{r}} \right]\le \delta
 \end{align}

The logarithmic dependence on $\delta$ captures the very strong tail bound that might be useful elsewhere. 
Our next goal is extracting the asymptotic velocity, which only requires the second moment bound and Chebyshev. Take the order of limit: pick $\epsilon >0$, then for any function approaching infinity abrbitrarily slowly, $\lim_{r\rightarrow\infty} f(r)=\infty$, define $\tau(r)$ such that
\begin{align}
     \tau = rf(r),
\end{align}
we obtain a linear light cone almost surely
\begin{align}
    \lim_{r_0\rightarrow \infty} \BP\bigg[\lV\frac{1}{2}[A_{r},O_{0}(\tau(r))]P\rV \ge \epsilon \bigg]= 0. 
\end{align}
Note we can also get a velocity for a fixed $\epsilon >0, \delta >0$:
\begin{align}
    u = 2ec\frac{\ln(D^{a/2}/\delta)}{\epsilon^2}.
\end{align}

\textbf{Case $3/2<\alpha < 2$:}
Moreover, we now find 
\begin{equation}
M = \sum_{q=1}^{n_*} 2^{q(4-2\alpha)/4} < r^{1-\alpha/2} \sum_{k^\prime=0}^\infty 2^{-k^\prime((4-2\alpha)/4} = \frac{r^{1-\alpha/2}}{1-2^{-(4-2\alpha)/4}}=: M'
\end{equation}
and that 
\begin{equation}
\frac{1}{2^{k_*\alpha/2}} \le \frac{2M}{r} \le  \frac{2M'}{r} \le \frac{2r^{-\alpha/2}}{1-2^{-(4-2\alpha)/4}} 
\end{equation}

For $k\le k_*$,
\begin{align}
D^{4r/p(k)}C_{k,p(k)} &= \left( D^{4r/p(k)N_k} e^{\frac{\eta_k^2R_kT}{pN_k}}e^3 2^{8+2\alpha}M'^3p \frac{\tau}{r2^{k\alpha/2}}\right)^{N_k}\\
&= \left( D^{4r/p(k)N_k} e^{\frac{\eta_k^2R_kT}{pN_k}}\frac{e^3 2^{8+2\alpha}}{(1-2^{-(4-2\alpha)/4})^3}p \frac{\tau}{r^{3\alpha/2 -2}2^{k\alpha/2}}\right)^{N_k}\\
&= \left( D^{\frac{8M'2^{\alpha/2}}{p_1}} c_1\frac{e^3 2^{8+3\alpha/2}}{(1-2^{-(4-2\alpha)/4})^3}p_1 \frac{\tau}{r^{3\alpha/2 -2}}\right)^{N_k}\\
&\le \left( D^{\frac{a}{p_1}} bp_1 \frac{\tau}{r^{3\alpha/2 -2}}\right)^{N_k}
\end{align}
where 
\begin{prop}
$e^{\frac{\eta_k^2R_kT}{pN_k}} \le c_1$
\end{prop}
\begin{proof}
\begin{align}
\frac{\eta_k^2R_kT}{pN_k} &\le \frac{\tau2^{2\alpha}M'}{p2^{k(3\alpha/2-2)}r} \\
&\le \frac{\tau2^{2\alpha}}{p2^{k(3\alpha/2-2)}}\frac{r^{-\alpha/2}}{1-2^{-(4-2\alpha)/4}} \\
&\le \frac{\tau}{r^{3\alpha/2-2}}\frac{2^{2\alpha}}{pr^{2-\alpha}2^{k(3\alpha/2-2)}}\frac{1}{1-2^{-(4-2\alpha)/4}}\\
&\le \frac{(1-2^{-(4-2\alpha)/4})^3}{p_1 e^3 2^{8+3\alpha/2}} \frac{2^{2\alpha}}{pr^{2-\alpha}2^{k(3\alpha/2-2)}}\frac{1}{1-2^{-(4-2\alpha)/4}}\\
&\le \frac{(1-2^{-(4-2\alpha)/4})^3}{ 2 e^3 2^{8+3\alpha/2}} \frac{2^{2\alpha}}{4}\frac{1}{1-2^{-(4-2\alpha)/4}}:=\ln(c_1)
\end{align}
where we used $p \ge 2, r^{2-\alpha}>1, 2^{k(3\alpha/2-2)}>2$.
\end{proof}
Hence 
\begin{align}
\vertiii{ \exists_{\pm r}\nexists_{\pm 2r}O^{\tau}}_{p_1}& = \vertiii{[1-\prod_k(1-\chi_k)]O^{\tau}}_{p_1}\\
&\le \sum_k \vertiii{\chi_k\prod_{k'>k}(1-\chi_{k'})O^{\tau}}_{p_1}\\
&\le \sum_k 2\exp(\frac{1}{1-2^{-\alpha/4}}) D^{2r/p(k)}\sqrt{C_{k,p(k)}}\\
&\le \exp(\frac{1}{1-2^{-\alpha/4}})\frac{2}{1-2^{-2\alpha+2}} \frac{D^{\frac{a}{p_1}} bp_1 \frac{\tau}{r^{3\alpha/2 -2}}}{1-D^{\frac{a}{p_1}} bp_1 \frac{\tau}{r^{3\alpha/2 -2}}}\\
&:= c_2 \frac{\sqrt{D^{\frac{a}{p_1}} bp_1 \frac{\tau}{r^{3\alpha/2 -2}}}}{1-\sqrt{D^{\frac{a}{p_1}} bp_1 \frac{\tau}{r^{3\alpha/2 -2}}}}
\end{align}
where $(1-2^{-2\alpha+2})$ came from the geometric series for $k'\le k_*$.
\begin{align}
     \BP[\lV \exists_{\pm r}\nexists_{\pm 2r}O^{\tau} \rV \ge \epsilon ] \le \BP[ \lV (\exists_{\pm r}\nexists_{\pm 2r}O^{\tau} )_{\le 2r} \rV_p\ge \epsilon ]&\le \frac{\BE[\lV (\exists_{\pm r}\nexists_{\pm 2r}O^{\tau}  )_{\le2r}\rV_p^p ]}{\epsilon^p}\\
     & \le ( \frac{c_2}{\epsilon}  \frac{\sqrt{D^{\frac{a}{p_1}} bp_1 \frac{\tau}{r^{3\alpha/2 -2}}}}{1-\sqrt{D^{\frac{a}{p_1}} bp_1 \frac{\tau}{r^{3\alpha/2 -2}}}})^{p_1}\\
     & \le ( \frac{c_2}{\epsilon}  \frac{\sqrt{D^{\frac{a}{p_1}} bp_1 \frac{\tau}{r^{3\alpha/2 -2}}}}{1-1/c_2})^{p_1}:=( \sqrt{D^{\frac{a}{p_1}} c'p_1 \frac{\tau}{\epsilon^2r^{3\alpha/2 -2}}})^{p_1}\\     
     &\le D^{a'r^{1-\alpha/2}} \exp(-c{\frac{\epsilon^2r^{3\alpha/2 -2}}{\tau}})\label{eq:long_range_deviation_decomposed_br_spectral}
\end{align}
where we set $\displaystyle p_1 = \frac{1}{e}\frac{\epsilon^2r^{3\alpha/2-2}}{c'\tau}\ge 2$.\footnote{Here we do not have to worry about the constraint $p \ge 2$, because when that happen, $p=2$, $D^{a'r^{1-\alpha/2}} \exp(-1)\ge 1$, i.e. the bound is vacuous anyway.}, and in the third inequality we used $c_2\sqrt{D^{\frac{a}{p_1}} bp_1 \frac{\tau}{r^{3\alpha/2 -2}}} \le 1 $ otherwise the bound becomes vacuous. In the last inequality we set $a':=\frac{2^{2+\alpha/2}}{1-2^{-(4-2\alpha)/4}}$ to emphasize the $r$ dependence and $c=1/2ec'$.

The strategy is to do a union bound by putting together concentration bounds for each $\lV \exists_{\pm r}\nexists_{\pm 2r}O^{\tau}  \rV := f(2^mr_0)$, tuning an error $\epsilon_{2^mr_0}$ corresponding to a failure probability $\delta_{2^mr_0}$.
\begin{align}
    \epsilon_{2^mr_0} &= \sqrt{
    \frac{\tau}{cr^{3\alpha/2-2}}(\ln(D)a'r^{1-\alpha/2} - \ln(\delta_{2^mr_0}) )
    }
    \\ \textrm{ensures\ } &\BP[f(2^mr_0) \ge \epsilon_{2^mr_0} ] \le \delta_{2^mr_0}
\end{align}
where we simply invert Eq.(\ref{eq:long_range_deviation_decomposed_br_spectral}) to get $\epsilon$ in terms of $\delta$. In order that the union bound works, set the failure probabilities to decay with
\begin{align}
\delta_{2^mr_0} = \delta_{r_0}\exp(-2^{m(1-\alpha/2)}).     
\end{align}
Then 
\begin{align}
\epsilon_{2^mr_0} &= \sqrt{
    \frac{\tau}{cr^{3\alpha/2-2}}\left(\ln(D)a'r^{1-\alpha/2} - \ln(\delta_{r_0})+2^{m(1-\alpha/2)} \right)}
\end{align}
The sum of error will be important later for plugging in the union bound.
\begin{align}
    \sum^{\infty}_{m = 0} \epsilon_{2^mr_0}  &= \sum^{\infty}_{m = 0} \sqrt{
    \frac{\tau}{cr^{3\alpha/2-2}}\left(\ln(D)a'r^{1-\alpha/2} - \ln(\delta_{r_0})+2^{m(1-\alpha/2)}\right)}\\   
    &\le\sqrt{
    \frac{\tau}{c_0r_0^{3\alpha/2-2}}\left(\ln(D)a'r_0^{1-\alpha/2} - \ln(\delta_{r_0})+1 \right)} 
     =: \epsilon'_{r_0}
\end{align}
where we used the geometric since the summand is decaying by $\sqrt{2^{3-2\alpha}}<1$ and $c_0=c(1-2^{3/2-\alpha})^2$.
We now use the union bound
\begin{align}
    \BP[\lV O^{\tau} _{\ge r_0} \rV \ge \epsilon'_{r_0}] \le \BP[\lV O^{\tau} _{\ge r_0} \rV \ge \sum^{\infty}_{m = 0} \epsilon_{r_0 + m}  &\le \BP[\sum_{m=0}^\infty \lV \exists_{\pm r}\nexists_{\pm 2r}O^{\tau}  \rV \ge \sum^{\infty}_{m = 0} \epsilon_{2^mr_0} ] \\
    &\le \BP[\bigcup_{m=0}^\infty \lV O^{\tau} _{2^mr_0} \rV \ge \epsilon_{2^mr_0} ] \\
    & \le \sum_{m=0}^\infty \BP[\lV O^{\tau} _{2^mr_0} \rV \ge \epsilon_{2^mr_0} ] \\
    & \le \sum_{m=0}^\infty \delta_{2^mr_0} = \sum_{m=0}^\infty \delta_{r_0}\exp(-2^{m(1-\alpha/2)}).\\
    &\le \frac{\delta_{r_0}}{e(1-1/e)(1-\alpha/2)}:=c_3\delta_{r_0}
\end{align}    
showing the concentration bound that
\begin{align}
    \BP(\lV O^\tau_{\ge r} \rV \ge \epsilon )\le c_3e D^{a'r^{1-\alpha/2}} \exp(-{c_0\frac{\epsilon^2r^{3\alpha/2 -2}}{\tau}})
\end{align}
and with failure probability $\delta = \delta_{r_0}c_3$,
\begin{align}
    \lV O^\tau_{\ge r} \rV\le \epsilon(\delta) = \sqrt{
    \frac{\tau}{c_0r^{3\alpha/2-2}}\left(\ln(D)a'r^{1-\alpha/2} - \ln(\delta/c_3e) \right)} 
\end{align}
In particular we can obtain asymptotic velocity. For any $\epsilon>0$, $\zeta>0$, let $\tau(r)$ such that 
\begin{align}
    \frac{r^{2\alpha-3}}{\tau(r)} = \frac{\ln(D)a'}{c_0\epsilon^2} +\zeta
\end{align}
Then the commutator norm vanishes outside the algebraic light cone $\tau(r)$ almost surely
\begin{align}
    \lim_{r\rightarrow \infty} \BP(\lV\frac{1}{2}[A_{r},O_{0}(\tau(r))]\rV \ge \epsilon )\le \lim_{r\rightarrow \infty}c_3e\exp(-\zeta c_0\epsilon^2r^{1-\alpha/2})=0. 
\end{align}



\end{document}